\documentclass[12pt,reqno]{amsart}

\usepackage{amsmath}
\usepackage{amssymb}
\usepackage{amsthm}
\usepackage{stmaryrd}
\usepackage{hyperref}
\usepackage{cleveref}
\usepackage{geometry}
\usepackage[linesnumbered,ruled]{algorithm2e}
\usepackage{array}       
\usepackage{arydshln}  
\usepackage{scalerel,stackengine}
\usepackage{float}    
\usepackage{diagbox}  
\usepackage{multirow} 
\usepackage{xcolor}   
\usepackage{subcaption}
\usepackage{graphicx}
\graphicspath{{fig/}}

\allowdisplaybreaks

\usepackage{color}
\newcommand{\ri}{{\rm i}}
\newcommand{\kr}{k_\rho}
\newcommand{\bs}[1]{\boldsymbol{#1}}
\newcommand{\bd}[1]{\mathbf{#1}}
\newcommand{\bb}[1]{\mathbb{#1}}
\newcommand{\J}[1]{\bb{J}_{#1}}

\newcommand{\Cont}[1]{\left \llbracket {#1} \right \rrbracket}

\theoremstyle{definition}
\newtheorem{theorem}{Theorem}

\newtheorem{prop}{Proposition}

\theoremstyle{remark}
\newtheorem{remark}{Remark}

\title{On the computation of Green's function in layered elastic media}
\author{Heng Yuan, Wenzhong Zhang, Bo Wang, Wei Cai}

\begin{document}
	
	\begin{abstract}
		We develop a recursive scheme for the computation of Green's function for time-harmonic elastic wave in three-dimensional layered media, based on a matrix basis representation in the Fourier domain and a classic interface reaction-transmission theory.
		With transversal rotational symmetry preserved, the scheme decouples the Green's function into SH waves and P-SV waves using five scalar functions, which are recursively solved at linear cost with respect to the number of layers, regardless of the layers the source and target particles locate in.
		The resulting scheme leads to a stable numerical method ready for Hankel transform evaluation in the spatial domain, constituting a foundation for the analysis and efficient numerical simulation of elastic wave scattering in layered media.
		
		Keywords: elastic wave equation, layered media, Green's functions, matrix basis
	\end{abstract}
	
	\maketitle
	
	\section{Introduction}
	The propagation of elastic waves in layered media is a fundamental physical problem with applications in seismology, geophysical prospecting \cite{aki2002quantitative,wei2022theory} and the development of sound-absorbing materials\cite{umnova2010propagation}, etc.
	For efficient computation with a reduced cost of meshing, numerical simulation of wave propagation is often built upon boundary integral equation (BIE) methods through the use of reciprocal theorems, which rely on the evaluation of Green's functions.
	
	In a homogeneous, isotropic and unbounded medium, the free-space Green's function for a point source was derived by Stokes \cite{stockes1849dynamical}.
	For time-harmonic free-space problems, the $3 \times 3$ tensor function is also known as the Kupradze matrix \cite{kupradze1976three}.
	For problems in stratified media with piecewise constant material parameters, the BIE methods either adopt free-space Green's function with additional integral equations for each material surface, or utilize the layer-media Green's function (LMGF) that naturally incorporate the transmission conditions across interfaces.
	In \cite{gonsalves1990direct}, for instance, BIE methods based on both types of Green's functions have been numerically verified for solving a half-space problem.
	For cases with multiple layers where using LMGF saves more integral equations, however, there has long been a gap in the literature between the study of elastic wave propagation in layered media and the practical use of LMGF in BIE methods.
	On the one hand, the Thomson--Haskell method \cite{thomson1950transmission,haskell1990dispersion}, the ray method \cite{muller1970exact}, the reflectivity method \cite{fuchs1971computation}, etc. developed layer-wise recursion schemes for the propagation of body waves in layered media. Subsequent works in \cite{haskell1960crustal,haskell1962crustal,knopoff1964matrix} further introduced a decoupling of horizontally polarized shear (SH) waves, and a mixture of compressional (P) waves with vertically polarized shear (SV) waves.
	On the other hand, early BIE methods such as the Aki--Larner method \cite{larner1970near}, the discrete wavenumber method (DWM) \cite{bouchon1976discrete,bouchon1979discrete,bouchon1981simple,bouchon1980calculation} and modal methods \cite{harvey1981seismogram}, etc. apply discretization and truncation on the frequency domain, approximating the point incitation of the Green's function by a finite superposition of planar waves.
	Typical issues lying within the gap include the scalability of the LMGF schemes for multiple layers, and the numerical stability for the tail evaluation of the inverse Fourier transform, or equivalently the underlying Hankel transform, as the formulation contains exponentially increasing terms on the tail.
	Besides, there is also a minor distinction between the representation required by body wave scattering and by the LMGF: \cite{knopoff1964matrix} reports a total of $6$ degrees of freedom (DoF) per layer in body wave scattering ($2$ for SH and $4$ for P-SV), meanwhile the actual DoFs for LMGF is $5$, to be elaborated later in this paper. Inconsistencies in DoF might also bring numerical issues when calculating the LMGF in the frequency domain.
	
	The goal of this work is to fill this gap by developing an efficient scheme to evaluate the elastic-wave LMGF in the frequency domain. The proposed scheme can handle multilayer configurations with varying source–target positions and control the exponential terms in the general reflection and transmission matrices, thereby paving the way for follow-up research on on the fast multipole method (FMM), which has proven effective in accelerating BIE methods for Maxwell's equations in layered media \cite{yuan2026fast,eibert2026multilevel}. 
	The mathematical foundation of our approach is a matrix basis representation \cite{bo2022maxwellDGF} of $3 \times 3$ tensors for Green's functions in the frequency domain, with an extension of the classic reflection-transmission type derivation from \cite{chew1999waves}.
	The matrix basis representation categorizes the transversal rotational symmetry of LMGF into $5$ DoFs, which linearly transform across interfaces.
	The explicit separation of rotational asymmetry allows neat conversions from two-dimensional inverse Fourier transforms into Hankel transforms with order up to two, which are also studied as Sommerfeld-type integrals.
	Then, a reformulation of the governing equations of the LMGF rediscovers the decoupling of SH waves and P-SV waves (c.f. \cite{knopoff1964matrix}), written using five scalar functions.
	Finally, the reaction-transmission interface handling is implemented in a recursive manner that consumes linear cost with respect to the number of layers, with a careful handling of exponential factors.
	The proposed scheme is justified by numerical tests in the spatial domain.
	
	The rest of the paper is organized as follows.
	In \Cref{sect_lmgf}, we briefly review the Green's function of the elastic wave equation in the free space, and then elaborate the matrix basis formulation of the LMGF, with a $5$ DoF decoupling of SH and P-SV waves.
	In \Cref{sect_solver}, a recursive scheme based on the reflection-transmission theory for addressing mixed-wavenumber interface problems is proposed, with discussion on the numerical evaluation in the spatial domain.
	Numerical tests in \Cref{sect_numer} including verification on interface conditions and a convergence test towards a homogeneous medium validate the numerical quality of proposed scheme. A final conclusion and discussion on our future work are given in \Cref{sect_conclusion}.

	\section{The Green's function of elastic wave equation in layered media}\label{sect_lmgf}
	
	In this section, we employ the classic potential functions and the matrix basis proposed in \cite{bo2022maxwellDGF} to reduce the computation of the LMGF into three Helmholtz problems in layered media with arbitrary layers. 
	
	\subsection{The free-space Green's function of elastic wave equation}\label{sect_fsgf}
	Suppose a homogeneous and isotropic solid material occupies the entire space, with density $\rho$ and Lamé constants $\lambda, \mu$, respectively. With an external source $\bd{b}(\bd{r})$, and assuming the displacement $\bd{u} = (u_1, u_2, u_3)^T$ is sufficiently small, the time harmonic elastic wave equation in solid is given by:
	\begin{equation}\label{elasticwaveeqform1}
		-\omega^2 \rho \bd{u} = \nabla \cdot \mathcal{T} + \bd{b},
	\end{equation}
	where the $3 \times 3$ tensor $\bs{\mathcal{T}}$ is known as the stress tensor with entries given by
	\begin{equation*}
		\mathcal{T}=\lambda(\nabla\cdot \bs u)\mathbb I+\mu(\nabla\bs u+(\nabla\bs u)^{\rm T}),
	\end{equation*}
	and $\mathbb I$ is the $3\times 3$ identity matrix. The time harmonic dependence $e^{-\ri \omega t}$ is adopted throughout this paper. Alternatively, the elastic wave equation \eqref{elasticwaveeqform1} can be written using matrix calculus notation as
	\begin{equation*}
		(\lambda + \mu) \nabla \nabla \cdot \bd{u} + \mu \nabla^2 \bd{u} + \omega^2 \rho \bd{u} = -\bd{b}.
	\end{equation*}
	
	Given a source at location $\bd{r}'$, the free space dyadic Green's function $\bb{G}^f(\bd{r};\bd{r}')$ of the elastic wave equation satisfies 
	\begin{equation}\label{eq_EWE_GF}
		(\lambda + \mu) \nabla \nabla \cdot \bb{G}^f + \mu \nabla^2 \bb{G}^f + \omega^2 \rho \bb{G}^f = -\delta(\bd{r} - \bd{r}') \bb{I},
	\end{equation}
	with well-known solution\cite[Chapter 8.3]{chew2022integral}
	\begin{equation}\label{freespacepart_EWE}
		\bb{G}^f(\bd{r};\bd{r}') = \dfrac{1}{\mu}\left(\bb{I}+\dfrac{\nabla\nabla}{(k^{s})^{2}}\right)G_{k^{s}}^f(\mathbf{r},\mathbf{r}')-\dfrac{1}{\gamma}\dfrac{\nabla\nabla}{(k^{c})^{2}}G_{k^{c}}^f(\mathbf{r},\mathbf{r}'),
	\end{equation}
	where $\gamma = \lambda + 2\mu$,
	\begin{equation}
		k^{s} = \sqrt{\frac{\omega^2 \rho}{\mu}}, \quad k^{c} = \sqrt{\frac{\omega^2 \rho}{\gamma}}
	\end{equation}
	are the wave numbers of the S-wave and the P-wave, respectively, and
	\begin{equation}\label{eq_gf}
		G_{k}^f(\mathbf{r},\mathbf{r}')=\dfrac{e^{\ri  k|\mathbf{r}-\mathbf{r}'|}}{4\pi|\mathbf{r}-\mathbf{r}'|}
	\end{equation}
	is the free-space Green's functions of the Helmholtz equation with wave number $k$. We assume both $\gamma, \mu > 0$ as happens to most solid materials. 
	
	\subsection{The elastic wave equation with point source in layered media}
\begin{figure}
		\centering
		\includegraphics[width=0.5\linewidth]{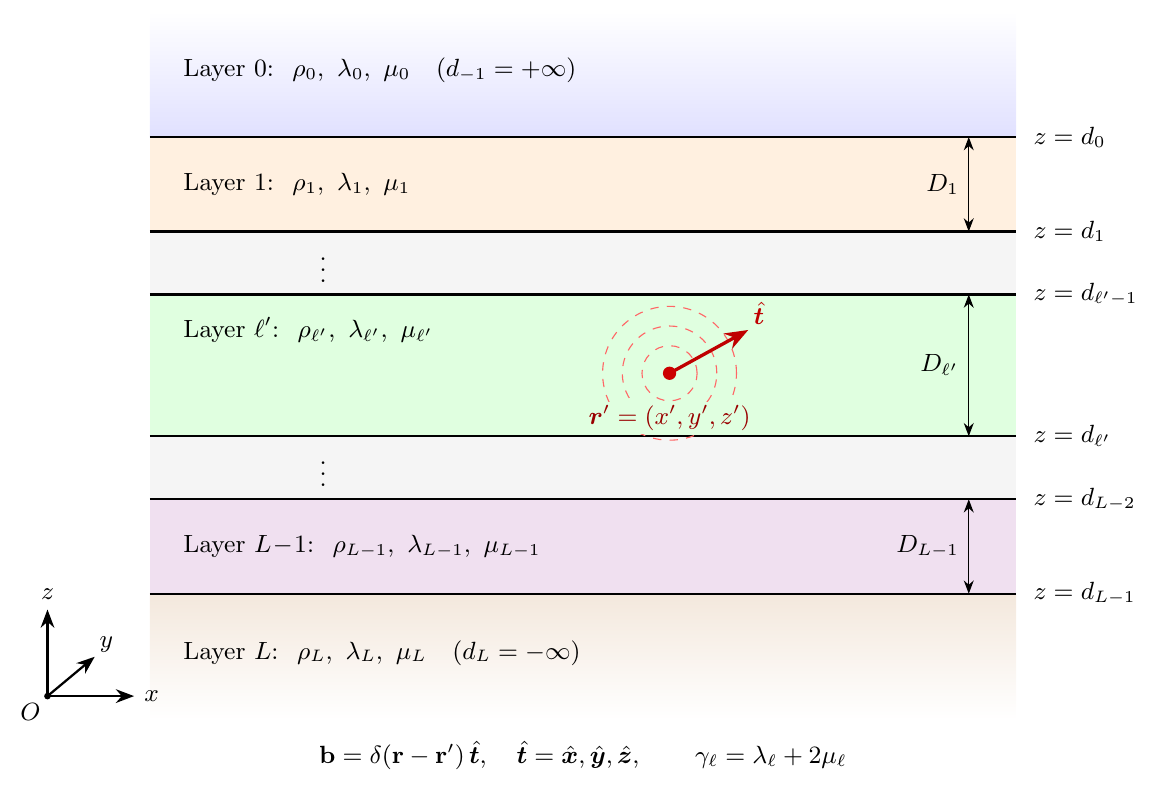}
		\caption{An illustration of the layered media structure}
		\label{fig:layeredmedia}
	\end{figure}

	Consider a multi-layered medium (see Fig.  \ref{fig:layeredmedia}) with $L+1$ layers along the z-direction, where the interface is located at $z = d_{\ell}$ for $\ell = 0, 1, \ldots, L-1$. We will also use the notation $d_{-1}=+\infty$, $d_{L}=-\infty$ throughout this paper. 
	Each layer is homogeneous with constant density $\rho_\ell$ and Lamé constants $\lambda_\ell, \mu_\ell$, $\ell = 0, \cdots, L$. Then, the P-wave modulus is given by
	\begin{equation}\label{eq_gamma}
		\gamma_\ell = \lambda_\ell + 2 \mu_\ell.
	\end{equation}
	The thicknesses of the layers are given by
	\begin{equation}\label{layer_height_coffD}
		D_{\ell} = d_{\ell-1} - d_{\ell}, \quad \ell = 1,\cdots, L-1.
	\end{equation}
	
	Suppose there is a point source at $\bd{r}^{\prime}=(x^{\prime},y^{\prime}, z^{\prime})$ in the $\ell^{\prime}$-th layer ($d_{\ell^{\prime}}<z^{\prime}<d_{\ell^{\prime}-1}$) of the layered medium with directed unit impulse force
	\begin{equation}
		\mathbf{b}=\delta(\mathbf{r}-\mathbf{r}')\hat{\bs t},\quad \hat{\bs t}=\hat{\bs x},\hat{\bs y},\hat{\bs z}.
	\end{equation} 
	The LMGF $\bd{G}^{\hat{\bs t}}(\bd r, \bd r')=[{G}^{\hat{\bs t}}_1, {G}^{\hat{\bs t}}_2, {G}^{\hat{\bs t}}_3]^{\rm T}$ for elastic wave equation satisfies
	\begin{equation}\label{eq_EWE_Gr}
		(\lambda_{\ell} + \mu_{\ell}) \nabla \nabla \cdot \bd{G}^{\hat{\bs t}}(\bd{r},\bd{r}') + \mu_{\ell} \nabla^2 \bd{G}^{\hat{\bs t}} (\bd{r},\bd{r}')+ \omega^2 \rho_{\ell} \bd{G}^{\hat{\bs t}}(\bd{r},\bd{r}') = -\delta(\bd{r} - \bd{r}') {\hat{\bs t}},
	\end{equation}
	at field point $\bd{r}=(x,y,z)$ in the $\ell$-th layer ($d_{\ell}<z<d_{\ell-1}$). Due to the presence of the layered media, the Green's function must satisfy transmission conditions
	\begin{equation}\label{eq_interfacecond}
		\Cont{\bd G^{\hat{\bs t}}}=\bd{0}_{3\times 1},\quad \Cont{\bd{n} \cdot \mathcal T^{\hat{\bs t}} } = \bd{0}_{3\times 1}, \quad \hat{\bs t}=\hat{\bs x},\hat{\bs y},\hat{\bs z},
	\end{equation}
	across the interfaces $z=d_{\ell}$ with the normal vector  $\bd{n} = \bd{e}_3 = (0, 0, 1)$, where 
	\begin{equation}\label{stresstensor}
		\mathcal{T}^{\hat{\bs t}}=\lambda(\nabla\cdot \bd G^{\hat{\bs t}})\mathbb I+\mu(\nabla\bd G^{\hat{\bs t}}+(\nabla\bd G^{\hat{\bs t}})^{\rm T})
	\end{equation}
	is the stress tensor of the Green's function $\bd G^{\hat{\bs t}}$, and $\llbracket\cdot\rrbracket$ represents the jump of the piece-wise smooth function across the interface $z=d_{\ell}$, i.e. 
	\begin{equation*}
		\llbracket f\rrbracket = \lim_{z\to d_{\ell}+0}f - \lim_{z\to d_{\ell}-0}f.
	\end{equation*}
	If no specific layer is of interest, we omit the subscript $\ell$ for simplicity. Equation \eqref{stresstensor}, for instance, means the expression valid in any given layer.
	

	\subsection{Potential functions inside each layer}
	As in the derivation of the dyadic Green's function in the free space \cite{chew2022integral}, we introduce the vector potential function $\mathbf G_{\bs A}^{\hat{\bs t}}=\nabla\times \mathbf G^{\hat{\bs t}}$ and the scalar potential function $G_B^{\hat{\bs t}}=\nabla\cdot \mathbf G^{\hat{\bs t}}$ inside each layer, 
	which satisfy the following equations:
	\begin{equation}\label{potential_EQ}
		\begin{split}
			\Delta\mathbf G_{\bs A}^{\hat{\bs t}}(\mathbf{r},\mathbf{r}')+(k_{\ell}^{s})^{2}\mathbf G_{\bs A}^{\hat{\bs t}}(\mathbf{r},\mathbf{r}')=-\frac{1}{\mu_{\ell}}\nabla\times(\delta(\mathbf{r}-\mathbf{r}')\hat{\bs t}),\quad d_{\ell}<z<d_{\ell-1},\\
			\Delta G_B^{\hat{\bs t}}(\mathbf{r},\mathbf{r}')+(k_{\ell}^{c})^{2}G_B^{\hat{\bs t}}(\mathbf{r},\mathbf{r}')=-\dfrac{1}{\gamma_{\ell}}\nabla\cdot(\delta(\mathbf{r}-\mathbf{r}')\hat{\bs t}),\quad d_{\ell}<z<d_{\ell-1},
		\end{split}
	\end{equation}
	where the two wave numbers are given by
	\begin{equation}
		k_{\ell}^{s} = \sqrt{\frac{\omega^2 \rho_{\ell}}{\mu_{\ell}}}, \quad k_{\ell}^{c} = \sqrt{\frac{\omega^2 \rho_{\ell}}{\gamma_{\ell}}}.
	\end{equation}
	Then, the potential functions are solved as
	\begin{equation}\label{secondpotential}
		\mathbf G_{\bs A}^{\hat{\bs t}}(\mathbf{r},\mathbf{r}')=\nabla\times {\bs\Phi}^{\hat{\bs t}}(\mathbf{r},\mathbf{r}'),\quad G_B^{\hat{\bs t}}(\mathbf{r},\mathbf{r}')=\nabla\cdot {\bs\Psi}^{\hat{\bs t}}(\mathbf{r},\mathbf{r}'),
	\end{equation}
	where the vector functions ${\bs\Phi}^{\hat{\bs t}}$ and ${\bs\Psi}^{\hat{\bs t}}$ satisfy the vector Helmholtz equations:
	\begin{equation}\label{potentialeqs}
		\begin{aligned}
			\Delta{\bs\Phi}^{\hat{\bs t}}(\mathbf{r},\mathbf{r}')+(k_{\ell}^{s})^{2}{\bs\Phi}^{\hat{\bs t}}(\mathbf{r},\mathbf{r}')=-\dfrac{1}{\mu_{\ell'}}\delta(\mathbf{r}-\mathbf{r}')\hat{\bs t},\quad d_{\ell}<z<d_{\ell-1},\\ 
			\Delta{\bs\Psi}^{\hat{\bs t}}(\mathbf{r},\mathbf{r}')+(k_{\ell}^{c})^{2}{\bs\Psi}^{\hat{\bs t}}(\mathbf{r},\mathbf{r}')=-\dfrac{1}{\gamma_{\ell'}}\delta(\mathbf{r}-\mathbf{r}')\hat{\bs t},\quad d_{\ell}<z<d_{\ell-1}.
		\end{aligned}
	\end{equation}
	Using the identity
	\begin{equation*}
		\Delta {\bd G}^{\hat{\bs t}}=\nabla\nabla\cdot {\bd G}^{\hat{\bs t}}-\nabla\times\nabla\times {\bd G}^{\hat{\bs t}}
	\end{equation*}
	and the definition of potential functions $\mathbf G_{\bs A}^{\hat{\bs t}}$ and $G_B^{\hat{\bs t}}$ in equation \eqref{eq_EWE_Gr}, we get similar representation as in the free space case, i.e.,
	\begin{equation}\label{potentialrep1}
		{\mathbf G}^{\hat{\bs t}}(\mathbf{r},\mathbf{r}')=-\frac{1}{(k_{\ell}^{c})^{2}}\nabla G_B^{\hat{\bs t}}(\mathbf{r},\mathbf{r}')+\frac{1}{(k_{\ell}^{s})^{2}}\nabla \times\mathbf G_{\bs A}^{\hat{\bs t}}(\mathbf{r},\mathbf{r}')-\frac{1}{\omega^2\rho_{\ell}}\delta(\mathbf{r}-\mathbf{r}')\hat{\bs t},\quad d_{\ell}<z<d_{\ell-1},
	\end{equation}
	for all $\ell=0, 1, \cdots, L$. Further, substituting \eqref{secondpotential} into \eqref{potentialrep1} and using the equations \eqref{potentialeqs} to cancel the delta function gives formula
	\begin{equation}\label{Elasticdynamics_layerDGf_hatt}
		\mathbf{G}^{\hat{\bs t}}(\mathbf{r},\mathbf{r}')=\left(\bb{I}+\dfrac{\nabla\nabla}{(k_{\ell}^{s})^{2}}\right){\bs\Phi}^{\hat{\bs t}}(\mathbf{r},\mathbf{r}')-\dfrac{\nabla\nabla}{(k_{\ell}^{c})^{2}}{\bs\Psi}^{\hat{\bs t}}(\mathbf{r},\mathbf{r}'),\quad d_{\ell}<z<d_{\ell-1},
	\end{equation}
	in each layer. Therefore, the dyadic Green function 
	\begin{equation}
		\bb{G}(\bd r, \bd r')=\left[\bd{G}^{\hat{\bs x}}, \bd{G}^{\hat{\bs y}}, \bd{G}^{\hat{\bs z}}\right]
	\end{equation}
	has the form
	\begin{equation}\label{Elasticdynamics_layerDGf_matrix}
		\bb{G}(\bd r, \bd r')=\left(\bb{I}+\dfrac{\nabla\nabla}{(k_{\ell}^{s})^{2}}\right){\bs\Phi}(\mathbf{r},\mathbf{r}')-\dfrac{\nabla\nabla}{(k_{\ell}^{c})^{2}}{\bs\Psi}(\mathbf{r},\mathbf{r}'),
	\end{equation}
	where the dyadic potential functions
	\begin{equation}
		{\bs\Phi}(\mathbf{r},\mathbf{r}')=\begin{bmatrix}
			{\bs\Phi}^{\hat{\bs x}} & {\bs\Phi}^{\hat{\bs y}} & {\bs\Phi}^{\hat{\bs z}}
		\end{bmatrix},\quad{\bs\Psi}(\mathbf{r},\mathbf{r}')=\begin{bmatrix}
			{\bs\Psi}^{\hat{\bs x}} & {\bs\Psi}^{\hat{\bs y}} & {\bs\Psi}^{\hat{\bs z}}
		\end{bmatrix}
	\end{equation}
	satisfy the dyadic Helmholtz equations:
	\begin{equation}\label{potentialeqs_matrix}
		\Delta{\bs\Phi}(\mathbf{r},\mathbf{r}')+(k_{\ell}^{s})^{2}{\bs\Phi}(\mathbf{r},\mathbf{r}')=-\dfrac{1}{\mu_{\ell}}\delta(\mathbf{r}-\mathbf{r}')\mathbb{I},\quad \Delta{\bs\Psi}(\mathbf{r},\mathbf{r}')+(k_{\ell}^{c})^{2}{\bs\Psi}(\mathbf{r},\mathbf{r}')=-\dfrac{1}{\gamma_{\ell}}\delta(\mathbf{r}-\mathbf{r}')\mathbb{I}
	\end{equation}
	in each layer and interface conditions across the interfaces $\{z=d_{\ell}\}_{\ell=0}^{L-1}$ which are given implicitly in \eqref{eq_interfacecond}.
	
	Next, we shall establish a representation for ${{\bs\Phi}}(\mathbf{r},\mathbf{r}')$ and ${{\bs\Psi}}(\mathbf{r},\mathbf{r}')$ by using the matrix basis proposed in our previous paper \cite{bo2022maxwellDGF}. This representation enables us to decouple the interface conditions in \eqref{eq_interfacecond} and then leads to three Helmholtz interface problems for the coefficients in the representation. 
	
	\subsection{Derivation of the LMGFs in the Fourier domain}\label{subsection_derivation_LMGF}
	Note that the point source $\delta(\bd{r} - \bd{r}') \bd{I}$ is trivial unless $\ell = \ell'$, i.e. when the target and the source are from the same layer, one can subtract the free-space Green's function from the equation to eliminate the singularity, namely,
	\begin{equation}\label{reactionLMGF}
		\bb{G}_{\ell\ell'}^r(\bd{r};\bd{r}') = \bb{G}(\bd{r};\bd{r}') - \delta_{\ell\ell'} \mathbb{G}^f(\bd{r};\bd{r}'),
	\end{equation}
	which we call the \emph{reaction field} of the LMGF. Here, $\delta_{\ell\ell'}$ is the Kronecker symbol. We have added subscripts $\ell,\ell'$ in the notation of reaction components to clearly indicate the source and target layers.  As in the free space case \eqref{freespacepart_EWE}, $\bb{G}^f(\bd r, \bd r')$ due to point source in the $\ell'$-th layer can be rewritten as
	\begin{equation}\label{freespacepart_EWE_ellp}
		\bb{G}^f(\bd r, \bd r')=\left(\bb{I}+\dfrac{\nabla\nabla}{(k_{\ell'}^{s})^{2}}\right){\bs\Phi}^f(\mathbf{r},\mathbf{r}')-\dfrac{\nabla\nabla}{(k_{\ell'}^{c})^{2}}{\bs\Psi}^f(\mathbf{r},\mathbf{r}')
	\end{equation}
	where 
	\begin{equation}\label{Xst_freeexpand}
		{{\bs\Phi}}^f(\mathbf{r},\mathbf{r}')=\dfrac{1}{\mu_{\ell'}}{G}_{k_{\ell'}^{s}}^f(\mathbf{r},\mathbf{r}')\bb{I},\quad
		{{\bs\Psi}}^f(\mathbf{r},\mathbf{r}')=\dfrac{1}{\gamma_{\ell'}}{G}_{k_{\ell'}^{c}}^f(\mathbf{r},\mathbf{r}')\bb{I}.
	\end{equation}
	Define reaction components 
	\begin{equation}\label{layeredDecomp}
		{\bs\Phi}_{\ell\ell'}={\bs\Phi}-\delta_{\ell\ell'}{\bs\Phi}^f,\quad{\bs\Psi}_{\ell\ell'}={\bs\Psi}-\delta_{\ell\ell'}{\bs\Psi}^f,
	\end{equation}
	in the $\ell$-th layer. They are the solutions of homogeneous Helmholtz equations
	\begin{equation}\label{potentialeqs_reaction}
		\Delta{\bs\Phi}_{\ell\ell'}(\mathbf{r},\mathbf{r}')+(k_{\ell}^{s})^{2}{\bs\Phi}_{\ell\ell'}(\mathbf{r},\mathbf{r}')=0,\quad \Delta{\bs\Psi}_{\ell\ell'}(\mathbf{r},\mathbf{r}')+(k_{\ell}^{c})^{2}{\bs\Psi}_{\ell\ell'}(\mathbf{r},\mathbf{r}')=0,
	\end{equation} 
	in each layer. From \eqref{Elasticdynamics_layerDGf_matrix}, \eqref{reactionLMGF} and \eqref{freespacepart_EWE_ellp},  we obtain
	\begin{equation}\label{reactioncomponent}
		\bb{G}_{\ell\ell'}^r(\bd{r};\bd{r}') =\left(\bb{I}+\dfrac{\nabla\nabla}{(k_{\ell}^{s})^{2}}\right){\bs\Phi}_{\ell\ell'}(\mathbf{r},\mathbf{r}')-\dfrac{\nabla\nabla}{(k_{\ell}^{c})^{2}}{\bs\Psi}_{\ell\ell'}(\mathbf{r},\mathbf{r}')
	\end{equation}

	
	
	In order to solve the Helmholtz equations in \eqref{potentialeqs_reaction} subject to the interface conditions \eqref{eq_interfacecond}, we exploit the shift‑invariant property in the horizontal plane and apply the 2‑D Fourier transform in the $x‑y$ plane with respect to the offsets $x'$ and $y'$, i.e.,
	\begin{equation}\label{eq_2DFT}
		\mathcal{F}[f](k_x,k_y)=\int_{\mathbb{R}^2}f(x,y,z)e^{-\ri k_x(x-x')-\ri k_y(y-y')}dxdy.
	\end{equation}
	In the frequency domain, let $(\kr,\alpha)$ be the polar coordinate pair of $(k_x, k_y)$, i.e.
	\begin{equation}\label{eq_kx_ky}
		\quad k_x = \kr \cos \alpha, \quad k_y = \kr \sin \alpha.
	\end{equation} 
	The Fourier transform of equations in \eqref{potentialeqs_reaction} are ODEs with respect to $z$:
	\begin{equation}\label{potentialeqs_matrix_freq_react}
		\begin{aligned}
			\partial_{zz}\hat{{\bs\Phi}}_{\ell\ell'}(k_x, k_y, z, z')+(k_{\ell z}^{s})^{2}\hat{{\bs\Phi}}_{\ell\ell'}(k_x, k_y, z, z')=0,\\ 
			\partial_{zz}\hat{{\bs\Psi}}_{\ell\ell'}(k_x, k_y, z, z')+(k_{\ell z}^{c})^{2}\hat{{\bs\Psi}}_{\ell\ell'}(k_x, k_y, z, z')=0,
		\end{aligned}
	\end{equation}
	where $\hat{{\bs\Phi}}_{\ell\ell'}$ and $\hat{{\bs\Psi}}_{\ell\ell'}$ are the Fourier transforms of ${\bs\Phi}_{\ell\ell'}$ and ${\bs\Psi}_{\ell\ell'}$, respectively,
	\begin{equation}
		k_{\ell z}^{s} = \sqrt{(k_{\ell}^{s})^{2} - \kr^2},\quad k_{\ell z}^{c} = \sqrt{(k_{\ell}^{c})^{2} - \kr^2}
	\end{equation}
	with the branch cut chosen such that $\Im(k_{\ell z}^{s}), \Im(k_{\ell z}^{c}) \ge 0$ for all $\ell=0,1,\cdots,L$. The general solutions for these ODEs in each layer have the form:
	\begin{equation}\label{react_field_expand_ud}
		\begin{aligned}
			\hat{{\bs\Phi}}_{\ell\ell'}(k_x,k_y,z,z')=&\hat{{\bs\Phi}}^{\uparrow}_{\ell\ell'}(k_x,k_y,z')\hat{e}_{\ell}^{\uparrow}(z)+\hat{{\bs\Phi}}^{\downarrow}_{\ell\ell'}(k_x,k_y,z')\hat{e}_{\ell}^{\downarrow}(z),\\
			\hat{{\bs\Psi}}_{\ell\ell'}(k_x,k_y,z,z')=&\hat{{\bs\Psi}}^{\uparrow}_{\ell\ell'}(k_x,k_y,z')\tilde{e}_{\ell}^{\uparrow}(z)+\hat{{\bs\Psi}}^{\downarrow}_{\ell\ell'}(k_x,k_y,z')\tilde{e}_{\ell}^{\downarrow}(z)
		\end{aligned}
	\end{equation}
	where $\hat{{\bs\Phi}}_{\ell\ell'}^{\ast},\hat{{\bs\Psi}}_{\ell\ell'}^{\ast},\ast=\uparrow,\downarrow$ are $3\times 3$ matrices independent of $z$, and the upward and downward propagating exponential functions are defined as
	\begin{equation}\label{up_down_exp}
		\hat{e}_{\ell}^{\uparrow}(z)=e^{\mathrm{i}k_{\ell z}^{s}(z-d_{\ell})},\quad \hat{e}_{\ell}^{\downarrow}(z)=e^{\mathrm{i}k_{\ell z}^{s}(d_{\ell-1}-z)},\quad
		\tilde{e}_{\ell}^{\uparrow}(z)=e^{\mathrm{i}k_{\ell z}^{c}(z-d_{\ell})},\quad \tilde{e}_{\ell}^{\downarrow}(z)=e^{\mathrm{i}k_{\ell z}^{c}(d_{\ell-1}-z)}.
	\end{equation}
	for $d_{\ell}<z<d_{\ell-1}$, and they are well-defined for all $\ell=0,1,\dots, L$, because of the convention $d_{-1}=+\infty$, $d_{L}=-\infty$ used throughout this paper. In particular, the branch cut $\Im(k_{\ell z}^{s}), \Im(k_{\ell z}^{c}) \ge 0$ implies 
	\begin{equation}
		\hat{e}_{0}^{\downarrow}(z)=\tilde{e}_{0}^{\downarrow}(z)=0,\quad\hat{e}_{L}^{\uparrow}(z)=\tilde{e}_{L}^{\uparrow}(z)=0,
	\end{equation}
	for any given $z$. Therefore, the general solutions \eqref{react_field_expand_ud} naturally satisfies the outgoing radiation conditions in the top layer $\ell = 0$ and the bottom layer $\ell = L$. This results in four additional free parameters $\hat{{\bs\Phi}}^{\downarrow}_{0\ell'},\hat{{\bs\Psi}}^{\downarrow}_{0\ell'},\hat{{\bs\Phi}}^{\uparrow}_{L\ell'},\hat{{\bs\Psi}}^{\uparrow}_{L\ell'}$, even though the terms to which they correspond do not exist. However, to simplify subsequent derivations, we set 
	\begin{equation*}
		\hat{{\bs\Phi}}^{\downarrow}_{0\ell'}(k_x,k_y,z')=\hat{{\bs\Psi}}^{\downarrow}_{0\ell'}(k_x,k_y,z')=\bs 0_{3\times3},\quad\hat{{\bs\Phi}}^{\uparrow}_{L\ell'}(k_x,k_y,z')=\hat{{\bs\Psi}}^{\uparrow}_{L\ell'}(k_x,k_y,z')=\bs 0_{3\times3}.
	\end{equation*}

	The unknown matrices $\hat{{\bs\Phi}}_{\ell\ell'}^{\ast},\hat{{\bs\Psi}}_{\ell\ell'}^{\ast},\ast=\uparrow,\downarrow$ shall be determined by the interface conditions \eqref{eq_interfacecond}. We shall adopt the theory of matrix basis (cf. \cite{bo2022maxwellDGF}) to decouple the interface conditions. The first $5$ matrices in the matrix basis are given by
	\begin{align}\label{mat-J}
		\begin{split}
			\J{1} & = \begin{bmatrix}
				1 & & \\
				& 1 & \\
				& & 0
			\end{bmatrix}, \enspace
			\J{2}  = \begin{bmatrix}
				0 & & \\
				& 0 & \\
				& & 1
			\end{bmatrix}, \enspace
			\J{3}  = \begin{bmatrix}
				0 & 0 & \ri k_x \\
				0 & 0 & \ri k_y \\
				0 & 0 & 0
			\end{bmatrix}, \\
			\J{4} & = \begin{bmatrix}
				0 & 0 & 0 \\
				0 & 0 & 0 \\
				\ri k_x & \ri k_y & 0
			\end{bmatrix}, \enspace
			\J{5}  = \begin{bmatrix}
				-k_x^2 & -k_x k_y & 0 \\
				-k_x k_y & -k_y^2 & 0 \\
				0 & 0 & 0
			\end{bmatrix}.
		\end{split}
	\end{align}
	By the expression \eqref{stresstensor}, we have
	\begin{align*}\label{eq_cont_T}
		\mathcal{T}_{31}^{\hat{\bs t}} = \mu \left( \frac{\partial  G^{\hat{\bs t}}_{3}}{\partial x} + \frac{\partial  G^{\hat{\bs t}}_{1}}{\partial z} \right), \quad
		\mathcal{T}_{32}^{\hat{\bs t}}= \mu \left( \frac{\partial G^{\hat{\bs t}}_{3}}{\partial y} + \frac{\partial G^{\hat{\bs t}}_{2}}{\partial z} \right), \quad
		\mathcal{T}_{33}^{\hat{\bs t}} = \gamma \frac{\partial G^{\hat{\bs t}}_{3}}{\partial z} + \lambda \left( \frac{\partial G^{\hat{\bs t}}_{1}}{\partial x} + \frac{\partial G^{\hat{\bs t}}_{2}}{\partial y} \right).
	\end{align*}
	Applying Fourier transform gives
	\begin{equation}\label{eq_cont_T_mat} 
		\widehat{\mathcal{T}}_{33}^{\hat{\bs t}} \hat{\bs{z}}=\left(\gamma\J{2}\partial_z+\lambda\J{4}\right)\widehat{\bd G}^{\hat{\bs t}},\quad
		\widehat{\mathcal{T}}_{31}^{\hat{\bs t}}\hat{\bs{x}}+\widehat{\mathcal{T}}^{\hat{\bs t}}_{32}\hat{\bs{y}}=\mu\left(\J{1}\partial_z+\J{3}\right)\widehat{\bd G}^{\hat{\bs t}}.
	\end{equation}
	Thus, the Fourier transform of the interface equations \eqref{eq_interfacecond} implies
	\begin{equation}\label{eq_interfacecont_T1}
		\Cont{\J{1} \widehat{\bb{G}}} = \bd{0},\quad\Cont{\J{2} \widehat{\bb{G}}} = \bd{0},\quad \Cont{\mu (\J{1} \partial_z +\J{3} ) \widehat{\bb{G}}} = \bd{0},\quad\Cont{(\gamma \J{2} \partial_z + \lambda \J{4} ) \widehat{\bb{G}}} = \bd{0}.
	\end{equation}
	By the interface conditions above, we have representations for the matrices $\hat{{\bs\Phi}}_{\ell\ell'}^{\ast},\hat{{\bs\Psi}}_{\ell\ell'}^{\ast},\ast=\uparrow,\downarrow$ using the basis \eqref{mat-J}. 
	\begin{theorem}\label{thm_reactfield_matrixbasis}
		The unknown tensors $\hat{{\bs\Phi}}_{\ell\ell'}^{\ast},\hat{{\bs\Psi}}_{\ell\ell'}^{\ast},\ast=\uparrow,\downarrow$ in the reaction field expansion \eqref{react_field_expand_ud} can be represented by the matrix basis defined in \eqref{mat-J}, i.e.
		\begin{equation}\label{react_field_matrixbasis_coeff}
			\hat{{\bs\Phi}}_{\ell\ell'}^{\ast}=\sum_{i=1}^{5}\alpha_{\ell\ell',i}^{\ast}(k_x,k_y,z')\J{i},\quad \hat{{\bs\Psi}}_{\ell\ell'}^{\ast}=\sum_{i=1}^{5}\beta_{\ell\ell',i}^{\ast}(k_x,k_y,z')\J{i},
		\end{equation}
		where $\alpha_{\ell\ell',i}^{*},\beta_{\ell\ell',i}^{*},i=1,\cdots,5$ are unknown constants with respect to $z$.
	\end{theorem}
	\begin{proof}
		The proof will be given in \Cref{sect_proof_reactfield_matrixbasis}. 
	\end{proof} 
	
	Define scalar coefficients 
	\begin{equation}\label{react_phipsi}
		\phi_{\ell\ell'i}^r=\alpha_{\ell\ell',i}^{\uparrow}\hat{e}_{\ell}^{\uparrow}(z)+\alpha_{\ell\ell',i}^{\downarrow}\hat{e}_{\ell}^{\downarrow}(z),\quad \psi_{\ell\ell'i}^r=\beta_{\ell\ell',i}^{\uparrow}\tilde{e}_{\ell}^{\uparrow}(z)+\beta_{\ell\ell',i}^{\downarrow}\tilde{e}_{\ell}^{\downarrow}(z),
	\end{equation}
	and substitutes \eqref{react_field_matrixbasis_coeff} into \eqref{react_field_expand_ud}.  We get 
	\begin{equation}\label{Xst_react_expand}
		\hat{{\bs\Phi}}_{\ell\ell'}=\sum_{i=1}^{5}\phi_{\ell\ell'i}^r(k_x,k_y,z,z')\J{i},\quad\hat{{\bs\Psi}}_{\ell\ell'}=\sum_{i=1}^{5}\psi_{\ell\ell'i}^r(k_x,k_y,z,z')\J{i},
	\end{equation}
	Define their free-space counterparts \eqref{Xst_freeexpand} as follows
	\begin{equation}\label{Xst_freeexpand_frequency}
		\begin{aligned}
			\hat{{\bs\Phi}}^f(k_{\rho},z,z')=\dfrac{1}{\mu_{\ell'}}\widehat{G}_{k_{\ell'}^{s}}^f(k_{\rho},z,z')\bb{I}=\sum_{i=1}^{5}\phi_i^f(k_{\rho},z,z')\J{i},\\
			\hat{{\bs\Psi}}^f(k_{\rho},z,z')=\dfrac{1}{\gamma_{\ell'}}\widehat{G}_{k_{\ell'}^{c}}^f(k_{\rho},z,z')\bb{I}=\sum_{i=1}^{5}\psi_i^f(k_{\rho},z,z')\J{i},
		\end{aligned}
	\end{equation}
	where $\widehat{G}_{{k}}^f(k_{\rho},z,z') = \dfrac{\ri}{2 {k}_{z}} e^{\ri {k}_{z} |z - z'|}$ is the Fourier transform for the Helmholtz fundamental solution \eqref{eq_gf} in free space, and
	\begin{equation}\label{coefficient_free}
		\phi_1^f=\phi_2^f=\dfrac{1}{\mu_{\ell'}}\widehat{G}_{k_{\ell'}^{s}}^f(k_{\rho},z,z'),\quad\psi_1^f=\psi_2^f=\dfrac{1}{\gamma_{\ell'}}\widehat{G}_{k_{\ell'}^{c}}^f(k_{\rho},z,z'),\quad\phi_i^f=\psi_i^f=0,i=3,4,5.
	\end{equation}
	By the definition \eqref{layeredDecomp}, we have
	\begin{equation}\label{Xst_layerexpand}
		\hat{{\bs\Phi}}=\sum_{i=1}^{5}\phi_{\ell\ell'i}(k_x,k_y,z,z')\J{i},\quad
		\hat{{\bs\Psi}}=\sum_{i=1}^{5}\psi_{\ell\ell'i}(k_x,k_y,z,z')\J{i}.
	\end{equation}
	in each layer, where the coefficients
	\begin{equation}\label{Xst_layerexpand_decomposition}
		\phi_{\ell\ell'i}=\delta_{\ell\ell'}\phi_{i}^f+\phi_{\ell\ell'i}^r,\quad\psi_{\ell\ell'i}=\delta_{\ell\ell'}\psi_i^f+\psi_{\ell\ell'i}^r,\quad i=1,2,\cdots,5
	\end{equation}
	satisfy the following Helmholtz equations in the frequency domain:
	\begin{equation}\label{FCsystem_old_phipsi}
		\begin{aligned}
			\partial_{zz}\phi_{\ell\ell'i}+(k_{\ell z}^{s})^{2}\phi_{\ell\ell'i}=-\dfrac{1}{\mu_{\ell'}}\delta(z-z'),\quad\partial_{zz}\psi_{\ell\ell'i}+(k_{\ell z}^{c})^{2}\psi_{\ell\ell'i}=-\dfrac{1}{\gamma_{\ell'}}\delta(z-z'),\quad i=1,2,\\
			\partial_{zz}\phi_{\ell\ell'i}+(k_{\ell z}^{s})^{2}\phi_{\ell\ell'i}=0,\quad\partial_{zz}\psi_{\ell\ell'i}+(k_{\ell z}^{c})^{2}\psi_{\ell\ell'i}=0,\quad i=3,4,5.
		\end{aligned}
	\end{equation}
	\begin{remark}
		In fact, we can further show that the coefficients $\phi_{\ell\ell',i},\psi_{\ell\ell',i},i=1,\cdots,5$ are rotationally invariant in the $k_x-k_y$ plane, i.e.,
		\begin{equation}\label{react_field_matrixbasis_coeff_better}
			\hat{{\bs\Phi}}_{\ell\ell'}=\sum_{i=1}^{5}\phi_{\ell\ell',i}(\kr,z')\J{i},\quad \hat{{\bs\Psi}}_{\ell\ell'}=\sum_{i=1}^{5}\psi_{\ell\ell',i}(\kr,z')\J{i},
		\end{equation}
		This will be shown during the calculation process in the end of this section.
	\end{remark}
	
	Note that the partial differential operators $\Delta$ and $\nabla\nabla$ in the frequency domain can be written as  
	\begin{equation}
		\mathcal{F}[\Delta]=-k_{\rho}^2+\partial_{zz},\quad\mathcal{F}[\nabla\nabla]=\J{5}+(\J{3}+\J{4})\partial_z+\J{2}\partial_{zz}.
	\end{equation}
	Taking Fourier transform on the representation \eqref{reactionLMGF}  and \eqref{reactioncomponent} gives
	\begin{equation}\label{Elasticdynamics_layerDGf_matrix_react}
		\begin{aligned}
			\widehat{\mathbb{G}}^f(k_x, k_y, z, z')=&\left[\bb{I}+(k_{\ell'}^{s})^{-2}\left(\J{5}+(\J{3}+\J{4})\partial_z+\J{2}\partial_{zz}\right)\right]\hat{{\bs\Phi}}^f\\
			&-(k_{\ell'}^{c})^{-2}\left(\J{5}+(\J{3}+\J{4})\partial_z+\J{2}\partial_{zz}\right)\hat{{\bs\Psi}}^f,\\
			\widehat{\mathbb{G}}_{\ell\ell'}^r(k_x, k_y, z, z')=&	\left[\bb{I}+(k_{\ell}^{s})^{-2}\left(\J{5}+(\J{3}+\J{4})\partial_z+\J{2}\partial_{zz}\right)\right]\hat{{\bs\Phi}}_{\ell\ell'}\\
			&-(k_{\ell}^{c})^{-2}\left(\J{5}+(\J{3}+\J{4})\partial_z+\J{2}\partial_{zz}\right)\hat{{\bs\Psi}}_{\ell\ell'}.
		\end{aligned}
	\end{equation}
	Summing up the representations in \eqref{Elasticdynamics_layerDGf_matrix_react} and substituting \eqref{Xst_layerexpand} into the result, we obtain 
	\begin{equation}\label{Elasticdynamics_layerDGf_Fourierv1}
		\begin{aligned}
			\widehat{\mathbb{G}}=&\phi_{\ell\ell'1}\J{1}+(k_{\ell}^{s})^{-2}\Big[k_{\rho}^2\left(\phi_{\ell\ell'2}-\partial_z\phi_{\ell\ell'3}\right)\J{2}+\partial_z\big(\phi_{\ell\ell'2}-\partial_z\phi_{\ell\ell'3}\big)\J{3}\\
			&+\left(\partial_z\phi_{\ell\ell'1}+k_{\rho}^2(\phi_{\ell\ell'4}-\partial_z\phi_{\ell\ell'5})\right)\J{4}+\Big(\phi_{\ell\ell'1}+\partial_z(\phi_{\ell\ell'4}-\partial_z\phi_{\ell\ell'5})\Big)\J{5}\Big]+\delta_{S}\\
			&-(k_{\ell}^{c})^{-2}\Big[\partial_z\left(\partial_z\psi_{\ell\ell'2}-k_{\rho}^2\psi_{\ell\ell'3}\right)\J{2}+\left(\partial_z\psi_{\ell\ell'2}-k_{\rho}^2\psi_{\ell\ell'3}\right)\J{3}\\&+\partial_z\left(\psi_{\ell\ell'1}+\partial_z\psi_{\ell\ell'4}-k_{\rho}^2\psi_{\ell\ell'5}\right)\J{4}
			+\left(\psi_{\ell\ell'1}+\partial_z\psi_{\ell\ell'4}-k_{\rho}^2\psi_{\ell\ell'5}\right)\J{5}\Big]
		\end{aligned}
	\end{equation}
	where $\delta_{S}:=-\dfrac{\delta(z-z')}{\mu_{\ell'}(k_{\ell}^{s})^{2}}\J{2}$. There are 10 unknown scalar functions $\phi_{\ell\ell'i},\psi_{\ell\ell'i},i=1,2,\cdots,5$ in the expression \eqref{Elasticdynamics_layerDGf_Fourierv1}, and it can be seen that there are redundant unknowns. Actually, define  
	\begin{equation}\label{reacioncomponentslayerwise}
		\begin{aligned}
			&\tilde{\phi}_{\ell\ell'1}=\phi_{\ell\ell'1},\quad\tilde{\phi}_{\ell\ell'2}=\dfrac{\phi_{\ell\ell'2}-\partial_z\phi_{\ell\ell'3}}{(k_{\ell}^{s})^{2}},\quad\tilde{\phi}_{\ell\ell'3}=\dfrac{\partial_z\phi_{\ell\ell'1}+k_{\rho}^2(\phi_{\ell\ell'4}-\partial_z\phi_{\ell\ell'5})}{(k_{\ell}^{s})^{2}},\\
			&\tilde{\psi}_{\ell\ell'1}=\dfrac{\partial_z\psi_{\ell\ell'2}-k_{\rho}^2\psi_{\ell\ell'3}}{(k_{\ell}^{c})^{2}},\quad\tilde{\psi}_{\ell\ell'2}=\dfrac{\psi_{\ell\ell'1}+\partial_z\psi_{\ell\ell'4}-k_{\rho}^2\psi_{\ell\ell'5}}{(k_{\ell}^{c})^{2}}.
		\end{aligned}
	\end{equation}
	and piecewise functions
	\begin{equation}\label{reacioncomponents}
		\tilde\phi_i(k_x, k_y, z, z')=\tilde\phi_{\ell\ell'i}(k_x, k_y, z, z'),\quad \tilde\psi_i(k_x, k_y, z, z')=\tilde\psi_{\ell\ell'i}(k_x, k_y, z, z'),\quad d_{\ell}<z<d_{\ell-1}.
	\end{equation}
	Using the equation \eqref{FCsystem_old_phipsi} that $\phi_{\ell\ell'1}$ satisfies in each layer to eliminate the term $\partial_{zz}$, we get
	\begin{equation*}
		(k_{\ell}^{s})^{-2}k_{\rho}^2\left(\phi_{\ell\ell'1}+\partial_z(\phi_{\ell\ell'4}-\partial_z\phi_{\ell\ell'5})\right)=\tilde{\phi}_{\ell\ell'1}+\partial_z\tilde{\phi}_{\ell\ell'3}+\dfrac{\delta(z-z')}{\mu_{\ell'}(k_{\ell}^{s})^{2}}
	\end{equation*}
	Therefore, the formulation \eqref{Elasticdynamics_layerDGf_Fourierv1} can be rewritten as
	\begin{equation}\label{Elasticdynamics_layerDGf_Fourierv2}
		\begin{aligned}
			\widehat{\mathbb{G}}=&\tilde{\phi}_{1}\J{1}+\left(k_{\rho}^2\tilde{\phi}_{2}-\partial_z\tilde{\psi}_{1}\right)\J{2}+\left(\partial_z\tilde{\phi}_{2}-\tilde{\psi}_{1}\right)\J{3}+\left(\tilde{\phi}_{3}-\partial_z\tilde{\psi}_{2}\right)\J{4}\\
			&+\left(k_{\rho}^{-2}[\tilde{\phi}_{1}+\partial_z\tilde{\phi}_{3}]-\tilde{\psi}_{2}\right)\J{5}+\dfrac{\delta(z-z')}{\omega^2\rho_{\ell'}}\left(k_{\rho}^{-2}\J{5}-\J{2}\right).
		\end{aligned}
	\end{equation}
	From the equations in \eqref{FCsystem_old_phipsi}, the definitions in \eqref{reacioncomponentslayerwise} and the fact $\mu_{\ell'}(k_{\ell'}^{s})^{2}=\gamma_{\ell'}(k_{\ell'}^{c})^{2}=\omega^2\rho_{\ell'}$, it can be easily verified that $\tilde{\phi}_{\ell\ell'k}, \tilde{\psi}_{\ell\ell'k}$ satisfy the Helmholtz equations
	\begin{equation}\label{FCsystem_phi123_psi12}
		\begin{aligned}
			\partial_{zz}\tilde{\phi}_{1}+(k_{\ell z}^{s})^{2}\tilde{\phi}_{1}=-\dfrac{1}{\mu_{\ell'}}\delta(z-z')\\
			\partial_{zz}\tilde{\phi}_{2}+(k_{\ell z}^{s})^{2}\tilde{\phi}_{2}=-\dfrac{1}{\omega^2\rho_{\ell'}}\delta(z-z'),\quad
			\partial_{zz}\tilde{\phi}_{3}+(k_{\ell z}^{s})^{2}\tilde{\phi}_{3}=-\dfrac{1}{\omega^2\rho_{\ell'}}\delta'(z-z')\\
			\partial_{zz}\tilde{\psi}_{1}+(k_{\ell z}^{c})^{2}\tilde{\psi}_{1}=-\dfrac{1}{\omega^2\rho_{\ell'}}\delta'(z-z'),\quad 
			\partial_{zz}\tilde{\psi}_{2}+(k_{\ell z}^{c})^{2}\tilde{\psi}_{2}=-\dfrac{1}{\omega^2\rho_{\ell'}}\delta(z-z')
		\end{aligned}
	\end{equation}
	in the $\ell$-th layer for all $\ell=0, 1, \cdots, L$. 
	\begin{remark}
		When all functions $\tilde{\phi}_1,\tilde{\phi}_2,\tilde{\phi}_3,\tilde{\psi}_1,\tilde{\psi}_2$ in \eqref{Elasticdynamics_layerDGf_Fourierv2} only take the free field part, we can infer the free field part of the elastic wave in the frequency domain, which can eliminate the term of Dirac delta function, and its inverse Fourier transform is the same as \eqref{freespacepart_EWE_ellp}.
	\end{remark}
	
	Now, the interface conditions \eqref{eq_interfacecont_T1} are revisited using the new expression \eqref{Elasticdynamics_layerDGf_Fourierv2}.
	We always assume that the source point $\bd{r}'$ is not on the interfaces. Using the product table \eqref{producttable}, we can calculate from \eqref{Elasticdynamics_layerDGf_Fourierv2} that
	\begin{align*}
		\J{1}\widehat{\mathbb{G}}
		=&\tilde{\phi}_{1}\J{1}+\left(\partial_z\tilde{\phi}_{2}-\tilde{\psi}_{1}\right)\J{3}+k_{\rho}^{-2}\left(\tilde{\phi}_{1}+\partial_z\tilde{\phi}_{3}-k_{\rho}^{2}\tilde{\psi}_{2}\right)\J{5}+\frac{\delta(z-z')}{\omega^2\rho_{\ell'}k_{\rho}^2}\J{5}\\
		\J{2}\widehat{\mathbb{G}}
		=&\left(k_{\rho}^2\tilde{\phi}_{2}-\partial_z\tilde{\psi}_{1}\right)\J{2}+\left(\tilde{\phi}_{3}-\partial_z\tilde{\psi}_{2}\right)\J{4}-\frac{\delta(z-z')}{\omega^2\rho_{\ell'}}\J{2}\\
		\mu_{\ell}\left(\J{1}\partial_z+\J{3}\right)\widehat{\mathbb{G}} 
		=&\mu_{\ell}\partial_z\tilde{\phi}_{1}\J{1}+\left[\mu_{\ell}\left(- (k_{\ell z}^{s})^{2}+k_{\rho}^2\right)\tilde{\phi}_{2}-2\mu_{\ell}\partial_z\tilde{\psi}_{1}-\frac{2\mu_{\ell}\delta(z-z')}{\omega^2\rho_{\ell'}}\right]\J{3}\\
		&+k_{\rho}^{-2}\left[\mu_{\ell} \partial_z\tilde{\phi}_{1}+\mu_{\ell}\left(- (k_{\ell z}^{s})^{2}+ k_{\rho}^{2}\right)\tilde{\phi}_{3}-2\mu_{\ell}k_{\rho}^{2}\partial_z\tilde{\psi}_{2}\right]\J{5},\\
		\left(\gamma_{\ell}\J{2}\partial_z+\lambda_{\ell}\J{4}\right)\widehat{\mathbb{G}} 
		=&\left[2\mu_{\ell} k_{\rho}^2\partial_z\tilde{\phi}_{2}+\left(\gamma_{\ell} (k_{\ell z}^{c})^{2}+\lambda_{\ell} k_{\rho}^2\right)\tilde{\psi}_{1}\right]\J{2}\\
		&+\left[2\mu_{\ell}\partial_z\tilde{\phi}_{3}+\left(\gamma_{\ell} (k_{\ell z}^{c})^{2}+\lambda_{\ell} k_{\rho}^2\right)\tilde{\psi}_{2}\right]\J{4}+\frac{2\mu_{\ell}\delta(z-z')}{\omega^2\rho_{\ell'}}\J{4},
	\end{align*}
	for $d_{\ell}<z<d_{\ell-1}$ and $\ell=0,1, \cdots, L$.
	Using these expressions into the interface conditions \eqref{eq_interfacecont_T1} and the independence of the basis matrices $\J{1}, \J{2}, \J{3}, \J{4}, \J{5}$ gives
	\begin{equation}\label{interface_solid-solidv4}
		\begin{aligned}
			\left\llbracket \tilde{\phi}_1 \right\rrbracket=0,\quad
			\left\llbracket \partial_z\tilde{\phi}_2-\tilde{\psi}_1 \right\rrbracket=0,\quad
			\left\llbracket \tilde{\phi}_1+\partial_z\tilde{\phi}_3-k_{\rho}^{2}\tilde{\psi}_2 \right\rrbracket=0\\
			\left\llbracket k_{\rho}^2\tilde{\phi}_2-\partial_z\tilde{\psi}_1 \right\rrbracket=0,\quad
			\left\llbracket \tilde{\phi}_3-\partial_z\tilde{\psi}_2 \right\rrbracket=0\\
			\left\llbracket \mu\partial_z\tilde{\phi}_1 \right\rrbracket=0,\quad
			\left\llbracket \mu \left(-(k_{z}^{s})^{2}+k_{\rho}^2\right)\tilde{\phi}_2-2\mu\partial_z\tilde{\psi}_1 \right\rrbracket=0\\
			\left\llbracket \mu\partial_z\tilde{\phi}_1+\mu \left(-(k_{z}^{s})^{2}+k_{\rho}^{2}\right)\tilde{\phi}_3-2\mu k_{\rho}^{2}\partial_z\tilde{\psi}_2 \right\rrbracket=0\\
			\left\llbracket 2\mu k_{\rho}^2\partial_z\tilde{\phi}_2+\left(\gamma (k_{z}^{c})^{2}+\lambda k_{\rho}^2\right)\tilde{\psi}_1 \right\rrbracket=0,\quad
			\left\llbracket 2\mu\partial_z\tilde{\phi}_3+\left(\gamma (k_{z}^{c})^{2}+\lambda k_{\rho}^2\right)\tilde{\psi}_2 \right\rrbracket=0
		\end{aligned}
	\end{equation}
	at each interface $z=d_{\ell},\ell=0,1,\cdots,L-1$. 
	Together with the equations \eqref{FCsystem_phi123_psi12}, we obtain three independent problems for three groups of coefficients $\{\tilde{\phi}_{1}\}, \{\tilde{\phi}_{2}, \tilde{\psi}_{1}\},\{\tilde{\phi}_{3},\tilde{\psi}_{2}\}$, i.e.,
	\begin{itemize}
		\item the Helmholtz equation for $\tilde{\phi}_{1}$ in layered media:
		\begin{equation}\label{FCsystem_phi1}
			\begin{cases}
				\partial_{zz}\tilde{\phi}_1+(k_{z}^{s})^{2}\tilde{\phi}_1=-\dfrac{1}{\mu_{\ell'}}\delta(z-z'),\quad z\neq d_{\ell},\;\ell=0, 1, \cdots, L-1,\\
				\left\llbracket \tilde{\phi}_1 \right\rrbracket=0,\quad\left\llbracket \mu\partial_z\tilde{\phi}_1 \right\rrbracket=0.
			\end{cases}
		\end{equation}
		\item the mixed-wavenumber Helmholtz equations for $\tilde{\phi}_{2}, \tilde{\psi}_{1}$ in layered media: 
		\begin{equation}\label{FCsystem_phi2_psi1}
			\begin{cases}
				\partial_{zz}\tilde{\phi}_2+(k_{z}^{s})^{2}\tilde{\phi}_2=-\dfrac{1}{\omega^2\rho_{\ell'}}\delta(z-z'),\quad z\neq d_{\ell},\;\ell=0, 1, \cdots, L-1,\\
				\partial_{zz}\tilde{\psi}_1+(k_{z}^{c})^{2}\tilde{\psi}_1=-\dfrac{1}{\omega^2\rho_{\ell'}}\delta'(z-z'),\quad z\neq d_{\ell},\;\ell=0, 1, \cdots, L-1,\\
				\left\llbracket 2\mu k_{\rho}^2\partial_z\tilde{\phi}_2+\left(\gamma (k_{z}^{c})^{2}+\lambda k_{\rho}^2\right)\tilde{\psi}_1 \right\rrbracket=0,\quad
				\left\llbracket k_{\rho}^2\tilde{\phi}_2-\partial_z\tilde{\psi}_1 \right\rrbracket=0,\\[7pt]
				\left\llbracket \mu \left(-(k_{z}^{s})^{2}+k_{\rho}^2\right)\tilde{\phi}_2-2\mu\partial_z\tilde{\psi}_1 \right\rrbracket=0,\quad
				\left\llbracket \partial_z\tilde{\phi}_2-\tilde{\psi}_1 \right\rrbracket=0.
			\end{cases}
		\end{equation}
		\item the mixed-wavenumber Helmholtz equations of $\tilde{\phi}_{3}, \tilde{\psi}_{2}$ in layered media:
		\begin{equation}\label{FCsystem_phi3_psi2}
			\begin{cases}
				\partial_{zz}\tilde{\phi}_3+(k_{z}^{s})^{2}\tilde{\phi}_3=-\dfrac{1}{\omega^2\rho_{\ell'}}\delta'(z-z'),\quad z\neq d_{\ell},\;\ell=0, 1, \cdots, L-1,\\
				\partial_{zz}\tilde{\psi}_2+(k_{z}^{c})^{2}\tilde{\psi}_2=-\dfrac{1}{\omega^2\rho_{\ell'}}\delta(z-z'),\quad z\neq d_{\ell},\;\ell=0, 1, \cdots, L-1,\\
				\left\llbracket 2\mu\partial_z\tilde{\phi}_3+\left(\gamma (k_{z}^{c})^{2}+\lambda k_{\rho}^2\right)\tilde{\psi}_2 \right\rrbracket=0,\quad
				\left\llbracket \tilde{\phi}_3-\partial_z\tilde{\psi}_2 \right\rrbracket=0,\\[7pt]
				\left\llbracket \mu \left(-(k_{z}^{s})^{2}+k_{\rho}^{2}\right)\tilde{\phi}_3-2\mu k_{\rho}^{2}\partial_z\tilde{\psi}_2 \right\rrbracket=0,\quad
				\left\llbracket \partial_z\tilde{\phi}_3-k_{\rho}^{2}\tilde{\psi}_2 \right\rrbracket=0.
			\end{cases}
		\end{equation}
	\end{itemize}
	In the top and bottom layers, outgoing radiation  conditions
	\begin{equation}\label{decaycond}
		\lim_{z\to \pm\infty}\tilde{\phi}_i(k_x,k_y,z,z')=0,\quad i=1,2,3;\quad \lim_{z\to \pm\infty}\tilde{\psi}_i(k_x,k_y,z,z')=0,\quad i=1,2,
	\end{equation}
	are imposed.
	It can be clearly seen from these problems that they are radially symmetric in the $k_x-k_y$ plane. Therefore, we have
	\begin{equation}
		\tilde{\phi}_i(k_x,k_y,z,z')=\tilde{\phi}_i(k_{\rho},z,z'),\quad \tilde{\psi}_i(k_x,k_y,z,z')=\tilde{\psi}_i(k_{\rho},z,z'),
	\end{equation}
	which verifies the formulations in \eqref{react_field_matrixbasis_coeff_better}.

	Therefore, the dyadic Green's function of the elastic wave equation in layered media can be decomposed into a combination of five unknown functions. 
	Then, the computation is transformed into the solution of one layered Helmholtz equation \eqref{FCsystem_phi1} and two mixed-wavenumber layered Helmholtz equations \eqref{FCsystem_phi2_psi1}-\eqref{FCsystem_phi3_psi2}.
	Among them, the solution of the layered Helmholtz equation \eqref{FCsystem_phi1} already has a stable recurrence formula (cf. \cite{bo2020tefmm}). In the next section, we focus on the solution of the mixed-wavenumber layered Helmholtz equations \eqref{FCsystem_phi2_psi1}-\eqref{FCsystem_phi3_psi2}. 
	
	\begin{remark}
		The structure of these three problems reveals an essential difference between the elastic wave in layered media and its free-space counterpart. In the free space, the representation \eqref{freespacepart_EWE} shows that the shear (S) and compressional (P) waves are completely decoupled: the potentials ${\bs\Phi}^f$ and ${\bs\Psi}^f$ satisfy independent Helmholtz equations with wave numbers $k^{s}$ and $k^{c}$, respectively, and each propagates without ever exciting the other. In layered media, the potentials still satisfy decoupled Helmholtz equations inside each layer, see \eqref{FCsystem_phi123_psi12}, but the interface conditions \eqref{eq_interfacecond} prescribe the continuity of the total displacement and traction, which are mixtures of both wave types. Consequently, as shown in \eqref{FCsystem_phi2_psi1} and \eqref{FCsystem_phi3_psi2}, the S-wave components $\tilde\phi_2, \tilde\phi_3$ and the P-wave components $\tilde\psi_1, \tilde\psi_2$ are coupled through the interface conditions, so that an incident P-wave (resp. S-wave) generally produces both reflected and transmitted P- and S-waves at every interface. This phenomenon, known as mode conversion (cf. \cite{aki2002quantitative,kennett2009seismic}), is absent in the free space. The only exception is $\tilde\phi_1$ in \eqref{FCsystem_phi1}, which corresponds to the horizontally polarized shear (SH) wave and satisfies a scalar Helmholtz interface problem of the same type as in the acoustic case; the remaining components correspond to the coupled P--SV waves and lead to mixed-wavenumber problems involving both $k_{\ell z}^{s}$ and $k_{\ell z}^{c}$. As a result, the reflection and transmission coefficients for the P--SV system are $2\times 2$ matrices rather than scalars, and their denominators may vanish at real values of $k_\rho$ larger than the bulk wave numbers. These zeros are associated with interface and surface waves, such as Rayleigh and Stoneley waves (cf. \cite{achenbach2012wave}), which have no free-space analogue. This coupling makes both the analysis and the numerical evaluation of the elastic LMGF considerably more involved than that of the Maxwell's equations in layered media.
	\end{remark}

	\section{Stable recursion for the computation of the Green's function in layered media}\label{sect_solver}
	As in the evaluation of the Green's function of Helmholtz equation in layered media, the LMGFs in the spatial domain eventually leads to some Sommerfeld-type integrals resulted from the inverse Fourier transform.
	These integrals are numerically computed along an integral path named as Sommerfeld integral path (SIP). Efficient and accurate algorithms for the computation of the Sommerfeld-type integrals have been a challenging problem in the engineering community for more than 30 years (cf. \cite{michalski1998extrapolation,michalski2016efficient}). 
	Moreover, accurate and efficient calculation of the LMGFs relies on a stable and efficient algorithm for the computation of the spectral domain LMGFs, i.e., $\{\tilde{\phi}_{1}\}, \{\tilde{\phi}_{2}, \tilde{\psi}_{1}\}\{\tilde{\phi}_{3},\tilde{\psi}_{2}\}$ in this paper. 
	
	For the problem \eqref{FCsystem_phi1}, a stable recursive formulation for any complex number of $k_{\rho}$ not on the branch cut and an arbitrary number of layers is well-known (cf. \cite{chew1999waves}). However, the other two mixed-wavenumber problems \eqref{FCsystem_phi2_psi1} and \eqref{FCsystem_phi3_psi2} physically have two types of waves coupled via interface conditions. In this section, we shall establish stable recursive formulations for the solutions of these problems.
	
	
	
	For the sake of completeness, we first summarize the recursive algorithm for the solution of \eqref{FCsystem_phi1} which are interpreted as the free-space component in the source layer as well as the \emph{reaction field} incited by the presence of other layers. The reaction field is further decomposed into $4$ components according to the vertical field propagation directions around the target and source points, i.e.
	\begin{equation}\label{solution_phi1}
		\tilde{\phi}_{1}(k_{\rho}, z, z')= \delta_{\ell\ell'} \tilde{\phi}_1^f(k_{\rho}, z, z')+\dfrac{1}{\mu_{\ell'}}\dfrac{\ri}{2k_{\ell'z}^{s}}\sum_{\ast,\star=\uparrow,\downarrow}\sigma_{\ell\ell'0}^{\ast\star}(\kr){Z}_{\ell\ell'}^{\ast\star}(k_{\rho}, z,z'),
	\end{equation}
	for $\bs r$ in the $\ell$-th layer and $\bs r'$ in the $\ell'$-th layer, 
	where $\tilde{\phi}_1^f=\phi_1^f$ is the free-space component defined in \eqref{coefficient_free}, and the $Z^{\ast\star}$-functions 
	\begin{equation}\label{zexponential}
		\begin{split}
			{Z}_{\ell\ell'}^{\uparrow\uparrow}(k_{\rho}, z,z')=&\begin{cases}
				\displaystyle e^{\ri (k_{\ell z}^{s}z-k_{\ell' z}^{s}z')},\quad \ell<\ell',\\
				\displaystyle e^{\ri (k_{\ell' z}^{s}\tau_{\ell'-1}(z')-k_{\ell z}^{s}\tau_{\ell}(z))},\quad\ell\geq \ell',
			\end{cases}\\
			{Z}_{\ell\ell'}^{\uparrow\downarrow}(k_{\rho}, z,z')=&\begin{cases}
				\displaystyle e^{\ri (k_{\ell z}^{s}z-k_{\ell' z}^{s}\tau_{\ell'}(z'))},\quad \ell\leq\ell',\\
				\displaystyle e^{\ri (k_{\ell' z}^{s}z'-k_{\ell z}^{s}\tau_{\ell }(z))},\quad \ell>\ell',
			\end{cases}\\
			{Z}_{\ell\ell'}^{\downarrow\uparrow}(k_{\rho}, z,z')=&\begin{cases}
				\displaystyle e^{\ri (k_{\ell z}^{s}\tau_{\ell-1}(z)-k_{\ell' z}^{s}z')},\quad\ell<\ell',\\
				\displaystyle e^{\ri(k_{\ell' z}^{s}\tau_{\ell'-1}(z')-k_{\ell z}^{s}z)},\quad\ell\geq \ell',
			\end{cases}\\
			{Z}_{\ell\ell'}^{\downarrow\downarrow}(k_{\rho}, z,z')=&\begin{cases}
				\displaystyle e^{\ri (k_{\ell z}^{s}\tau_{\ell-1}(z)-k_{\ell' z}^{s}\tau_{\ell'}(z'))},\quad\ell\leq \ell',\\
				e^{\ri (k_{\ell' z}^{s}z'-k_{\ell z}^{s}z)},\quad \ell>\ell',
			\end{cases}
		\end{split}
	\end{equation}
	are exponential functions indicating the different cases of vertical field propagation directions, 
	\begin{equation}
		\label{z-reflection}
		\tau_{\ell}(z)=2d_{\ell}-z,
	\end{equation}
	is the image coordinates of $z$ with respect to the interface $z=d_{\ell}$.
	The densities $\sigma_{\ell\ell'0}^{\ast\star}(k_{\rho})$ can be calculated recursively as shown in \cite{yuan2026computation} Appendix A.
	
	
	\subsection{Recursive formula for the solution of the problem \texorpdfstring{\eqref{FCsystem_phi2_psi1}}{}}\label{sect_solver_phi2_psi1}
	For the sake of brevity, we define
	exponential matrices:
	\begin{equation}\label{diagonalmatrix_z}
		{\mathbb{E}}_{\ell}^{\uparrow}(z)=\begin{bmatrix}
			e^{\ri k_{\ell z}^{c}z}&0\\0&e^{\ri k_{\ell z}^{s}z}
		\end{bmatrix},\quad
		{\mathbb{E}}_{\ell}^{\downarrow}(z)=\begin{bmatrix}
			e^{-\ri k_{\ell z}^{c}z}&0\\0&e^{-\ri k_{\ell z}^{s}z}
		\end{bmatrix},
	\end{equation}
	and particularly 
	\begin{equation}\label{matrix_coeff_E}
		\mathbb{E}_{\ell}:=\begin{bmatrix}
			e^{\ri k_{\ell z}^{c}D_{\ell}} & 0\\
			0 & e^{\ri k_{\ell z}^{s}D_{\ell}}
		\end{bmatrix}=\mathbb{E}_{\ell}^{\uparrow}(D_{\ell})=\mathbb{E}_{\ell}^{\downarrow}(-D_{\ell}).
	\end{equation}
	It can be clearly seen that all $\{\mathbb{E}_{\ell}\}_{\ell=1}^{L}$ decay exponentially as $k_{\rho}\rightarrow\infty$ suppose the aforementioned branch cuts $\Im(k_{\ell z}^{s}), \Im(k_{\ell z}^{c}) \ge 0$ are adopted. In particular, the fact $D_0=+\infty$, $D_{L}=+\infty$ implies that $\mathbb E_{0}=\mathbb E_{L}=0$.

	Recall the definitions in \eqref{reacioncomponentslayerwise}-\eqref{reacioncomponents} and \eqref{coefficient_free}, we have
	\begin{equation*}
		\begin{aligned}
			\tilde{\psi}_1^f(k_{\rho},z,z')=&\frac{\ri \partial_ze^{\ri k_{\ell'z}^{c}|z-z'|}}{2(k_{\ell'}^{c})^{2}\gamma_{\ell'}k_{\ell'z}^{c}}=\dfrac{1}{2\omega^2\rho_{\ell'}}\big[-H(z-z')e^{\ri k_{\ell'z}^{c}(z-z')}+H(z'-z)e^{\ri k_{\ell'z}^{c}(z'-z)}\big],\\
			\tilde{\phi}_2^f(k_{\rho},z,z')=&\frac{\ri e^{\ri k_{\ell'z}^{s}|z-z'|}}{2(k_{\ell'}^{s})^{2}\mu_{\ell'}k_{\ell'z}^{s}}=\dfrac{\ri}{2\omega^2\rho_{\ell'}k_{\ell'z}^{s}}\big[H(z-z')e^{\ri k_{\ell'z}^{s}(z-z')}+H(z'-z)e^{\ri k_{\ell'z}^{s}(z'-z)}\big],
		\end{aligned}
	\end{equation*}
	where 
	$$H(x)=\begin{cases}
		\displaystyle 0,\quad x<0,\\
		\displaystyle \frac{1}{2},\quad x=0,\\
		\displaystyle 1,\quad x>0,
	\end{cases}$$
	is the Heaviside step function. 
	According to the superposition principle, the reaction components $\tilde{\phi}_{2}^r=\tilde{\phi}_{2}-\tilde{\phi}_{2}^f, \tilde{\psi}_{1}^r=\tilde{\psi}_{1}-\tilde{\psi}_{1}^f$ satisfy the homogeneous equations:
	\begin{equation}\label{react_ode}
		\partial_{zz}\tilde{\psi}_1^r+(k_{\ell z}^{c})^{2}\tilde{\psi}_1^r=0,\quad \partial_{zz}\tilde{\phi}_2^r+(k_{\ell z}^{s})^{2}\tilde{\phi}_2^r=0,
	\end{equation}
	and the interface conditions
	\begin{equation}\label{interface_condition_x23v1}
		\left\llbracket  {\bs a}\tilde{\psi}_1+\bs b 	\partial_z\tilde{\phi}_2\right\rrbracket=0,\quad\left\llbracket \bs d \partial_{z}\tilde{\psi}_1+ {\bs c}\tilde{\phi}_2\right\rrbracket=0,\quad  z=d_{\ell},\quad\ell=0,1,\cdots,L-1
	\end{equation}
	where the sliced interface parameter vectors $\bs a,\bs b,\bs c,\bs d$ in each layer $\ell$ are defined as
	\begin{equation}
		{\bs a}_{\ell}=\begin{bmatrix}
			\gamma_{\ell} (k_{\ell z}^{c})^{2}+\lambda_{\ell} k_{\rho}^2\\
			-1 
		\end{bmatrix},\quad {\bs b}_{\ell}=\begin{bmatrix}
			2\mu_{\ell} k_{\rho}^2\\
			1
		\end{bmatrix},\quad \bs c_{\ell}=\begin{bmatrix}
			\mu_{\ell}((k_{\ell z}^{s})^{2}-k_{\rho}^2) \\
			-k_{\rho}^2 
		\end{bmatrix},\quad \bs d_{\ell}=\begin{bmatrix}
			2\mu_{\ell}\\
			1
		\end{bmatrix}.
	\end{equation}
	Obviously, the ordinary differential equations \eqref{react_ode} have general solutions 
	\begin{equation}\label{reactfourerformula2v}
		\tilde{\psi}_{\ell \ell'1}^r(k_x, k_y, z,z')=A_{\ell}^{\psi}(k_x, k_y, z')e^{\ri k_{\ell z}^{c}(z-d_{\ell})}+B_{\ell}^{\psi}(k_x, k_y,z')e^{-\ri k_{\ell z}^{c}(z-d_{\ell-1})},
	\end{equation}
	\begin{equation}\label{reactfourerformula3v}
		\tilde{\phi}_{\ell \ell'2}^r(k_x, k_y, z,z')=A_{\ell}^{\phi}(k_x, k_y,z')e^{\ri k_{\ell z}^{s}(z-d_{\ell})}+B_{\ell}^{\phi}(k_x, k_y,z')e^{-\ri k_{\ell z}^{s}(z-d_{\ell-1})},
	\end{equation}
	in each layer $0\leq \ell\leq L$. It is worth pointing out that the convention $d_{-1}=+\infty$, $d_{L}=-\infty$ implies that
	$$
	e^{-\ri k_{0 z}^{s}(z-d_{-1})}\equiv 0,\quad e^{\ri k_{L z}^{s}(z-d_{L})}\equiv 0,
	$$
	for any given $z$. Therefore, the coefficients $$B^{\psi}_0(k_x, k_y,z'), \quad B_{0}^{\phi}(k_x, k_y,z') ,\quad A^{\psi}_L(k_x, k_y,z'),\quad A_{L}^{\phi} (k_x, k_y,z'),$$
	do not contribute to the reaction components $\{\tilde{\psi}_{\ell \ell'1}^r(k_x, k_y, z,z'), \tilde{\phi}_{\ell \ell'2}^r(k_x, k_y, z,z')\}$. Therefore, we set
	\begin{equation}\label{decaycondcoeff}
		B^{\psi}_0(k_x, k_y,z')=B_{0}^{\phi}(k_x, k_y,z')  \equiv 0,\quad A^{\psi}_L(k_x, k_y,z')=A_{L}^{\phi} (k_x, k_y,z') \equiv 0,
	\end{equation}
	for simplicity. 
	From these general solutions \eqref{reactfourerformula2v}-\eqref{reactfourerformula3v}, we can calculate that
	\begin{equation}\label{densityoninterfaces}
		\begin{split}
			&{\bs a}_{\ell'}\tilde{\psi}_1^f+{\bs b}_{\ell'}\partial_z\tilde{\phi}_2^f\Big|_{z=d_{\ell'-1}-0}={\mathbb A}_{\ell'}^{+}\mathbb{E}_{\ell'}^{\downarrow}(z'-d_{\ell'-1}){\bs A}_{\ell'}^f,\\
			&{\bs d}_{\ell'}\partial_z\tilde{\psi}_1^f+{\bs c}_{\ell'}\tilde{\phi}_2^f\Big|_{z=d_{\ell'-1}-0}={\mathbb B}_{\ell'}^{+}\mathbb{E}_{\ell'}^{\downarrow}(z'-d_{\ell'-1}){\bs A}_{\ell'}^f\\
			&{\bs a}_{\ell'}\tilde{\psi}_1^f+{\bs b}_{\ell'}\partial_z\tilde{\phi}_2^f\Big|_{z=d_{\ell'}+0}=        {\mathbb A}_{\ell'}^{-}\mathbb{E}_{\ell'}^{\uparrow}(z'-d_{\ell'}){\bs B}_{\ell}^f,\\
			&{\bs d}_{\ell'}\partial_z\tilde{\psi}_1^f+{\bs c}_{\ell'}\tilde{\phi}_2^f\Big|_{z=d_{\ell'}+0}= {\mathbb B}_{\ell'}^{-}\mathbb{E}_{\ell'}^{\uparrow}(z'-d_{\ell'}){\bs B}_{\ell}^f,
		\end{split}
	\end{equation}
	and
	\begin{equation}\label{vectorvaluelimitLlayer}
		\begin{aligned}
			{\bs a}_{\ell}\tilde{\psi}_1^r+{\bs b}_{\ell}\partial_z\tilde{\phi}_2^r\Big|_{z=d_{\ell}+0}=&{\mathbb A}_{\ell}^{+}{\bs A}_{\ell}^r+{\mathbb A}_{\ell}^{-}\mathbb{E}_{\ell}{\bs B}_{\ell}^r,\\
			{\bs d}_{\ell}\partial_z\tilde{\psi}_1^r+{\bs c}_{\ell}\tilde{\phi}_2^r\Big|_{z=d_{\ell}+0}=&{\mathbb B}_{\ell}^{+}{\bs A}_{\ell}^r+{\mathbb B}_{\ell}^{-}\mathbb{E}_{\ell}{\bs B}_{\ell}^r,\\
			{\bs a}_{\ell+1}\tilde{\psi}_1^r+{\bs b}_{\ell+1}\partial_z\tilde{\phi}_2^r\Big|_{z=d_{\ell}-0}=&{\mathbb A}_{\ell+1}^{+}\mathbb{E}_{\ell+1}{\bs A}_{\ell+1}^r+{\mathbb A}_{\ell+1}^{-}{\bs B}_{\ell+1}^r,\\
			{\bs d}_{\ell+1}\partial_z\tilde{\psi}_1^r+{\bs c}_{\ell+1}\tilde{\phi}_2^r\Big|_{z=d_{\ell}-0}=&{\mathbb B}_{\ell+1}^{+}\mathbb{E}_{\ell+1}{\bs A}_{\ell+1}^r+{\mathbb B}_{\ell+1}^{-}{\bs B}_{\ell+1}^r,
		\end{aligned}
	\end{equation}
	where the coefficient vectors of the upward and downward propagating wave are defined as
	\begin{equation}\label{freeandreact_component_coff}
		\begin{aligned}
			{\bs A}_{\ell'}^f:=\begin{bmatrix}
				\dfrac{-1}{2\omega^2\rho_{\ell'}}\\[7pt]
				\dfrac{\ri}{2\omega^2\rho_{\ell'}k_{\ell'z}^{s}}
			\end{bmatrix},\quad{\bs B}_{\ell'}^f:=\begin{bmatrix}
				\dfrac{1}{2\omega^2\rho_{\ell'}}\\[7pt]
				\dfrac{\ri}{2\omega^2\rho_{\ell'}k_{\ell'z}^{s}}
			\end{bmatrix},\quad{\bs A}_{\ell}^r:=\begin{bmatrix}
				A_{\ell}^{\psi}\\A_{\ell}^{\phi}
			\end{bmatrix},\quad{\bs B}_{\ell}^r:=\begin{bmatrix}
				B_{\ell}^{\psi}\\B_{\ell}^{\phi}
			\end{bmatrix},
		\end{aligned}
	\end{equation}
	and the interface parameter matrices in equations are abbreviated as 
	\begin{equation}\label{interface_coeffMatrix}
		{\mathbb A}_{\ell}^{\pm}:=\begin{pmatrix}
			{\bs a}_{\ell} & \pm \ri k_{\ell z}^{s}{\bs b}_\ell
		\end{pmatrix},\quad {\mathbb B}_{\ell}^{\pm}:=\begin{pmatrix}
			\pm \ri k_{\ell z}^{c}{\bs d}_\ell & {\bs c}_{\ell}
		\end{pmatrix},\quad \ell=0, 1,\cdots,L.
	\end{equation}
	The constraints \eqref{decaycondcoeff} gives $\bs A_{L}^r=0, \bs B_0^r=0$.
	Therefore, substituting \eqref{densityoninterfaces}-\eqref{vectorvaluelimitLlayer} into the interface conditions \eqref{interface_condition_x23v1}, we obtain the following linear system 
	\begin{equation}\label{layer_down_interface2_ellge0v}
		\begin{aligned}
			\mathbb{A}_{\ell}^{+}{\bs A}_{\ell}^r+\mathbb{A}_{\ell}^{-}\mathbb{E}_{\ell}{\bs B}_{\ell}^r=\mathbb{A}_{\ell+1}^{+}\mathbb{E}_{\ell+1}{\bs A}_{\ell+1}^r+\mathbb{A}_{\ell+1}^{-}{\bs B}_{\ell+1}^r+\bs S_{\ell}^{(1)},\\
			\mathbb{B}_{\ell}^{+}{\bs A}_{\ell}^r+\mathbb{B}_{\ell}^{-}\mathbb{E}_{\ell}{\bs B}_{\ell}^r=\mathbb{B}_{\ell+1}^{+}\mathbb{E}_{\ell+1}{\bs A}_{\ell+1}^r+\mathbb{B}_{\ell+1}^{-}{\bs B}_{\ell+1}^r+\bs S_{\ell}^{(2)},
		\end{aligned}
	\end{equation}
	where 
	\begin{equation*}
		\bs S_{\ell}^{(1)}=\begin{cases}
			\displaystyle   \mathbb{A}_{\ell'}^{+}\mathbb{E}_{\ell'}^{\downarrow}(z'-d_{\ell'-1}){\bs A}_{\ell'}^f, & \ell=\ell'-1,\\
			\displaystyle   -\mathbb{A}_{\ell'}^{-}\mathbb{E}_{\ell'}^{\uparrow}(z'-d_{\ell'}){\bs B}_{\ell'}^f& \ell=\ell',\\
			\displaystyle  \bs 0,&{\rm else},
		\end{cases}\quad \bs S_{\ell}^{(2)}=\begin{cases}
			\displaystyle   \mathbb{B}_{\ell'}^{+}\mathbb{E}_{\ell'}^{\downarrow}(z'-d_{\ell'-1}){\bs A}_{\ell'}^f, & \ell=\ell'-1,\\
			\displaystyle   -\mathbb{B}_{\ell'}^{-}\mathbb{E}_{\ell'}^{\uparrow}(z'-d_{\ell'}){\bs B}_{\ell'}^f& \ell=\ell',\\
			\displaystyle  \bs 0,& {\rm else}.
		\end{cases}
	\end{equation*}
	Further by cross-eliminating terms ${\bs B}_{\ell+1}^r, {\bs A}_{\ell}^r$, we can obtain recursive equations
	\begin{equation}\label{reactrecurrence}
		\begin{bmatrix}
			{\bs A}_{\ell}^r\\
			{\bs B}_{\ell+1}^r
		\end{bmatrix}=\begin{bmatrix}
			\mathbb{R}_{\ell,\ell+1} &  \mathbb{T}_{\ell+1,\ell}\\
			\mathbb{T}_{\ell,\ell+1}& \mathbb{R}_{\ell+1,\ell}
		\end{bmatrix}\begin{bmatrix}
			\mathbb{E}_{\ell} {\bs B}_{\ell}^r\\
			\mathbb{E}_{\ell+1}  {\bs A}_{\ell+1}^r
		\end{bmatrix}+\bs S_{\ell},\quad \ell=0, 1, \cdots, L-1,
	\end{equation}
	where  $\bs S_{\ell}=0$ except
	\begin{equation}
		\bs S_{\ell'-1}=\begin{bmatrix}  \mathbb{T}_{\ell',\ell'-1} \\ \mathbb{R}_{\ell',\ell'-1}
		\end{bmatrix}\mathbb{E}_{\ell'}^{\downarrow}(z'-d_{\ell'-1})\bs A_{\ell'}^f,\quad \bs S_{\ell'}=\begin{bmatrix}  \mathbb{R}_{\ell',\ell'+1}  \\ \mathbb{T}_{\ell',\ell'+1}
		\end{bmatrix}\mathbb{E}_{\ell'}^{\uparrow}(z'-d_{\ell'})\bs B_{\ell'}^f
	\end{equation}
	and
	\begin{equation}\label{Def_reflectR_transmissionT}
		\begin{aligned}
			\mathbb{R}_{\ell,\ell+1}=&-\left[(\mathbb A_{\ell+1}^+)^{-1}\mathbb A_{\ell}^++(\mathbb B_{\ell+1}^+)^{-1}\mathbb B_{\ell}^+\right]^{-1}
			\left[(\mathbb A_{\ell+1}^+)^{-1}\mathbb A_{\ell}^-+(\mathbb B_{\ell+1}^+)^{-1}\mathbb B_{\ell}^-\right],\\
			\mathbb{R}_{\ell+1,\ell}=&-\left[(\mathbb A_{\ell}^-)^{-1}\mathbb A_{\ell+1}^-+(\mathbb B_{\ell}^-)^{-1}\mathbb B_{\ell+1}^-\right]^{-1}
			\left[(\mathbb A_{\ell}^-)^{-1}\mathbb A_{\ell+1}^++(\mathbb B_{\ell}^-)^{-1}\mathbb B_{\ell+1}^+\right],\\
			\mathbb{T}_{\ell+1,\ell}=&2\left[(\mathbb A_{\ell+1}^+)^{-1}\mathbb A_{\ell}^++(\mathbb B_{\ell+1}^+)^{-1}\mathbb B_{\ell}^+\right]^{-1},\quad\mathbb{T}_{\ell,\ell+1}=2\left[(\mathbb A_{\ell}^-)^{-1}\mathbb A_{\ell+1}^-+(\mathbb B_{\ell}^-)^{-1}\mathbb B_{\ell+1}^-\right]^{-1},
		\end{aligned}
	\end{equation}
	are the reflection and transmission matrices. They are well defined as long as $k_{\rho}\neq k_{\ell}^{s}, k_{\ell}^{c}$ for all $\ell=0, 1, \cdots, L$. Their singular points leads to Stoneley-, Rayleigh- and guided-modes.
	\begin{remark}
		The reflection and transmission matrices above have similar form as the reflection and transmission coefficients defined for scalar Helmholtz equation in layered media (cf. \cite{chew1999waves}).
	\end{remark}

	The reflection and transmission system \eqref{reactrecurrence} can be decomposed into upward and downward wave propagation as summarized in the following proposition.
	\begin{prop}
		From equations \eqref{reactrecurrence}, we have recursions
		\begin{align}
			&{\bs B}_{\ell+1}^r=\mathbb{W}_{\ell+1}^-\mathbb{T}_{\ell,\ell+1}\big[\mathbb{E}_{\ell}{\bs B}_{\ell}^r+\delta_{\ell\ell'}\mathbb{E}_{\ell'}^{\uparrow}(z'-d_{\ell'}){\bs B}_{\ell'}^f\big],\label{uprecursion1}\\
			&{\bs A}_{\ell+1}^r=\widetilde{\mathbb R}_{\ell+1,\ell+2}\mathbb{E}_{\ell+1}{\bs B}_{\ell+1}^r,\label{reflectR_intro2}
		\end{align}
		for $\ell=\ell',\ell'+1,\cdots,L-1,$ and
		\begin{align}
			&{\bs A}_{\ell-1}^r=\mathbb{W}_{\ell-1}^+\mathbb{T}_{\ell,\ell-1}\big[\mathbb{E}_{\ell}{\bs A}_{\ell}^r+\delta_{\ell\ell'}\mathbb{E}_{\ell'}^{\downarrow}(z'-d_{\ell'-1}){\bs A}_{\ell'}^f\big],\label{reflectR_intro4}\\
			&{\bs B}_{\ell-1}^r=\widetilde{\mathbb R}_{\ell-1,\ell-2}\mathbb{E}_{\ell-1}{\bs A}_{\ell-1}^r,\label{reflectR_intro3}
		\end{align}
		for $\ell=\ell',\ell'-1, \cdots, 1,$ with starting values
		\begin{equation}\label{layer_ellp_AB_react}
			\begin{aligned}
				{\bs A}_{\ell'}^r=&\mathbb{W}_{\ell'}^+\widetilde{\mathbb R}_{\ell',\ell'+1}\left[\mathbb{E}_{\ell'}^{\uparrow}(z'-d_{\ell'}){\bs B}_{\ell'}^f+\mathbb{E}_{\ell'}\widetilde{\mathbb R}_{\ell',\ell'-1}\mathbb{E}_{\ell'}^{\downarrow}(z'-d_{\ell'-1}){\bs A}_{\ell'}^f\right],\\
				{\bs B}_{\ell'}^r=&\mathbb{W}_{\ell'}^-\widetilde{\mathbb R}_{\ell',\ell'-1}\left[\mathbb{E}_{\ell'}^{\downarrow}(z'-d_{\ell'-1}){\bs A}_{\ell'}^f+\mathbb{E}_{\ell'}\widetilde{\mathbb R}_{\ell',\ell'+1}\mathbb{E}_{\ell'}^{\uparrow}(z'-d_{\ell'}){\bs B}_{\ell'}^f\right].
			\end{aligned}
		\end{equation}
		Here 
		\begin{equation}\label{layerdef_Q_ABf}
			\mathbb{W}_{\ell}^{\pm}:=\begin{cases}
				\left[\mathbb{I}-{\mathbb{R}}_{\ell,\ell\pm1}\mathbb{E}_{\ell}\widetilde{\mathbb{R}}_{\ell,\ell\mp1}\mathbb{E}_{\ell}\right]^{-1},\quad \ell\neq \ell',\\
				\left[\mathbb{I}-\widetilde{\mathbb{R}}_{\ell',\ell'\pm1}\mathbb{E}_{\ell'}\widetilde{\mathbb{R}}_{\ell',\ell'\mp1}\mathbb{E}_{\ell'}\right]^{-1},\quad \ell=\ell',
			\end{cases}
		\end{equation}
		and the generalized reflection matrices are defined recursively as follows:
		\begin{itemize}
			\item For $\ell=L-1,L-2,\cdots, 0$: 
			\begin{equation}\label{generalFs_coeff_Rdown}
				\widetilde{\mathbb R}_{\ell,\ell+1}=\mathbb{R}_{\ell,\ell+1}+\mathbb{T}_{\ell+1,\ell}\mathbb{E}_{\ell+1}\widetilde{\mathbb R}_{\ell+1,\ell+2}\mathbb{E}_{\ell+1}\mathbb{W}_{\ell+1}^-\mathbb{T}_{\ell,\ell+1},
			\end{equation}
			\item For $\ell=0, 1, \cdots, L-1$:
			\begin{equation}\label{generalFs_coeff_Rup}
				\widetilde{\mathbb{R}}_{\ell+1,\ell}={\mathbb{R}}_{\ell+1,\ell}+{\mathbb{T}}_{\ell,\ell+1}\mathbb{E}_{\ell}\widetilde{\mathbb{R}}_{\ell,\ell-1}\mathbb{E}_{\ell}\mathbb{W}_{\ell}^+{\mathbb{T}}_{\ell+1,\ell}.
			\end{equation}
		\end{itemize}
		with zero initial values $\widetilde{\mathbb{R}}_{0,-1}=\widetilde{\mathbb{R}}_{L,L+1}=0$.
	\end{prop}
	\begin{proof}
		We shall prove the conclusion by induction.    Taking $\ell=L-1$ in the recursion \eqref{reactrecurrence} and using the fact ${\bs A}_{L}^r=\mathbf{0}$ gives
		\begin{equation*}
			{\bs A}_{L-1}^r=\mathbb{R}_{L-1,L}\mathbb{E}_{L-1}{\bs B}_{L-1}^r,\quad
			{\bs B}_{L}^r=\mathbb{T}_{L-1,L}\mathbb{E}_{L-1}{\bs B}_{L-1}^r
		\end{equation*}
		Set $\widetilde{\mathbb{R}}_{L-1,L}:=\mathbb{R}_{L-1,L}$, then we have 
		\begin{equation*}
			{\bs A}_{L-1}^r=\widetilde{\mathbb{R}}_{L-1,L}\mathbb{E}_{L-1}{\bs B}_{L-1}^r.
		\end{equation*}
		For any $\ell\in \{L-2,L-3,\cdots,\ell'+1\}$, suppose
		\begin{equation}\label{pre}
			{\bs A}_{\ell+1}^r=\widetilde{\mathbb R}_{\ell+1,\ell+2}\mathbb{E}_{\ell+1}{\bs B}_{\ell+1}^r,
		\end{equation}
		and substituting it into the second equation in \eqref{reactrecurrence} gives
		\begin{equation*}
			{\bs B}_{\ell+1}^r=\left[\mathbb{I}-\mathbb{R}_{\ell+1,\ell}\mathbb{E}_{\ell+1}\widetilde{\mathbb R}_{\ell+1,\ell+2}\mathbb{E}_{\ell+1}\right]^{-1}\mathbb{T}_{\ell,\ell+1}\mathbb{E}_{\ell}{\bs B}_{\ell}^r.
		\end{equation*}
		Substituting it back into \eqref{pre} gives
		\begin{equation}\label{gc2}
			{\bs A}_{\ell+1}^r=\widetilde{\mathbb R}_{\ell+1,\ell+2}\mathbb{E}_{\ell+1}\left[\mathbb{I}-\mathbb{R}_{\ell+1,\ell}\mathbb{E}_{\ell+1}\widetilde{\mathbb R}_{\ell+1,\ell+2}\mathbb{E}_{\ell+1}\right]^{-1}\mathbb{T}_{\ell,\ell+1}\mathbb{E}_{\ell}{\bs B}_{\ell}^r.
		\end{equation}
		Using the expression \eqref{gc2} in the first equation in \eqref{reactrecurrence} gives
		\begin{equation}\label{gc3}
			{\bs A}_{\ell}^r=\widetilde{\mathbb R}_{\ell,\ell+1}\mathbb{E}_{\ell}{\bs B}_{\ell}^r,
		\end{equation}
		which close the induction for $\ell=L-1,L-2,\cdots,\ell'+1$.

		For the special case $\ell=\ell'$, inserting the formula \eqref{gc3} with $\ell=\ell'+1$ into the second equation in \eqref{reactrecurrence} gives
		\begin{equation}\label{gc2v}
			{\bs B}_{\ell'+1}^r=(\mathbb I-\mathbb{R}_{\ell'+1,\ell'}\mathbb{E}_{\ell'+1}\widetilde{\mathbb R}_{\ell'+1,\ell'+2}\mathbb{E}_{\ell'+1})^{-1}\mathbb{T}_{\ell',\ell'+1}[\mathbb{E}_{\ell'}{\bs B}_{\ell'}^r+\mathbb{E}_{\ell'}^{\uparrow}(z'-d_{\ell'}){\bs B}_{\ell'}^f].
		\end{equation}
		Therefore, we complete the proof for \eqref{uprecursion1} and \eqref{reflectR_intro2}. Similarly, we can also prove \eqref{reflectR_intro4}, \eqref{reflectR_intro3}.
		
		Now, we present the derivation for \eqref{layer_ellp_AB_react}. By \eqref{reflectR_intro2} and \eqref{gc2v}, we have
		$$\bs A_{\ell'+1}^r=\widetilde{\mathbb R}_{\ell'+1,\ell'+2}\mathbb{E}_{\ell'+1}\mathbb W_{\ell'+1}^-\mathbb{T}_{\ell',\ell'+1}[\mathbb{E}_{\ell'}{\bs B}_{\ell'}^r+\mathbb{E}_{\ell'}^{\uparrow}(z'-d_{\ell'}){\bs B}_{\ell'}^f]$$
		Substituting it into the first equation in \eqref{reactrecurrence} for $\ell=\ell'$, we get
		\begin{equation*}
			{\bs A}_{\ell'}^r=\widetilde{\mathbb R}_{\ell',\ell'+1}[\mathbb{E}_{\ell'}{\bs B}_{\ell'}^r+\mathbb{E}_{\ell'}^{\uparrow}(z'-d_{\ell'}){\bs B}_{\ell'}^f].
		\end{equation*}
		Similarly, we can also prove 
		\begin{equation*}
			{\bs B}_{\ell'}^r=\widetilde{\mathbb R}_{\ell',\ell'-1}[\mathbb{E}_{\ell'}{\bs A}_{\ell'}^r+\mathbb{E}_{\ell'}^{\downarrow}(z'-d_{\ell'-1}){\bs A}_{\ell'}^f].
		\end{equation*}
		Then, the solution of the system consists of the above two equations gives \eqref{layer_ellp_AB_react}.
	\end{proof}

	Define upward propagators
	\begin{equation*}
		\mathbb{S}_{\ell,\ell+1}:=\mathbb W_{\ell+1}^-\mathbb{T}_{\ell,\ell+1}\mathbb{E}_{\ell},\quad \ell=0, 1, \cdots, L-1.
	\end{equation*}
	Then, the recurrence formulation \eqref{uprecursion1}  for $\{\bs B_{\ell}^r\}_{\ell=\ell'}^L$ can be reformulated into
	\begin{equation}\label{recurrencesourcelayer}
		{\bs B}_{\ell+1}^r=\mathbb{S}_{\ell,\ell+1}{\bs B}_{\ell}^r+\delta_{\ell\ell'}\mathbb{S}_{\ell',\ell'+1}\mathbb{E}_{\ell'}^{\uparrow}(z'-d_{\ell'-1}){\bs B}_{\ell'}^f,\quad\ell=\ell',\ell'+1,\cdots,L-1,
	\end{equation}
	Obviously the above matrices $\mathbb{S}_{\ell,\ell+1}$ are bounded. Define downward generalized transmission coefficient matrices
	\begin{equation}\label{generaltransmatdown}
		\widetilde{\mathbb T}_{\ell'\ell}=\mathbb{S}_{\ell-1,\ell}\mathbb{S}_{\ell-2,\ell-1}\cdots\mathbb{S}_{\ell',\ell'+1},\quad\ell=\ell'+1, \ell'+2,\cdots,L.
	\end{equation}
	By the representation \eqref{layer_ellp_AB_react} and the identity
	\begin{equation*}
		\mathbb{W}_{\ell}^{\pm}=\mathbb{I}+\mathbb{W}_{\ell}^{\pm}\widetilde{\mathbb{R}}_{\ell,\ell\pm1}\mathbb{E}_{\ell}\widetilde{\mathbb{R}}_{\ell,\ell\mp1}\mathbb{E}_{\ell},\quad \ell=0,1,\cdots, L,
	\end{equation*}
	the recurrence formula \eqref{recurrencesourcelayer} can be written as
	\begin{equation}\label{B_ellgeellp}
		\begin{split}
			{\bs B}_{\ell}^r=&\widetilde{\mathbb{T}}_{\ell'\ell}{\bs B}_{\ell'}^r+\widetilde{\mathbb{T}}_{\ell'\ell}\mathbb{E}_{\ell'}^{\uparrow}(z'-d_{\ell'-1}){\bs B}_{\ell'}^f,\\
			=&\widetilde{\mathbb{T}}_{\ell'\ell}\left[(\mathbb{W}_{\ell'}^-\widetilde{\mathbb R}_{\ell',\ell'-1}\mathbb{E}_{\ell'}\widetilde{\mathbb R}_{\ell',\ell'+1}\mathbb{E}_{\ell'}+\mathbb I)\mathbb{E}_{\ell'}^{\uparrow}(z'-d_{\ell'-1}){\bs B}_{\ell'}^f+\mathbb{W}_{\ell'}^-\widetilde{\mathbb R}_{\ell',\ell'-1}\mathbb{E}_{\ell'}^{\downarrow}(z'-d_{\ell'-1}){\bs A}_{\ell'}^f\right]\\
			=&\widetilde{\mathbb{T}}_{\ell'\ell}\mathbb{W}_{\ell'}^-\big[\widetilde{\mathbb R}_{\ell',\ell'-1}\mathbb{E}_{\ell'}^{\downarrow}(z'-d_{\ell'-1}){\bs A}_{\ell'}^f+\mathbb{E}_{\ell'}^{\uparrow}(z'-d_{\ell'-1}){\bs B}_{\ell'}^f\big]
		\end{split}
	\end{equation}
	for $\ell=\ell'+1,\cdots,L$. 
	Further, applying the reflection formula \eqref{reflectR_intro2} gives
	\begin{equation}\label{B_ellgeellpv2}
		\bs A_{\ell}^{r}=\widetilde{\mathbb R}_{\ell,\ell+1}\mathbb{E}_{\ell}\widetilde{\mathbb{T}}_{\ell'\ell}\mathbb{W}_{\ell'}^-\big[\widetilde{\mathbb R}_{\ell',\ell'-1}\mathbb{E}_{\ell'}^{\downarrow}(z'-d_{\ell'-1}){\bs A}_{\ell'}^f+\mathbb{E}_{\ell'}^{\uparrow}(z'-d_{\ell'-1}){\bs B}_{\ell'}^f\big]
	\end{equation}
	for $\ell=\ell'+1,\ell'+2,\cdots,L$.
	
	Mimicking the derivation for $\ell=\ell'-1, \ell'-2, \cdots, 0$, we have
	\begin{equation}\label{layer_coff_AB_exceptellpv1}
		\begin{aligned}
			\bs A_{\ell}^r=\widetilde{\mathbb{T}}_{\ell'\ell}\mathbb{W}_{\ell'}^+\big[\widetilde{\mathbb R}_{\ell',\ell'+1}\mathbb{E}_{\ell'}^{\uparrow}(z'-d_{\ell'}){\bs B}_{\ell'}^f+\mathbb{E}_{\ell'}^{\downarrow}(z'-d_{\ell'}){\bs A}_{\ell'}^f\big],\quad \ell=\ell'-1, \ell'-2, \cdots, 0.
		\end{aligned}
	\end{equation}
	where the upward general transmission matrices and the upward propagator are defined as
	\begin{equation}\label{generaltransmatup}
		\begin{aligned}
			\mathbb{S}_{\ell+1,\ell}:=\mathbb W_{\ell}^+\mathbb{T}_{\ell+1,\ell}\mathbb{E}_{\ell+1},\quad  \widetilde{\mathbb T}_{\ell'\ell}:=\mathbb{S}_{\ell+1,\ell}\mathbb{S}_{\ell+2,\ell+1}\cdots\mathbb{S}_{\ell',\ell'-1},
		\end{aligned}
	\end{equation}
	for $\ell=\ell'-1, \ell'-2,\cdots, 0$. 
	Applying reflection formula \eqref{reflectR_intro3} gives
	\begin{equation}\label{layer_coff_AB_exceptellpv2}
		\bs B_{\ell}^r=\widetilde{\mathbb R}_{\ell,\ell-1}\mathbb{E}_{\ell}\widetilde{\mathbb{T}}_{\ell'\ell}\mathbb{W}_{\ell'}^+\big[\widetilde{\mathbb R}_{\ell',\ell'+1}\mathbb{E}_{\ell'}^{\uparrow}(z'-d_{\ell'}){\bs B}_{\ell'}^f+\mathbb{E}_{\ell'}^{\downarrow}(z'-d_{\ell'}){\bs A}_{\ell'}^f\big],
	\end{equation}
	for $\ell=\ell'-1, \ell'-2, \cdots, 0$.
	
	Define density matrices  as follows
	\begin{itemize}
		\item In the source layer ($\ell=\ell'$)
		\begin{equation}\label{densitymatrixinsourcelayer}
			\begin{aligned}
				\mathbb{D}_{\ell'\ell'}^{\uparrow\uparrow}:=&\mathbb{W}_{\ell'}^{+}\widetilde{\mathbb{R}}_{\ell',\ell'+1}\mathbb E_{\ell'}\widetilde{\mathbb{R}}_{\ell',\ell'-1},\quad
				\mathbb{D}_{\ell'\ell'}^{\uparrow\downarrow}:=\mathbb{W}_{\ell'}^{+}\widetilde{\mathbb R}_{\ell',\ell'+1},\\
				\mathbb{D}_{\ell'\ell'}^{\downarrow\uparrow}:=&\mathbb{W}_{\ell'}^{-}\widetilde{\mathbb{R}}_{\ell',\ell'-1},\quad
				\mathbb{D}_{\ell'\ell'}^{\downarrow\downarrow}:=\mathbb{W}_{\ell'}^{-}\widetilde{\mathbb R}_{\ell',\ell'-1}\mathbb E_{\ell'}\widetilde{\mathbb{R}}_{\ell',\ell'+1},
			\end{aligned}
		\end{equation}
		\item In the layer above the source layer ($\ell<\ell'$)
		\begin{equation}\label{densitymatrixinuptargetlayer}
			\begin{aligned}
				\mathbb{D}_{\ell\ell'}^{\uparrow\uparrow}:=&\widetilde{\mathbb{T}}_{\ell'\ell}\mathbb{W}_{\ell'}^{+}\mathbb E_{\ell'}^{-1},\quad
				\mathbb{D}_{\ell\ell'}^{\uparrow\downarrow}:=\widetilde{\mathbb{T}}_{\ell'\ell}\mathbb{W}_{\ell'}^{+}\widetilde{\mathbb R}_{\ell',\ell'+1},\\
				\mathbb{D}_{\ell\ell'}^{\downarrow\uparrow}:=&\widetilde{\mathbb R}_{\ell,\ell-1}\mathbb{E}_{\ell}\widetilde{\mathbb{T}}_{\ell'\ell}\mathbb{W}_{\ell'}^+\mathbb E_{\ell'}^{-1},\quad
				\mathbb{D}_{\ell\ell'}^{\downarrow\downarrow}:=\widetilde{\mathbb R}_{\ell,\ell-1}\mathbb{E}_{\ell}\widetilde{\mathbb{T}}_{\ell'\ell}\mathbb{W}_{\ell'}^+\widetilde{\mathbb R}_{\ell',\ell'+1},
			\end{aligned}
		\end{equation}
		\item In the layer below the source layer ($\ell>\ell'$)
		\begin{equation}\label{densitymatrixindowntargetlayer}
			\begin{aligned}
				\mathbb{D}_{\ell\ell'}^{\uparrow\uparrow}:=&\widetilde{\mathbb R}_{\ell,\ell+1}\mathbb{E}_{\ell}\widetilde{\mathbb{T}}_{\ell'\ell}\mathbb{W}_{\ell'}^-\widetilde{\mathbb R}_{\ell',\ell'-1},\quad\mathbb{D}_{\ell\ell'}^{\uparrow\downarrow}:=\widetilde{\mathbb R}_{\ell,\ell+1}\mathbb{E}_{\ell}\widetilde{\mathbb{T}}_{\ell'\ell}\mathbb{W}_{\ell'}^-\mathbb E_{\ell'}^{-1},\\  
				\mathbb{D}_{\ell\ell'}^{\downarrow\uparrow}:=&\widetilde{\mathbb{T}}_{\ell'\ell}\mathbb{W}_{\ell'}^-\widetilde{\mathbb R}_{\ell',\ell'-1},\quad
				\mathbb{D}_{\ell\ell'}^{\downarrow\downarrow}:=\widetilde{\mathbb{T}}_{\ell'\ell}\mathbb{W}_{\ell'}^-\mathbb E_{\ell'}^{-1}.
			\end{aligned}
		\end{equation}
	\end{itemize} 
	Then, using the formulas \eqref{B_ellgeellp}-\eqref{B_ellgeellpv2},  \eqref{layer_coff_AB_exceptellpv1}-\eqref{layer_coff_AB_exceptellpv2} and \eqref{layer_ellp_AB_react} in \eqref{reactfourerformula2v}-\eqref{reactfourerformula3v}, we obtain
	\begin{equation}\label{psi1_phi2_react}
		\begin{split}
			\begin{bmatrix}
				\tilde{\psi}_{\ell\ell'1}^r\\\tilde{\phi}_{\ell\ell'2}^r
			\end{bmatrix}=&{\mathbb{E}}_{\ell}^{\uparrow}(z-d_{\ell})\bs A_{\ell}^{r}+{\mathbb{E}}_{\ell}^{\downarrow}(z-d_{\ell-1}) \bs B_{\ell}^{r}\\
			=&{\mathbb{E}}_{\ell}^{\uparrow}(z-d_{\ell})\mathbb D_{\ell\ell'}^{\uparrow\uparrow}\mathbb{E}_{\ell'}^{\downarrow}(z'-d_{\ell'-1}){\bs A}_{\ell'}^f+{\mathbb{E}}_{\ell}^{\uparrow}(z-d_{\ell})\mathbb D_{\ell\ell'}^{\uparrow\downarrow}\mathbb{E}_{\ell'}^{\uparrow}(z'-d_{\ell'}){\bs B}_{\ell'}^f\\
			+&{\mathbb{E}}_{\ell}^{\downarrow}(z-d_{\ell-1})\mathbb D_{\ell\ell'}^{\downarrow\uparrow}\mathbb{E}_{\ell'}^{\downarrow}(z'-d_{\ell'-1}){\bs A}_{\ell'}^f+{\mathbb{E}}_{\ell}^{\downarrow}(z-d_{\ell-1})\mathbb D_{\ell\ell'}^{\downarrow\downarrow}\mathbb{E}_{\ell'}^{\uparrow}(z'-d_{\ell'}){\bs B}_{\ell'}^f,
		\end{split}
	\end{equation}
	for $\ell=0,1,\cdots,L$.
	As in \cite{yuan2026fast}, the superscripts of the two arrows indicate the propagating direction of the waves in the vicinity of the target and source points, respectively. In particular, the definition $\widetilde{\mathbb{R}}_{0,-1}=\widetilde{\mathbb{R}}_{L,L+1}=\bs 0$ leads to
	\begin{equation}
		\mathbb{D}_{LL}^{\uparrow\downarrow}=\mathbb{D}_{00}^{\downarrow\uparrow}=\mathbb{D}_{0\ell'}^{\downarrow\downarrow}=\mathbb{D}_{0\ell'}^{\downarrow\uparrow}=\mathbb{D}_{L\ell'}^{\uparrow\uparrow}=\mathbb{D}_{L\ell'}^{\uparrow\downarrow}=\mathbb{D}_{\ell L}^{\uparrow\downarrow}=\mathbb{D}_{\ell L}^{\downarrow\downarrow}=\mathbb{D}_{\ell0}^{\downarrow\uparrow}=\mathbb{D}_{\ell0}^{\uparrow\uparrow}=\bs 0.
	\end{equation}

	\subsection{Recursive algorithm for the solution of problem \texorpdfstring{\eqref{FCsystem_phi3_psi2}}{}}\label{sect_solver_phi3_psi2}
	Similarly to the above derivation, we first consider the free-space component $\tilde{\phi}_{3}^f, \tilde{\psi}_{2}^f$ which satisfies:
	\begin{equation*}
		\begin{aligned}
			\partial_{zz}\tilde{\phi}_3^f+(k_{\ell z}^{s})^{2}\tilde{\phi}_3^f=-\dfrac{1}{\omega^2\rho_{\ell'}}\delta_{\ell\ell'}\delta'(z-z'),\quad
			\partial_{zz}\tilde{\psi}_2^f+(k_{\ell z}^{c})^{2}\tilde{\psi}_2^f=-\dfrac{1}{\omega^2\rho_{\ell'}}\delta_{\ell\ell'}\delta(z-z').
		\end{aligned}
	\end{equation*}
	The solution are given by
	\begin{equation}\label{freespacesecondgroup}
		\begin{aligned}
			\tilde{\phi}_3^f(k_{\rho},z,z')=&\dfrac{1}{2\omega^2\rho_{\ell'}}\left[-H(z-z')e^{\ri k_{\ell'z}^{s}(z-z')}+H(z'-z)e^{\ri k_{\ell'z}^{s}(z'-z)}\right],\\
			\tilde{\psi}_2^f(k_{\rho},z,z')=&\dfrac{\ri}{2\omega^2\rho_{\ell'}k_{\ell'z}^{c}}\left[H(z-z')e^{\ri k_{\ell'z}^{c}(z-z')}+H(z'-z)e^{\ri k_{\ell'z}^{c}(z'-z)}\right].
		\end{aligned}
	\end{equation}
	By the principle of superposition, the reaction components $\tilde{\phi}_{3}^r=\tilde{\phi}_{3}-\tilde{\phi}_{3}^f, \tilde{\psi}_{2}^r=\tilde{\psi}_{2}-\tilde{\psi}_{2}^f$ satisfy the homogeneous Helmholtz equations:
	\begin{equation*}
		\partial_{zz}\tilde{\psi}_2^r+(k_{\ell z}^{c})^{2}\tilde{\psi}_2^r=0,\quad \partial_{zz}\tilde{\phi}_3^r+(k_{\ell z}^{s})^{2}\tilde{\phi}_3^r=0,
	\end{equation*}
	inside each layer and interface conditions 
	\begin{equation*}
		\left\llbracket  \tilde{\bs a}\tilde{\psi}_2+\tilde{\bs b} 	\partial_z\tilde{\phi}_3\right\rrbracket=0,\quad\left\llbracket \tilde{\bs d }\partial_{z}\tilde{\psi}_2+ \tilde{\bs c}\tilde{\phi}_3\right\rrbracket=0,\quad  z=d_{\ell},\quad\ell=0,1,\cdots,L-1,
	\end{equation*}
	where the layered parameter vectors $\bs a,\bs b,\bs c,\bs d$ in each $\ell$-th layer  are defined as
	\begin{equation}\label{interface_coeffv}
		\tilde{\bs a}_{\ell}=\begin{bmatrix}
			\gamma_{\ell}(k_{\ell z}^{c})^{2}+\lambda_{\ell}k_{\rho}^2 \\ -k_{\rho}^2
		\end{bmatrix},\quad\tilde{\bs b}_{\ell}=\begin{bmatrix}
			2\mu_{\ell} \\ 1
		\end{bmatrix},\quad\tilde{\bs c}_{\ell}=\begin{bmatrix}
			\mu_{\ell} ((k_{\ell z}^{s})^{2}-k_{\rho}^2) \\ -1
		\end{bmatrix},\quad\tilde{\bs d}_{\ell}=\begin{bmatrix}
			2\mu_{\ell} k_{\rho}^2 \\ 1
		\end{bmatrix}.
	\end{equation}
	
	As in the derivation for $\tilde\phi_2, \tilde \psi_1$, we define the upward and downward coefficient matrices of the free-space field as follows
	\begin{equation*}
		\widetilde{\bs A}_{\ell'}^f:=\frac{1}{2\omega^2\rho_{\ell'}}\begin{bmatrix}
			\dfrac{\ri}{k_{\ell'z}^{c}}\\[7pt]
			-1
		\end{bmatrix},\quad\widetilde{\bs B}_{\ell'}^f:=\frac{1}{2\omega^2\rho_{\ell'}}\begin{bmatrix}
			\dfrac{\ri}{k_{\ell'z}^{c}}\\[7pt]
			1
		\end{bmatrix}.
	\end{equation*}
	Then, from the representation \eqref{freespacesecondgroup}, we calculate that:
	\begin{equation*}
		\begin{split}
			\begin{bmatrix}
				\tilde{\psi}_2^f(k_{\rho},d_{\ell'-1},z')\\
				\tilde{\phi}_3^f(k_{\rho},d_{\ell'-1},z')
			\end{bmatrix}=\mathbb{E}_{\ell'}^{\downarrow}(z'-d_{\ell'-1})\widetilde{\bs A}_{\ell'}^f,\quad
			\begin{bmatrix}
				\tilde{\psi}_2^f(k_{\rho},d_{\ell'},z')\\
				\tilde{\phi}_3^f(k_{\rho},d_{\ell'},z')
			\end{bmatrix}=\mathbb{E}_{\ell'}^{\uparrow}(z'-d_{\ell'})\widetilde{\bs B}_{\ell'}^f.
		\end{split}
	\end{equation*}
	Following the derivation in the last subsection,  we obtain the solution of \eqref{FCsystem_phi3_psi2} as follows
	\begin{equation}\label{react_psi2phi3}
		\begin{aligned}
			\begin{bmatrix}
				\tilde{\psi}_{2}^r\\\tilde{\phi}_{3}^r
			\end{bmatrix}=&{\mathbb{E}}_{\ell}^{\uparrow}(z-d_{\ell})\mathbb P_{\ell\ell'}^{\uparrow\uparrow}\mathbb{E}_{\ell'}^{\downarrow}(z'-d_{\ell'-1})\widetilde{\bs A}_{\ell'}^f+{\mathbb{E}}_{\ell}^{\uparrow}(z-d_{\ell})\mathbb P_{\ell\ell'}^{\uparrow\downarrow}\mathbb{E}_{\ell'}^{\uparrow}(z'-d_{\ell'})\widetilde{\bs B}_{\ell'}^f\\
			+&{\mathbb{E}}_{\ell}^{\downarrow}(z-d_{\ell-1})\mathbb P_{\ell\ell'}^{\downarrow\uparrow}\mathbb{E}_{\ell'}^{\downarrow}(z'-d_{\ell'-1})\widetilde{\bs A}_{\ell'}^f+{\mathbb{E}}_{\ell}^{\downarrow}(z-d_{\ell-1})\mathbb P_{\ell\ell'}^{\downarrow\downarrow}\mathbb{E}_{\ell'}^{\uparrow}(z'-d_{\ell'})\widetilde{\bs B}_{\ell'}^f,
		\end{aligned}
	\end{equation}
	where the density matrices $\mathbb{P}_{\ell\ell'}^{\ast\star}(k_{\rho})$ and corresponding general reflection, transmission matrices are defined in the same form as \eqref{densitymatrixinsourcelayer}-\eqref{densitymatrixindowntargetlayer}, \eqref{generalFs_coeff_Rdown}-\eqref{generalFs_coeff_Rup}, \eqref{generaltransmatdown} and \eqref{generaltransmatup} but with new interface coefficient vectors $\tilde{\bs a}_{\ell},\tilde{\bs b}_{\ell},\tilde{\bs c}_{\ell},\tilde{\bs d}_{\ell}$ defined in \eqref{interface_coeffv}.
	
	\subsection{Dyadic Green's functions in physical domain}
	The previous derivation in this section has given the analytical form for the solutions $\tilde{\phi}_1,\tilde{\phi}_2,\tilde{\phi}_3,\tilde{\psi}_1,\tilde{\psi}_2$ of the interface problem \eqref{FCsystem_phi1}-\eqref{FCsystem_phi3_psi2}. Combined with the formula \eqref{Elasticdynamics_layerDGf_Fourierv2} of the elastic wave dyadic Green's function in the Fourier domain, the Sommerfeld integral form of the elastic wave dyadic Green's function in the time domain can be obtained through the inverse Fourier transform.
	
	It is obvious that there are exponential decay terms reside in the density matrices ${\mathbb{D}}_{\ell\ell'}^{\ast\star},\mathbb{P}_{\ell\ell'}^{\ast\star}$ due to the $\mathbb E_{\ell}$ term in the propagators $\mathbb S_{\ell,\ell+1}, \mathbb S_{\ell+1,\ell}$. By simply manipulating the terms, we are able to explicitly factorize the exponential terms out which facilitate the computation of the dyadic Green's function. For this purpose, we first introduce the following bounded transformations 
	\begin{equation}\label{boundidentity_up2}
		\begin{split}
			&\mathcal{P}_{\ell}^{\pm}(\mathbb{F}):=\mathbb{E}_{\ell}^{\uparrow}(\pm d_{\ell})\mathbb{F}\mathbb{E}_{\ell}^{\downarrow}(\pm d_{\ell})=\begin{bmatrix}
				F_{11}&F_{12}e^{\pm\ri (k_{\ell z}^{c}-k_{\ell z}^{s})d_{\ell}}\\
				F_{21}e^{\pm\ri (k_{\ell z}^{s}-k_{\ell z}^{c})d_{\ell}}&F_{22}
			\end{bmatrix}\\
			&\mathcal{Q}_{\ell}^{\pm}(\mathbb{F}):=\mathbb{E}_{\ell}^{\uparrow}(\pm d_{\ell-1})\mathbb{F}\mathbb{E}_{\ell}^{\downarrow}(\pm d_{\ell-1})=\begin{bmatrix}
				F_{11}&F_{12}e^{\pm\ri (k_{\ell z}^{c}-k_{\ell z}^{s})d_{\ell-1}}\\
				F_{21}e^{\pm\ri (k_{\ell z}^{s}-k_{\ell z}^{c})d_{\ell-1}}&F_{22}
			\end{bmatrix}
		\end{split}
	\end{equation}
	for any bounded $2\times2$ matrices $\mathbb{F}=(F_{ij})_{2\times 2}$. From asymptotic
	\begin{equation}\label{wavesqrt_limitfact}
		\pm \ri (k_{\ell z}^{c}-k_{\ell z}^{s}) = \frac{\mp((k_{\ell}^{c})^{2}-(k_{\ell}^{s})^{2})}{\sqrt{\kr^2 - (k_{\ell}^{c})^{2}} + \sqrt{\kr^2 - (k_{\ell}^{s})^{2}}} = \mathcal{O}(\kr^{-1}),\quad \kr \to \infty,
	\end{equation}
	we have $\mathcal{P}_{\ell}(\mathbb{F}),\mathcal{Q}_{\ell}(\mathbb{F}) \sim \bb{F}$ as $\kr \to \infty$.

	From the definition \eqref{densitymatrixinsourcelayer}-\eqref{densitymatrixindowntargetlayer}, we can see that the exponential decay of ${\mathbb{D}}_{\ell\ell'}^{\ast\star},\mathbb{P}_{\ell\ell'}^{\ast\star}$ is governed by $\mathbb 
	E_{\ell'}$ when $\ell=\ell'$ and by  decay factors $\{\mathbb E_{\ell+1},\mathbb E_{\ell+2},\cdots,\mathbb E_{\ell'}\}$ and $\{\mathbb E_{\ell-1},\mathbb E_{\ell-2},\cdots,\mathbb E_{\ell'}\}$  for $\ell<\ell'$ and $\ell>\ell'$, respectively. Direct calculation shows that
	\begin{equation}
		\mathbb{E}_{\ell}^{\downarrow}(d_{\ell})\widetilde{\mathbb{T}}_{\ell'\ell}\mathbb{E}_{\ell'}^{\uparrow}(d_{\ell'})\sim \mathcal O(1),\quad \ell<\ell',\quad
		\mathbb{E}_{\ell}^{\uparrow}(d_{\ell-1})\widetilde{\mathbb{T}}_{\ell'\ell}\mathbb{E}_{\ell'}^{\downarrow}(d_{\ell'-1})\sim \mathcal O(1),\quad\ell>\ell',
	\end{equation}
	as $\kr \to \infty$. Define new density matrices as follows:
	\begin{itemize}
		\item In the source layer ($\ell=\ell'$)
		\begin{equation}\label{densitymatrixinsourcelayerv2}
			\begin{aligned}
				\widehat{\mathbb{D}}_{\ell'\ell'}^{\uparrow\uparrow}:=&\mathcal{P}_{\ell'}^{+}(\mathbb{W}_{\ell'}^{+}\widetilde{\mathbb{R}}_{\ell',\ell'+1})\mathcal{Q}_{\ell'}^{+}(\widetilde{\mathbb{R}}_{\ell',\ell'-1}),\quad\widehat{\mathbb{D}}_{\ell'\ell'}^{\uparrow\downarrow}:=\mathcal{P}_{\ell'}^{-}(\mathbb{W}_{\ell'}^{+}\widetilde{\mathbb R}_{\ell',\ell'+1})
				\\
				\widehat{\mathbb{D}}_{\ell'\ell'}^{\downarrow\uparrow}:=&\mathcal{Q}_{\ell'}^{+}(\mathbb{W}_{\ell'}^{-}\widetilde{\mathbb R}_{\ell',\ell'-1}),\quad
				\widehat{\mathbb{D}}_{\ell'\ell'}^{\downarrow\downarrow}:=\mathcal{Q}_{\ell'}^{-}(\mathbb{W}_{\ell'}^{-}\widetilde{\mathbb{R}}_{\ell',\ell'-1})\mathcal{P}_{\ell'}^{-}(\widetilde{\mathbb{R}}_{\ell',\ell'+1}),
			\end{aligned}
		\end{equation}
		\item In the layer above the source layer ($\ell<\ell'$)
		\begin{equation}\label{densitymatrixinuptargetlayerv2}
			\begin{aligned}
				\widehat{\mathbb{D}}_{\ell\ell'}^{\uparrow\uparrow}:=&\mathbb{E}_{\ell}^{\downarrow}(d_{\ell})\widetilde{\mathbb{T}}_{\ell'\ell}\mathbb{E}_{\ell'}^{\uparrow}(d_{\ell'})\mathcal{P}_{\ell'}^{-}(\mathbb{W}_{\ell'}^{+}),\\\widehat{\mathbb{D}}_{\ell\ell'}^{\uparrow\downarrow}:=&\mathbb{E}_{\ell}^{\downarrow}(d_{\ell})\widetilde{\mathbb{T}}_{\ell'\ell}\mathbb{E}_{\ell'}^{\uparrow}(d_{\ell'})\mathcal{P}_{\ell'}^{-}(\mathbb{W}_{\ell'}^{+}\widetilde{\mathbb R}_{\ell',\ell'+1}),\\
				\widehat{\mathbb{D}}_{\ell\ell'}^{\downarrow\uparrow}:=&\mathcal{Q}_{\ell}^{-}(\widetilde{\mathbb R}_{\ell,\ell-1})\mathbb{E}_{\ell}^{\downarrow}(d_{\ell})\widetilde{\mathbb{T}}_{\ell'\ell}\mathbb{E}_{\ell'}^{\uparrow}(d_{\ell'})\mathcal{P}_{\ell'}^{-}(\mathbb{W}_{\ell'}^{+}),\\
				\widehat{\mathbb{D}}_{\ell\ell'}^{\downarrow\downarrow}:=&\mathcal{Q}_{\ell}^{-}(\widetilde{\mathbb R}_{\ell,\ell-1})\mathbb{E}_{\ell}^{\downarrow}(d_{\ell})\widetilde{\mathbb{T}}_{\ell'\ell}\mathbb{E}_{\ell'}^{\uparrow}(d_{\ell'})\mathcal{P}_{\ell'}^{-}(\mathbb{W}_{\ell'}^{+}\widetilde{\mathbb R}_{\ell',\ell'+1}),
			\end{aligned}
		\end{equation}
		\item In the layer below the source layer ($\ell>\ell'$)
		\begin{equation}\label{densitymatrixindowntargetlayerv2}
			\begin{aligned}
				\widehat{\mathbb{D}}_{\ell\ell'}^{\uparrow\uparrow}:=&\mathcal{P}_{\ell}^{+}(\widetilde{\mathbb R}_{\ell,\ell+1})\mathbb{E}_{\ell}^{\uparrow}(d_{\ell-1})\widetilde{\mathbb{T}}_{\ell'\ell}\mathbb{E}_{\ell'}^{\downarrow}(d_{\ell'-1})\mathcal{Q}_{\ell'}^{+}(\mathbb{W}_{\ell'}^{-}),\\
				\widehat{\mathbb{D}}_{\ell\ell'}^{\uparrow\downarrow}:=&\mathcal{P}_{\ell}^{+}(\widetilde{\mathbb R}_{\ell,\ell+1})\mathbb{E}_{\ell}^{\uparrow}(d_{\ell-1})\widetilde{\mathbb{T}}_{\ell'\ell}\mathbb{E}_{\ell'}^{\downarrow}(d_{\ell'-1})\mathcal{Q}_{\ell'}^{+}(\mathbb{W}_{\ell'}^{-}\widetilde{\mathbb R}_{\ell',\ell'-1}),\\
				\widehat{\mathbb{D}}_{\ell\ell'}^{\downarrow\uparrow}:=&\mathbb{E}_{\ell}^{\uparrow}(d_{\ell-1})\widetilde{\mathbb{T}}_{\ell'\ell}\mathbb{E}_{\ell'}^{\downarrow}(d_{\ell'-1})\mathcal{Q}_{\ell'}^{+}(\mathbb{W}_{\ell'}^{-}\widetilde{\mathbb R}_{\ell',\ell'-1}),\\
				\widehat{\mathbb{D}}_{\ell\ell'}^{\downarrow\downarrow}:=&\mathbb{E}_{\ell}^{\uparrow}(d_{\ell-1})\widetilde{\mathbb{T}}_{\ell'\ell}\mathbb{E}_{\ell'}^{\downarrow}(d_{\ell'-1})\mathcal{Q}_{\ell'}^{+}(\mathbb{W}_{\ell'}^{-}).
			\end{aligned}
		\end{equation}
	\end{itemize} 
	Note that
	$$\mathbb E^{\downarrow}(d_{\ell})\mathbb E^{\uparrow}(d_{\ell-1})=\mathbb E^{\uparrow}(d_{\ell-1})\mathbb E^{\downarrow}(d_{\ell})=\mathbb E_{\ell}.$$
	We have
	\begin{equation}\label{densitymatrixinsourcelayerv1}
		\begin{aligned}
			\widehat{\mathbb{D}}_{\ell'\ell'}^{\uparrow\uparrow}=&\mathbb{E}_{\ell'}^{\uparrow}(d_{\ell'}){\mathbb{D}}_{\ell'\ell'}^{\uparrow\uparrow}\mathbb{E}_{\ell'}^{\downarrow}(d_{\ell'-1}),\quad\widehat{\mathbb{D}}_{\ell'\ell'}^{\uparrow\downarrow}=\mathcal{P}_{\ell'}^{-}({\mathbb{D}}_{\ell'\ell'}^{\uparrow\downarrow}),
			\\
			\widehat{\mathbb{D}}_{\ell'\ell'}^{\downarrow\uparrow}=&\mathcal{Q}_{\ell'}^{+}({\mathbb{D}}_{\ell'\ell'}^{\downarrow\uparrow}),\quad
			\widehat{\mathbb{D}}_{\ell'\ell'}^{\downarrow\downarrow}=\mathbb{E}_{\ell'}^{\downarrow}(d_{\ell'-1}){\mathbb{D}}_{\ell'\ell'}^{\downarrow\downarrow}\mathbb{E}_{\ell'}^{\uparrow}(d_{\ell'}),\\
			\widehat{\mathbb{D}}_{\ell\ell'}^{\uparrow\uparrow}=&\mathbb{E}_{\ell}^{\downarrow}(d_{\ell}){\mathbb{D}}_{\ell\ell'}^{\uparrow\uparrow}\mathbb{E}_{\ell'}^{\uparrow}(d_{\ell'-1}),\quad\widehat{\mathbb{D}}_{\ell\ell'}^{\uparrow\downarrow}=\mathbb{E}_{\ell}^{\downarrow}(d_{\ell}){\mathbb{D}}_{\ell\ell'}^{\uparrow\downarrow}\mathbb{E}_{\ell'}^{\uparrow}(d_{\ell'}),\quad \ell<\ell',\\
			\widehat{\mathbb{D}}_{\ell\ell'}^{\downarrow\uparrow}=&\mathbb{E}_{\ell}^{\downarrow}(d_{\ell-1}){\mathbb{D}}_{\ell\ell'}^{\downarrow\uparrow}\mathbb{E}_{\ell'}^{\uparrow}(d_{\ell'-1}),\quad
			\widehat{\mathbb{D}}_{\ell\ell'}^{\downarrow\downarrow}=\mathbb{E}_{\ell}^{\downarrow}(d_{\ell-1}){\mathbb{D}}_{\ell\ell'}^{\downarrow\downarrow}\mathbb{E}_{\ell'}^{\uparrow}(d_{\ell'}),\quad \ell<\ell',\\
			\widehat{\mathbb{D}}_{\ell\ell'}^{\uparrow\uparrow}=&\mathbb{E}_{\ell}^{\uparrow}(d_{\ell}){\mathbb{D}}_{\ell\ell'}^{\uparrow\uparrow}\mathbb{E}_{\ell'}^{\downarrow}(d_{\ell'-1}),\quad
			\widehat{\mathbb{D}}_{\ell\ell'}^{\uparrow\downarrow}=\mathbb{E}_{\ell}^{\uparrow}(d_{\ell}){\mathbb{D}}_{\ell\ell'}^{\uparrow\downarrow}\mathbb{E}_{\ell'}^{\downarrow}(d_{\ell'}),\quad \ell>\ell',\\
			\widehat{\mathbb{D}}_{\ell\ell'}^{\downarrow\uparrow}=&\mathbb{E}_{\ell}^{\uparrow}(d_{\ell-1}){\mathbb{D}}_{\ell\ell'}^{\downarrow\uparrow}\mathbb{E}_{\ell'}^{\downarrow}(d_{\ell'-1}),\quad
			\widehat{\mathbb{D}}_{\ell\ell'}^{\downarrow\downarrow}=\mathbb{E}_{\ell}^{\uparrow}(d_{\ell-1}){\mathbb{D}}_{\ell\ell'}^{\downarrow\downarrow}\mathbb{E}_{\ell'}^{\downarrow}(d_{\ell'}),\quad \ell>\ell'.
		\end{aligned}
	\end{equation}
	are all $\mathcal O(1)$ matrices and \eqref{psi1_phi2_react} can be rewritten as
	\begin{equation}\label{react_psi1phhi2_sourcelayer}
		\begin{split}
			\begin{bmatrix}
				\tilde{\psi}_{\ell'\ell'1}^r\\\tilde{\phi}_{\ell'\ell'2}^r
			\end{bmatrix}=&\mathbb{E}_{\ell'}^{\uparrow}(z)\widehat{\mathbb{D}}_{\ell'\ell'}^{\uparrow\downarrow}\mathbb{E}_{\ell'}^{\uparrow}(-\tau_{\ell'}(z')){\bs B}_{\ell'}^f
			+\mathbb{E}_{\ell'}^{\uparrow}(-\tau_{\ell'}(z))\widehat{\mathbb{D}}_{\ell'\ell'}^{\uparrow\uparrow}\mathbb{E}_{\ell'}^{\downarrow}(-\tau_{\ell'-1}(z')){\bs A}_{\ell'}^f\\
			+&\mathbb{E}_{\ell'}^{\downarrow}(-\tau_{\ell'-1}(z))\widehat{\mathbb{D}}_{\ell'\ell'}^{\downarrow\downarrow}\mathbb{E}_{\ell'}^{\uparrow}(-\tau_{\ell'}(z')){\bs B}_{\ell'}^f
			+\mathbb{E}_{\ell'}^{\downarrow}(z)\widehat{\mathbb{D}}_{\ell'\ell'}^{\downarrow\uparrow}\mathbb{E}_{\ell'}^{\downarrow}(-\tau_{\ell'-1}(z')){\bs A}_{\ell'}^f,
		\end{split}
	\end{equation}
	and
	\begin{equation}\label{react_psi1phhi2_sourcelayerdown}
		\begin{split}
			\begin{bmatrix}
				\tilde{\psi}_{\ell\ell'1}^r\\\tilde{\phi}_{\ell\ell'2}^r
			\end{bmatrix}=&{\mathbb{E}}_{\ell}^{\uparrow}(z)\widehat{\mathbb{D}}_{\ell\ell'}^{\uparrow\downarrow}\mathbb{E}_{\ell'}^{\uparrow}(-\tau_{\ell'}(z')){\bs B}_{\ell'}^f+{\mathbb{E}}_{\ell}^{\uparrow}(z)\widehat{\mathbb{D}}_{\ell\ell'}^{\uparrow\uparrow}\mathbb{E}_{\ell'}^{\downarrow}(z'){\bs A}_{\ell'}^f\big]\\
			+&\mathbb{E}_{\ell}^{\downarrow}(-\tau_{\ell-1}(z))\widehat{\mathbb{D}}_{\ell\ell'}^{\downarrow\downarrow}\mathbb{E}_{\ell'}^{\uparrow}(-\tau_{\ell'}(z')){\bs B}_{\ell'}^f+\mathbb{E}_{\ell}^{\downarrow}(-\tau_{\ell-1}(z))\widehat{\mathbb{D}}_{\ell\ell'}^{\downarrow\uparrow}\mathbb{E}_{\ell'}^{\downarrow}(z'){\bs A}_{\ell'}^f,\quad \ell<\ell',\\
			\begin{bmatrix}
				\tilde{\psi}_{\ell\ell'1}^r\\\tilde{\phi}_{\ell\ell'2}^r
			\end{bmatrix}=&{\mathbb{E}}_{\ell}^{\uparrow}(-\tau_{\ell}(z))\widehat{\mathbb{D}}_{\ell\ell'}^{\uparrow\downarrow}\mathbb{E}_{\ell'}^{\uparrow}(z'){\bs B}_{\ell'}^f+{\mathbb{E}}_{\ell}^{\uparrow}(-\tau_{\ell}(z))\widehat{\mathbb{D}}_{\ell\ell'}^{\uparrow\uparrow}\mathbb{E}_{\ell'}^{\downarrow}(-\tau_{\ell'-1}(z')){\bs A}_{\ell'}^f\\
			+&{\mathbb{E}}_{\ell}^{\downarrow}(z)\widehat{\mathbb{D}}_{\ell\ell'}^{\downarrow\downarrow}\mathbb{E}_{\ell'}^{\uparrow}(z'){\bs B}_{\ell'}^f+{\mathbb{E}}_{\ell}^{\downarrow}(z)\widehat{\mathbb{D}}_{\ell\ell'}^{\downarrow\uparrow}\mathbb{E}_{\ell'}^{\downarrow}(-\tau_{\ell'-1}(z')){\bs A}_{\ell'}^f,\quad\ell>\ell'.
		\end{split}
	\end{equation}

	Similar new expressions for $\{\tilde{\phi}_{3},\tilde{\psi}_{2}\}$ are given as follows:
	\begin{equation}\label{react_psi2phhi3_sourcelayer}
		\begin{split}
			\begin{bmatrix}
				\tilde{\psi}_{\ell'\ell'2}^r\\\tilde{\phi}_{\ell'\ell'3}^r
			\end{bmatrix}=&\mathbb{E}_{\ell'}^{\uparrow}(z)\widehat{\mathbb{P}}_{\ell'\ell'}^{\uparrow\downarrow}\mathbb{E}_{\ell'}^{\uparrow}(-\tau_{\ell'}(z'))\tilde{\bs B}_{\ell'}^f
			+\mathbb{E}_{\ell'}^{\uparrow}(-\tau_{\ell'}(z))\widehat{\mathbb{P}}_{\ell'\ell'}^{\uparrow\uparrow}\mathbb{E}_{\ell'}^{\downarrow}(-\tau_{\ell'-1}(z'))\tilde{\bs A}_{\ell'}^f\\
			+&\mathbb{E}_{\ell'}^{\downarrow}(-\tau_{\ell'-1}(z))\widehat{\mathbb{P}}_{\ell'\ell'}^{\downarrow\downarrow}\mathbb{E}_{\ell'}^{\uparrow}(-\tau_{\ell'}(z'))\tilde{\bs B}_{\ell'}^f
			+\mathbb{E}_{\ell'}^{\downarrow}(z)\widehat{\mathbb{P}}_{\ell'\ell'}^{\downarrow\uparrow}\mathbb{E}_{\ell'}^{\downarrow}(-\tau_{\ell'-1}(z'))\tilde{\bs A}_{\ell'}^f
		\end{split}
	\end{equation}
	and
	\begin{equation}\label{react_psi2phhi3_sourcelayerdown}
		\begin{split}
			\begin{bmatrix}
				\tilde{\psi}_{\ell\ell'2}^r\\\tilde{\phi}_{\ell\ell'3}^r
			\end{bmatrix}=&{\mathbb{E}}_{\ell}^{\uparrow}(-\tau_{\ell}(z))\widehat{\mathbb{P}}_{\ell\ell'}^{\uparrow\downarrow}\mathbb{E}_{\ell'}^{\uparrow}(z')\tilde{\bs B}_{\ell'}^f+{\mathbb{E}}_{\ell}^{\uparrow}(-\tau_{\ell}(z))\widehat{\mathbb{P}}_{\ell\ell'}^{\uparrow\uparrow}\mathbb{E}_{\ell'}^{\downarrow}(-\tau_{\ell'-1}(z'))\tilde{\bs A}_{\ell'}^f\\
			+&{\mathbb{E}}_{\ell}^{\downarrow}(z)\widehat{\mathbb{P}}_{\ell\ell'}^{\downarrow\downarrow}\mathbb{E}_{\ell'}^{\uparrow}(z')\tilde{\bs B}_{\ell'}^f+{\mathbb{E}}_{\ell}^{\downarrow}(z)\widehat{\mathbb{P}}_{\ell\ell'}^{\downarrow\uparrow}\mathbb{E}_{\ell'}^{\downarrow}(-\tau_{\ell'-1}(z'))\tilde{\bs A}_{\ell'}^f,\quad\ell>\ell',\\
			\begin{bmatrix}
				\tilde{\psi}_{\ell\ell'2}^r\\\tilde{\phi}_{\ell\ell'3}^r
			\end{bmatrix}=&{\mathbb{E}}_{\ell}^{\uparrow}(z)\widehat{\mathbb{P}}_{\ell\ell'}^{\uparrow\downarrow}\mathbb{E}_{\ell'}^{\uparrow}(-\tau_{\ell'}(z'))\tilde{\bs B}_{\ell'}^f+{\mathbb{E}}_{\ell}^{\uparrow}(z)\widehat{\mathbb{P}}_{\ell\ell'}^{\uparrow\uparrow}\mathbb{E}_{\ell'}^{\downarrow}(z')\tilde{\bs A}_{\ell'}^f\big]\\
			+&\mathbb{E}_{\ell}^{\downarrow}(-\tau_{\ell-1}(z))\widehat{\mathbb{P}}_{\ell\ell'}^{\downarrow\downarrow}\mathbb{E}_{\ell'}^{\uparrow}(-\tau_{\ell'}(z'))\tilde{\bs B}_{\ell'}^f+\mathbb{E}_{\ell}^{\downarrow}(-\tau_{\ell-1}(z))\widehat{\mathbb{P}}_{\ell\ell'}^{\downarrow\uparrow}\mathbb{E}_{\ell'}^{\downarrow}(z')\tilde{\bs A}_{\ell'}^f,\quad\ell<\ell'.
		\end{split}
	\end{equation}
	where $\widehat{\mathbb{P}}_{\ell\ell'}^{\ast\star}$ is defined in the same form as in \eqref{densitymatrixinsourcelayerv2}-\eqref{densitymatrixindowntargetlayerv2} with $\widetilde{\mathbb{T}}_{\ell\ell'}, \widetilde{\mathbb{R}}_{\ell\ell'}, \mathbb W_{\ell}^{\pm}$ replaced by their counterpart w.r.t.  interface coefficient vectors $\tilde{\bs a}_{\ell},\tilde{\bs b}_{\ell},\tilde{\bs c}_{\ell},\tilde{\bs d}_{\ell}$ defined in \eqref{interface_coeffv}.

	In order to formulate expressions for the coefficients $\tilde{\phi}_1^r,\tilde{\phi}_2^r,\tilde{\phi}_3^r,\tilde{\psi}_1^r,\tilde{\psi}_2^r$, we define four double wavenumber exponential functions as follows:
	\begin{equation}\label{zexponential_def}
		\begin{split}
			{Z}_{\ell\ell'}^{\uparrow\uparrow}(k_{\rho}, z,z';k^{c},k^{s})=&\begin{cases}
				\displaystyle e^{\ri (k_{\ell z}^{c}z-k_{\ell' z}^{s}z')},\quad \ell<\ell'\\
				\displaystyle e^{\ri (k_{\ell' z}^{s}\tau_{\ell'-1}(z')-k_{\ell z}^{c}\tau_{\ell}(z))},\quad\ell\geq \ell'
			\end{cases}\\
			{Z}_{\ell\ell'}^{\uparrow\downarrow}(k_{\rho}, z,z';k^{c},k^{s})=&\begin{cases}
				\displaystyle e^{\ri (k_{\ell z}^{c}z-k_{\ell' z}^{s}\tau_{\ell'}(z'))},\quad \ell\leq\ell'\\
				\displaystyle e^{\ri (k_{\ell' z}^{s}z'-k_{\ell z}^{c}\tau_{\ell }(z))},\quad \ell>\ell'
			\end{cases}\\
			{Z}_{\ell\ell'}^{\downarrow\uparrow}(k_{\rho}, z,z';k^{c},k^{s})=&\begin{cases}
				\displaystyle e^{\ri (k_{\ell z}^{c}\tau_{\ell-1}(z)-k_{\ell' z}^{s}z')},\quad\ell<\ell',\\
				\displaystyle e^{\ri(k_{\ell' z}^{s}\tau_{\ell'-1}(z')-k_{\ell z}^{c}z)},\quad\ell\geq \ell',
			\end{cases}\\
			{Z}_{\ell\ell'}^{\downarrow\downarrow}(k_{\rho}, z,z';k^{c},k^{s})=&\begin{cases}
				\displaystyle e^{\ri (k_{\ell z}^{c}\tau_{\ell-1}(z)-k_{\ell' z}^{s}\tau_{\ell'}(z'))},\quad\ell\leq \ell'\\
				e^{\ri (k_{\ell' z}^{s}z'-k_{\ell z}^{c}z)},\quad \ell>\ell',
			\end{cases}
		\end{split}
	\end{equation}
	where $k^{c},k^{s}$ are the two piecewise constant wave numbers. Their derivatives w.r.t. $z$ are given by
	\begin{equation}
		\partial_z{Z}_{\ell\ell'}^{\ast\star}(k_{\rho}, z,z';k^{c},k^{s})=\ri\tau^{\ast}k_{\ell z}^{c}{Z}_{\ell\ell'}^{\ast\star}(k_{\rho}, z,z';k^{c},k^{s}),
	\end{equation}
	where $\tau^{\uparrow}=1\ ,\ \tau^{\downarrow}=-1$ are sign variables.
	According to the values of $k^{c},k^{s}$, we further denote by
	\begin{equation}
		\begin{aligned}
			{Z}_{\ell\ell',PP}^{\ast\star}(k_{\rho}, z,z'):={Z}_{\ell\ell'}^{\ast\star}(k_{\rho}, z,z';k^{c},k^{c}),\quad{Z}_{\ell\ell',PS}^{\ast\star}(k_{\rho}, z,z'):={Z}_{\ell\ell'}^{\ast\star}(k_{\rho}, z,z';k^{c},k^{s}),\\
			{Z}_{\ell\ell',SP}^{\ast\star}(k_{\rho}, z,z'):={Z}_{\ell\ell'}^{\ast\star}(k_{\rho}, z,z';k^{s},k^{c}),\quad{Z}_{\ell\ell',SS}^{\ast\star}(k_{\rho}, z,z'):={Z}_{\ell\ell'}^{\ast\star}(k_{\rho}, z,z';k^{s},k^{s}).
		\end{aligned}
	\end{equation}
	Obviously, the exponential function \eqref{zexponential} defined earlier is equal to ${Z}_{\ell\ell',SS}^{\ast\star}(k_{\rho}, z,z')$.

	Denote by $\sigma_{\ell\ell'k}^{\ast\star}$, $\tilde\sigma_{\ell\ell'k}^{\ast\star}$ the entries of the density matrices $\widehat{\mathbb{D}}_{\ell\ell'}^{\ast\star},\widehat{\mathbb{P}}_{\ell\ell'}^{\ast\star}$, i.e.,
	\begin{equation}
		\widehat{\mathbb{D}}_{\ell\ell'}^{\ast\star}(k_{\rho})=\begin{bmatrix}
			\sigma_{\ell\ell'1}^{\ast\star}(k_{\rho}) & \sigma_{\ell\ell'2}^{\ast\star}(k_{\rho})\\
			\sigma_{\ell\ell'3}^{\ast\star}(k_{\rho}) & \sigma_{\ell\ell'4}^{\ast\star}(k_{\rho})
		\end{bmatrix},\quad\widehat{\mathbb{P}}_{\ell\ell'}^{\ast\star}(k_{\rho})=\begin{bmatrix}
			\tilde{\sigma}_{\ell\ell'1}^{\ast\star}(k_{\rho}) & \tilde{\sigma}_{\ell\ell'2}^{\ast\star}(k_{\rho})\\
			\tilde{\sigma}_{\ell\ell'3}^{\ast\star}(k_{\rho}) & \tilde{\sigma}_{\ell\ell'4}^{\ast\star}(k_{\rho})
		\end{bmatrix},\quad\ast,\star=\uparrow,\downarrow.
	\end{equation}
	Then, \eqref{solution_phi1},\eqref{react_psi1phhi2_sourcelayer}-\eqref{react_psi1phhi2_sourcelayerdown} and \eqref{react_psi2phhi3_sourcelayer}-\eqref{react_psi2phhi3_sourcelayerdown} can be rewritten as
	\begin{align}
		\tilde{\phi}_{\ell\ell'1}^r=&\dfrac{1}{\mu_{\ell'}}\dfrac{\ri}{2k_{\ell'z}^{s}}\sum_{\ast,\star=\uparrow,\downarrow}\sigma_{\ell\ell'0}^{\ast\star}(k_{\rho}){Z}_{\ell\ell',SS}^{\ast\star}(k_{\rho}, z,z'),\label{layerreact_phi1}\\
		\tilde{\psi}_{\ell\ell'1}^r=&\dfrac{1}{2\omega^2\rho_{\ell'}}\sum_{\ast,\star=\uparrow,\downarrow}\left[-\tau^{\star}\sigma_{\ell\ell'1}^{\ast\star}(k_{\rho}){Z}_{\ell\ell',PP}^{\ast\star}(k_{\rho}, z,z')+\dfrac{\ri}{k_{\ell'z}^{s}}\sigma_{\ell\ell'2}^{\ast\star}(k_{\rho}){Z}_{\ell\ell',PS}^{\ast\star}(k_{\rho}, z,z')\right],\label{layerreact_psi1}\\
		\tilde{\phi}_{\ell\ell'2}^r=&\dfrac{1}{2\omega^2\rho_{\ell'}}\sum_{\ast,\star=\uparrow,\downarrow}\left[-\tau^{\star}\sigma_{\ell\ell'3}^{\ast\star}(k_{\rho}){Z}_{\ell\ell',SP}^{\ast\star}(k_{\rho}, z,z')+\dfrac{\ri}{k_{\ell'z}^{s}}\sigma_{\ell\ell'4}^{\ast\star}(k_{\rho}){Z}_{\ell\ell',SS}^{\ast\star}(k_{\rho}, z,z')\right],\label{layerreact_phi2}\\
		\tilde{\psi}_{\ell\ell'2}^r=&\dfrac{1}{2\omega^2\rho_{\ell'}}\sum_{\ast,\star=\uparrow,\downarrow}\left[\dfrac{\ri}{k_{\ell'z}^{c}}\tilde{\sigma}_{\ell\ell'1}^{\ast\star}(k_{\rho}){Z}_{\ell\ell',PP}^{\ast\star}(k_{\rho}, z,z')-\tau^{\star}\tilde{\sigma}_{\ell\ell'2}^{\ast\star}(k_{\rho}){Z}_{\ell\ell',PS}^{\ast\star}(k_{\rho}, z,z')\right],\label{layerreact_psi2}\\
		\tilde{\phi}_{\ell\ell'3}^r=&\dfrac{1}{2\omega^2\rho_{\ell'}}\sum_{\ast,\star=\uparrow,\downarrow}\left[\dfrac{\ri}{k_{\ell'z}^{c}}\tilde{\sigma}_{\ell\ell'3}^{\ast\star}(k_{\rho}){Z}_{\ell\ell',SP}^{\ast\star}(k_{\rho}, z,z')-\tau^{\star}\tilde{\sigma}_{\ell\ell'4}^{\ast\star}(k_{\rho}){Z}_{\ell\ell',SS}^{\ast\star}(k_{\rho}, z,z')\right].\label{layerreact_phi3}
	\end{align}
	From \eqref{Elasticdynamics_layerDGf_Fourierv2}, we have
	\begin{equation}\label{Elasticdynamics_layerDGf_Fourierv2_reaction}
		\begin{aligned}
			\widehat{\mathbb{G}}_{\ell\ell'}^r=&\tilde{\phi}_{1}^r\J{1}+\left(k_{\rho}^2\tilde{\phi}_{2}^r-\partial_z\tilde{\psi}_{1}^r\right)\J{2}+\left(\partial_z\tilde{\phi}_{2}^r-\tilde{\psi}_{1}^r\right)\J{3}+\left(\tilde{\phi}_{3}^r-\partial_z\tilde{\psi}_{2}^r\right)\J{4}\\
			&+\left(k_{\rho}^{-2}[\tilde{\phi}_{1}^r+\partial_z\tilde{\phi}_{3}^r]-\tilde{\psi}_{2}^r\right)\J{5}.
		\end{aligned}
	\end{equation}
	Substituting \eqref{layerreact_phi1}-\eqref{layerreact_phi3} into \eqref{Elasticdynamics_layerDGf_Fourierv2_reaction} gives
	\begin{equation}\label{reactfieldspectraldomain}
		\begin{aligned}
			\widehat{\mathbb{G}}_{\ell\ell'}^{r}=&\dfrac{1}{2}\sum_{\ast,\star=\uparrow,\downarrow}\Big[\bs\Theta_{\ell\ell',PP}^{\ast\star}(k_x,k_y){Z}_{\ell\ell',PP}^{\ast\star}(k_{\rho}, z,z')+\bs\Theta_{\ell\ell',PS}^{\ast\star}(k_x,k_y){Z}_{\ell\ell',PS}^{\ast\star}(k_{\rho}, z,z')\\&+\bs\Theta_{\ell\ell',SP}^{\ast\star}(k_x,k_y){Z}_{\ell\ell',SP}^{\ast\star}(k_{\rho}, z,z')+\bs\Theta_{\ell\ell',SS}^{\ast\star}(k_x,k_y){Z}_{\ell\ell',SS}^{\ast\star}(k_{\rho}, z,z')\Big],
		\end{aligned}
	\end{equation}
	where
	\begin{equation*}
		\begin{aligned}
			\bs\Theta_{\ell\ell',PP}^{\ast\star}(k_x,k_y):=&\dfrac{-1}{\omega^2\rho_{\ell'}}\left(-\left[\J{3}+\ri\tau^{\ast}k_{\ell z}^{c}\J{2}\right]\tau^{\star}\sigma_{\ell\ell'1}^{\ast\star}(k_{\rho})+\left[\J{5}+\ri\tau^{\ast}k_{\ell z}^{c}\J{4}\right]\dfrac{\ri\tilde{\sigma}_{\ell\ell'1}^{\ast\star}(k_{\rho})}{k_{\ell'z}^{c}}\right),\\
			\bs\Theta_{\ell\ell',PS}^{\ast\star}(k_x,k_y):=&\dfrac{-1}{\omega^2\rho_{\ell'}}\left(\left[\J{3}+\ri\tau^{\ast}k_{\ell z}^{c}\J{2}\right]\dfrac{\ri\sigma_{\ell\ell'2}^{\ast\star}(k_{\rho})}{k_{\ell'z}^{s}}-\left[\J{5}+\ri\tau^{\ast}k_{\ell z}^{c}\J{4}\right]\tau^{\star}\tilde{\sigma}_{\ell\ell'2}^{\ast\star}(k_{\rho})\right),\\
			\bs\Theta_{\ell\ell',SP}^{\ast\star}(k_x,k_y):=&\dfrac{1}{\omega^2\rho_{\ell'}}\left(-\left[k_{\rho}^2\J{2}+\ri\tau^{\ast}k_{\ell z}^{s}\J{3}\right]\tau^{\star}\sigma_{\ell\ell'3}^{\ast\star}(k_{\rho})+\left[\J{4}+\ri\tau^{\ast}k_{\ell z}^{s}k_{\rho}^{-2}\J{5}\right]\dfrac{\ri\tilde{\sigma}_{\ell\ell'3}^{\ast\star}(k_{\rho})}{k_{\ell'z}^{c}}\right),\\
			\bs\Theta_{\ell\ell',SS}^{\ast\star}(k_x,k_y):=&\dfrac{1}{\omega^2\rho_{\ell'}}\left(\left[k_{\rho}^2\J{2}+\ri\tau^{\ast}k_{\ell z}^{s}\J{3}\right]\dfrac{i\sigma_{\ell\ell'4}^{\ast\star}(k_{\rho})}{k_{\ell'z}^{s}}-\left[\J{4}+\ri\tau^{\ast}k_{\ell z}^{s}k_{\rho}^{-2}\J{5}\right]\tau^{\star}\tilde{\sigma}_{\ell\ell'4}^{\ast\star}(k_{\rho})\right)\\
			&+\dfrac{\ri}{\mu_{\ell'}k_{\ell'z}^{s}}\left[\J{1}+k_{\rho}^{-2}\J{5}\right]\sigma_{\ell\ell'0}^{\ast\star}(k_{\rho}).
		\end{aligned}
	\end{equation*}
	Applying inverse Fourier transform on \eqref{reactfieldspectraldomain}, we obtain dyadic Green's function in physical domain, i.e., 
	\begin{equation}
		\begin{split}
			{\mathbb{G}}_{\ell\ell'}^{r}(\bs r, \bs r')={\mathbb{G}}_{\ell\ell',PP}^{r}(\bs r, \bs r')+{\mathbb{G}}_{\ell\ell',PS}^{r}(\bs r, \bs r')+{\mathbb{G}}_{\ell\ell',SP}^{r}(\bs r, \bs r')+{\mathbb{G}}_{\ell\ell',SS}^{r}(\bs r, \bs r'),
		\end{split}
	\end{equation}
	where 
	$$
	{\mathbb{G}}_{\ell\ell',AB}^{r}(\bs r, \bs r')=\dfrac{1}{8\pi^2}\sum_{\ast,\star=\uparrow,\downarrow}\int_{\mathbb R^2}e^{\ri k_x(x-x')+\ri k_y(y-y')}{Z}_{\ell\ell',AB}^{\ast\star}(k_{\rho}, z,z')\bs\Theta_{\ell\ell',AB}^{\ast\star}(k_x,k_y){\rm d}k_x{\rm d}k_y,
	$$
	for $A, B\in {P, S}$.
	
	\section{Numerical examples}\label{sect_numer}
	
	In this section, some numerical results are presented to validate the formula and recursion derived for the computation of the layered-medium elastodynamic dyadic Green's function. We test the displacement field generated by a point-force source in a five-layer medium. The homogeneous-medium limit of the layered Green's function and the transmission conditions across all material interfaces are examined.
	\begin{table}[htbp]
		\centering
		\caption{Material parameters in each layer.}
		\label{tab:layer_params}
		\begin{tabular}{c|ccccc}
			\hline
			& \(\ell=0\) & \(\ell=1\) & \(\ell=2\) & \(\ell=3\) & \(\ell=4\) \\
			\hline
			\(\rho_\ell\)    & 0.7 & 1.5 & 1.2 & 0.8 & 2.1 \\
			\(\mu_\ell\)    & 1.2 & 0.8 & 1.0 & 2.3 & 1.5 \\
			\(\lambda_\ell\) & 1.0 & 1.0 & 1.0 & 1.0 & 1.0 \\
			\hline
		\end{tabular}
	\end{table}
	
\begin{figure}[ht]
		\centering
		\begin{subfigure}{0.31\textwidth}
			\includegraphics[width=\linewidth]{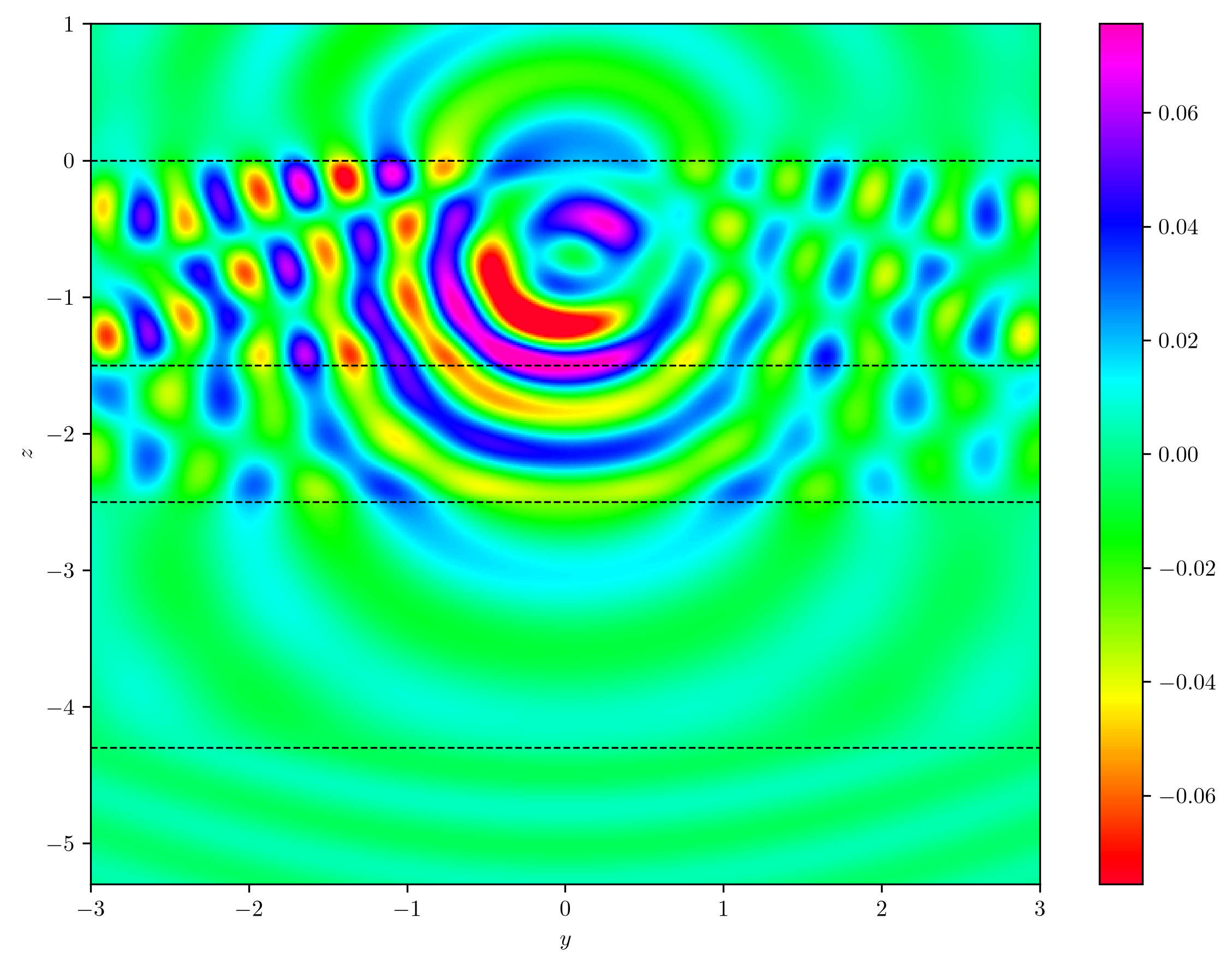}
			\caption{${\rm Re}(u_x)$}
		\end{subfigure}
		\quad
		\begin{subfigure}{0.31\textwidth}
			\includegraphics[width=\linewidth]{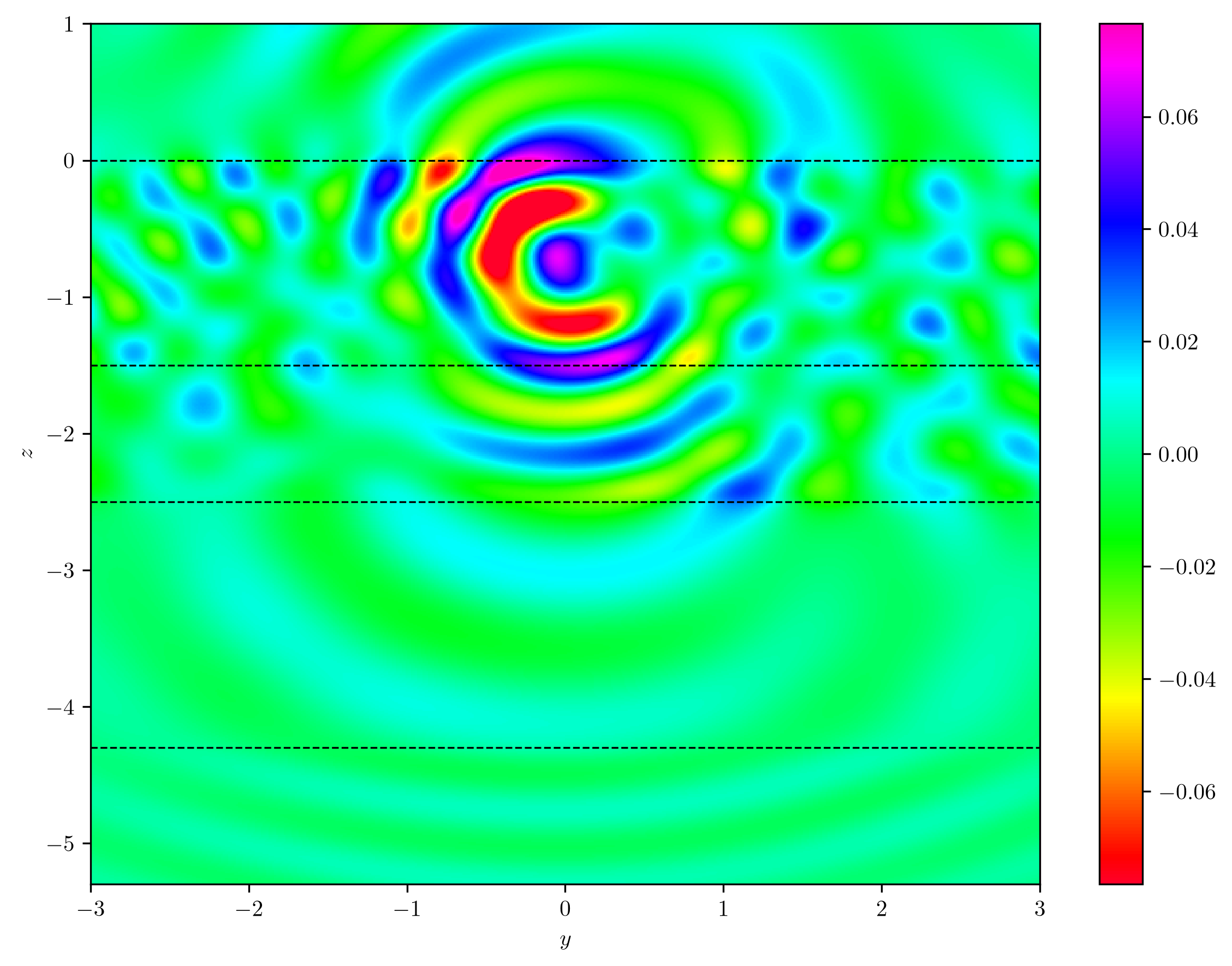}
			\caption{${\rm Re}(u_y)$}
		\end{subfigure}
		\quad
		\begin{subfigure}{0.31\textwidth}
			\includegraphics[width=\linewidth]{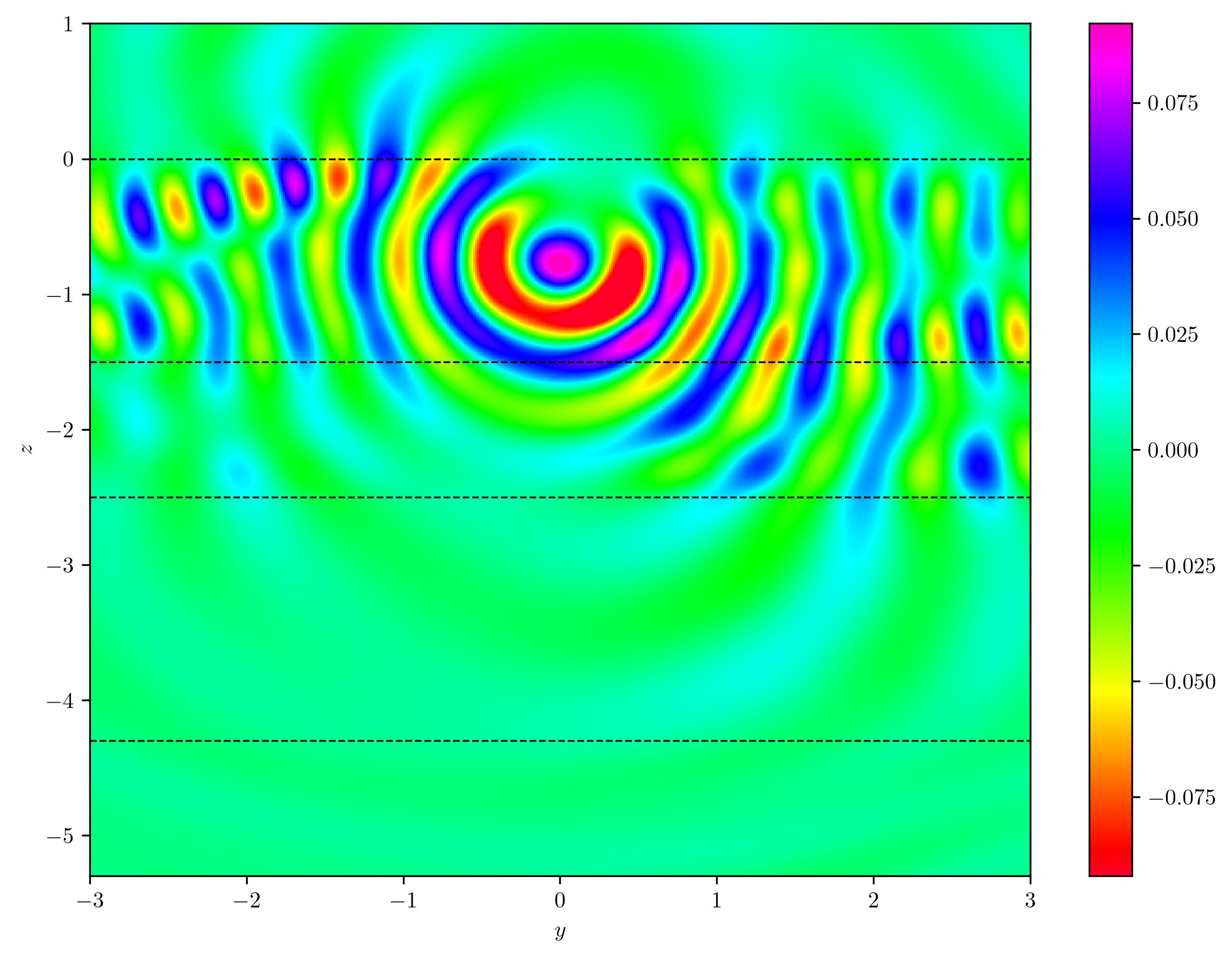}
			\caption{${\rm Re}(u_z)$}
		\end{subfigure}\\
		\begin{subfigure}{0.31\textwidth}
			\includegraphics[width=\linewidth]{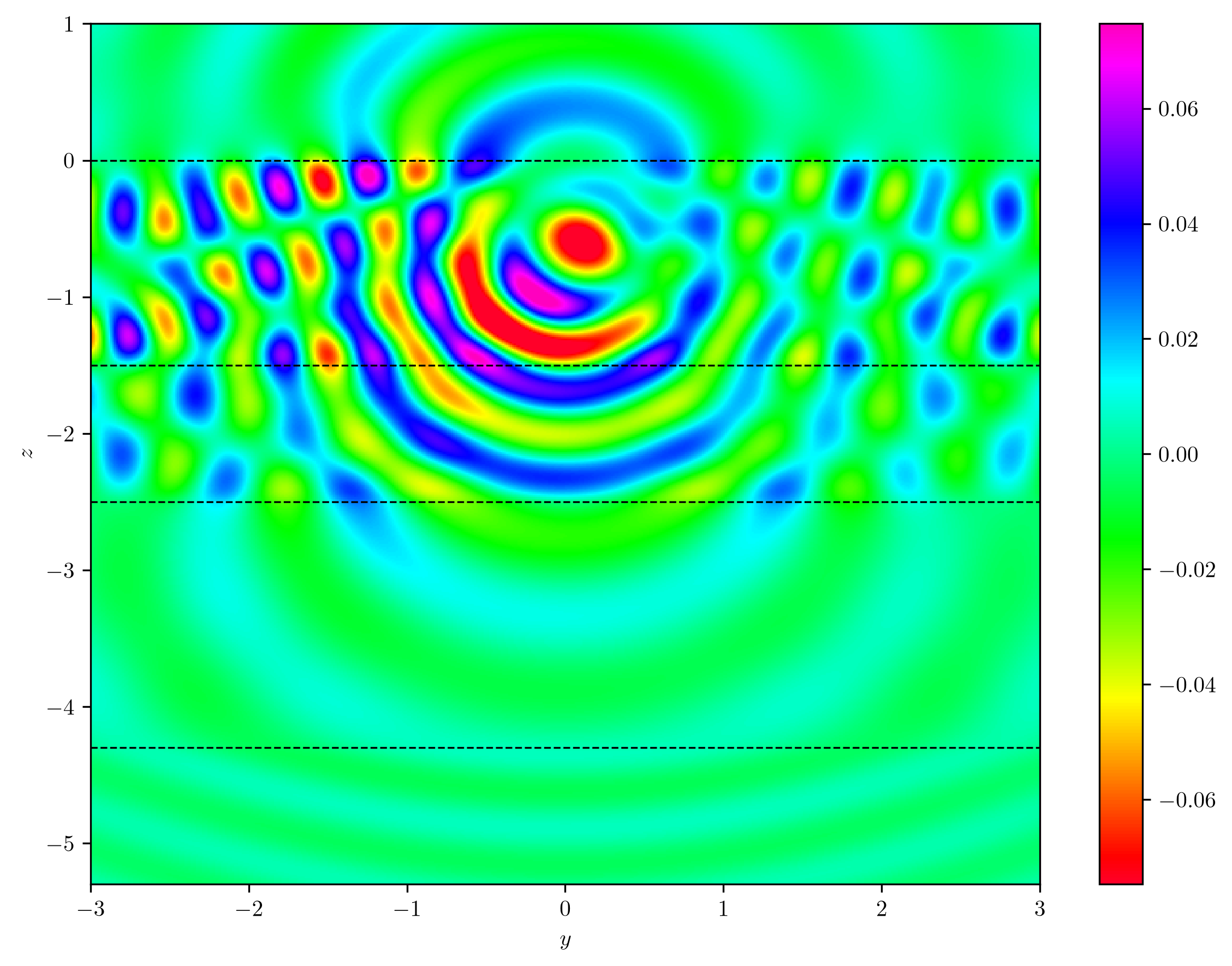}
			\caption{${\rm Im}(u_x)$}
		\end{subfigure}
		\quad
		\begin{subfigure}{0.31\textwidth}
			\includegraphics[width=\linewidth]{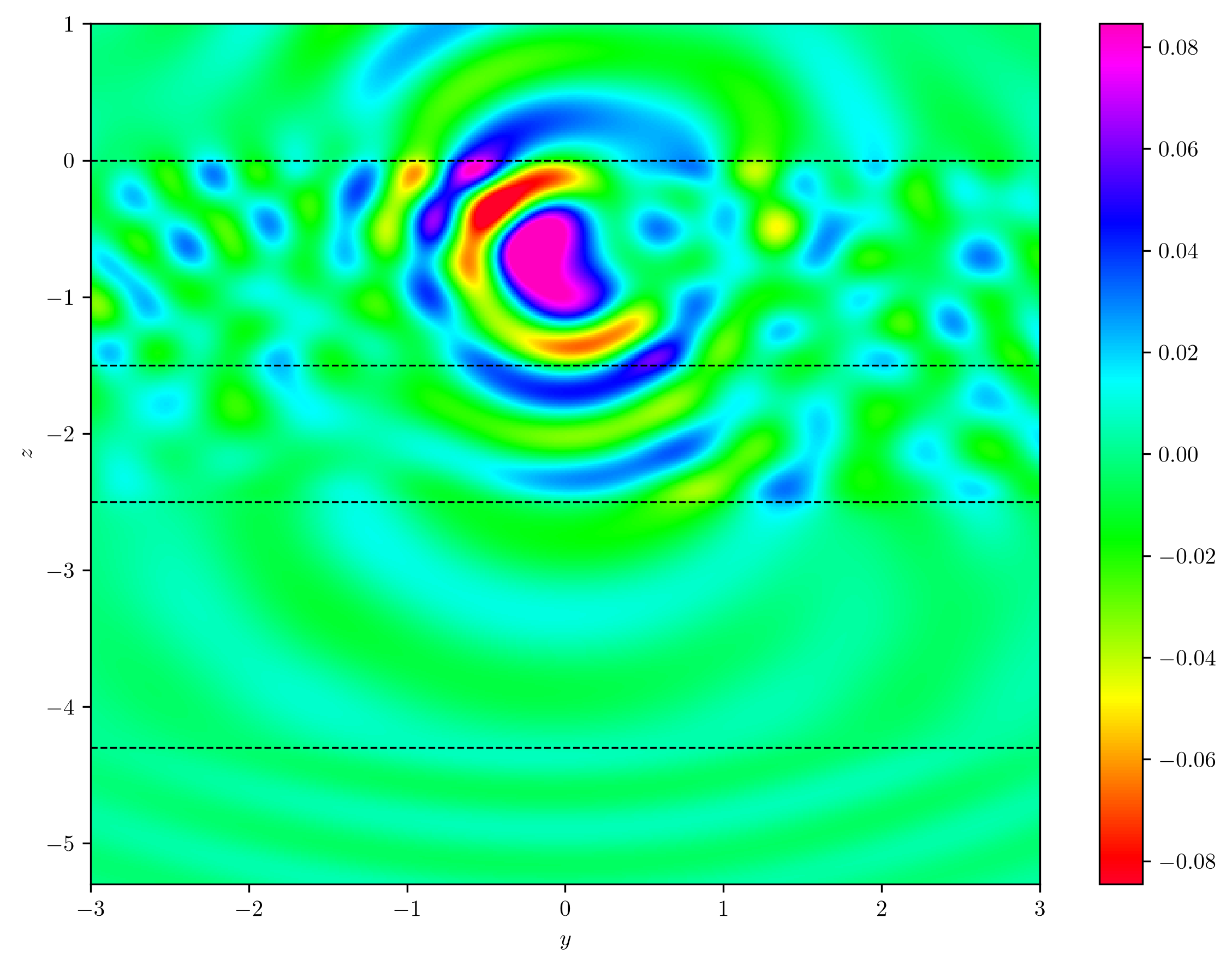}
			\caption{${\rm Im}(u_y)$}
		\end{subfigure}
		\quad
		\begin{subfigure}{0.31\textwidth}
			\includegraphics[width=\linewidth]{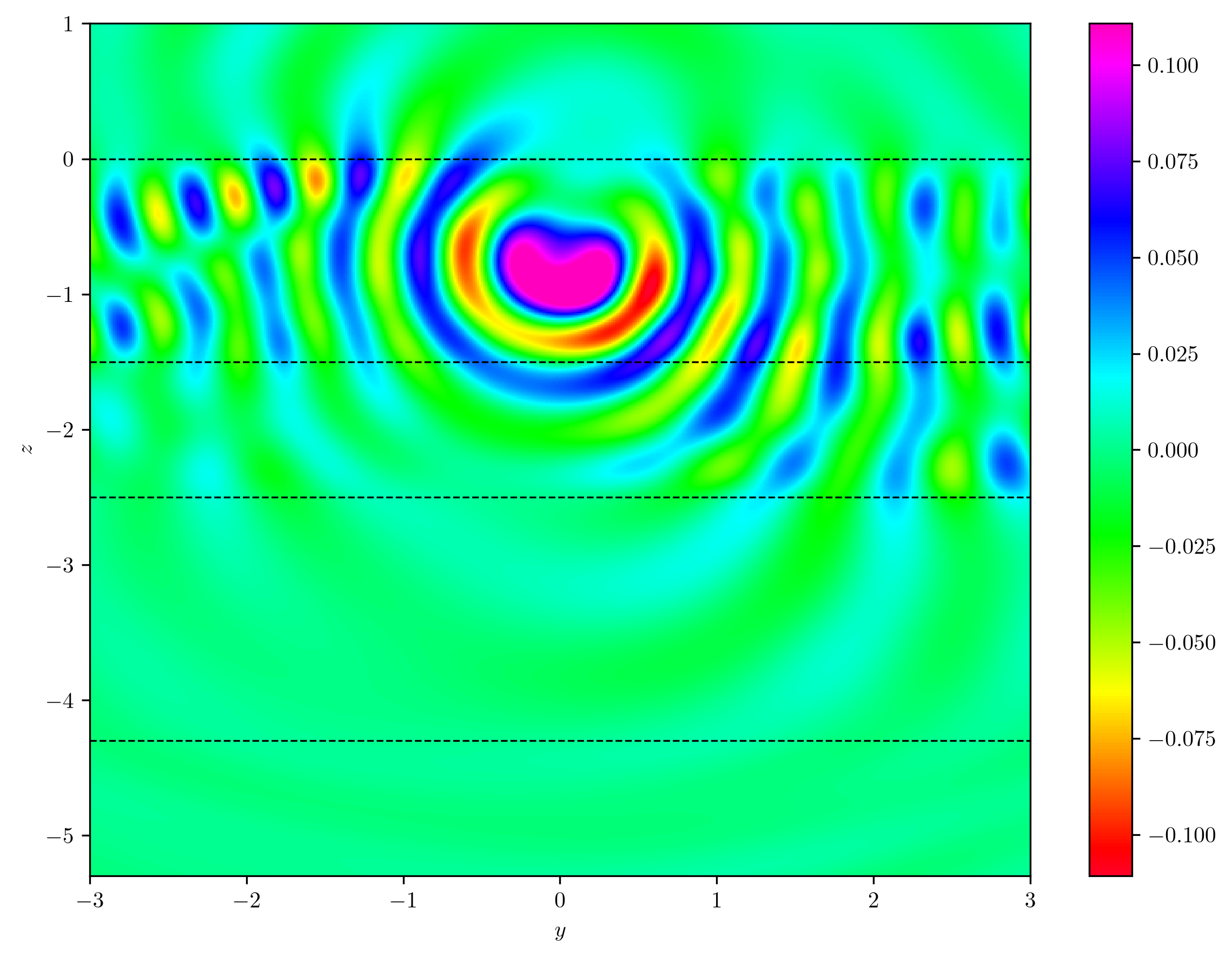}
			\caption{${\rm Im}(u_z)$}
		\end{subfigure}
		\caption{Real and imaginary parts of the displacement field generated by a point-force source located at $\bs r'=(0,0,-0.75)$ in the second layer.}
		\label{displacement_u}
	\end{figure}

	Unless otherwise specified, the numerical experiments are conducted in a five-layer medium with four interfaces located at
	\begin{equation*}
		\{d_{\ell}\}_{\ell=0}^{3}=\{0,-1.5,-2.5,-4.3\},
	\end{equation*}
	material parameters given in Table. \ref{tab:layer_params}, and the angular frequency is set to $\omega=10$. 
	In all the numerical experiments, the point source is located at $\bs r'=(0,0,-0.75)$ in layer $\ell'=1$ with strength given by
	$\bs\alpha'=\left(\frac{1}{2},\frac{1}{2},\frac{\sqrt{2}}{2}\right)^{T}$.
	The displacement field is given by
	\[
	\bs u(\bs r)=\mathbb G(\bs r,\bs r')\bs\alpha'.
	\]
	We calculate $\bs u(\bs r)$ on the plane $\{(x, y, z): x=0.5\}$ over the region $(y, z)\in [-3, 3]\times[-5.3, 1]$.
	The real and imaginary parts of the three displacement components are depicted 
	in Fig.~\ref{displacement_u}.

	\subsection{Convergence to the homogeneous-medium limit}
	
	If all layers have identical material parameters, the layered-medium
	dyadic Green's function reduces to the free-space dyadic Green's function
	of the corresponding homogeneous elastic medium. We use this limiting property as a numerical consistency test.
	
	In this test, we retain the geometry, angular frequency, and source location specified above, but consider a family of material parameters approaching those of a homogeneous reference medium.
	The reference parameters are set to $\rho_{\ell'}=1$, $\mu_{\ell'}=1.2$, and $\lambda_{\ell'}=1$. The mass density is fixed at $\rho_{\ell}=\rho_{\ell'}$ in all layers, while the Lam\'e parameters are perturbed from their reference values according to
	\begin{equation*}
		\lambda_{\ell}=\lambda_{\ell'}\left(1+\zeta_{\ell\ell'}\epsilon\right),
		\qquad
		\mu_{\ell}=\mu_{\ell'}\left(1-\zeta_{\ell\ell'}\epsilon\right),
		\qquad
		\ell=0,1,\ldots,L,
	\end{equation*}
	where
	\begin{equation}
		\zeta_{\ell\ell'}
		=1-\delta_{\ell\ell'}+\frac{\ell-\ell'}{L},
	\end{equation}
	$\delta_{\ell\ell'}$ is the Kronecker symbol. Therefore,
	\[
	\lambda_{\ell}\rightarrow\lambda_{\ell'},
	\qquad
	\mu_{\ell}\rightarrow\mu_{\ell'},
	\qquad
	\epsilon\rightarrow0,
	\]
	and the layered medium approaches a homogeneous medium.
	
	We take $\epsilon=10^{-4},10^{-5},10^{-6},10^{-7},10^{-8}$ and evaluate the Green tensor on the plane $\{(x, y, z): x=0.5\}$ over the region $(y, z)\in [-3, 3]\times[-5.3, 1]$ covered by a uniform $100\times100$ grid.
	For each target point $\bs r$, the relative error is measured in the
	Frobenius norm as
	\begin{equation*}
		E(\bs r):=\frac{\left\|\mathbb G(\bs r,\bs r')-\mathbb G^{f}(\bs r,\bs r')\right\|_{F}}{\left\|\mathbb G^{f}(\bs r,\bs r')\right\|_{F}},
	\end{equation*}
	where $\mathbb G^{f}$ denotes the free-space elastodynamic dyadic Green's
	function associated with the reference material. The maximum relative error
	within each target layer is reported in
	Table~\ref{tab:layered_free_convergence}. We can see clearly that the maximum relative
	error in every target layer decreases approximately linearly with $\epsilon$. It
	numerically confirm that the layered-medium dyadic Green's function
	converges to its free-space counterpart as the material contrasts between
	the layers vanish.
	\begin{table}[H]
		\centering
		\begin{tabular}{|c|c|c|c|c|c|}
			\hline
			\diagbox{Layer}{$\epsilon$}
			& $10^{-4}$ & $10^{-5}$
			& $10^{-6}$ & $10^{-7}$ & $10^{-8}$ \\[1.2ex]
			\hline
			$\ell=0$
			& $7.17\times10^{-4}$
			& $7.17\times10^{-5}$
			& $7.17\times10^{-6}$
			& $7.17\times10^{-7}$
			& $7.17\times10^{-8}$ \\
			\hline
			$\ell=1$
			& $2.52\times10^{-4}$
			& $2.52\times10^{-5}$
			& $2.52\times10^{-6}$
			& $2.52\times10^{-7}$
			& $2.52\times10^{-8}$ \\
			\hline
			$\ell=2$
			& $1.04\times10^{-3}$
			& $1.04\times10^{-4}$
			& $1.04\times10^{-5}$
			& $1.04\times10^{-6}$
			& $1.04\times10^{-7}$ \\
			\hline
			$\ell=3$
			& $2.17\times10^{-3}$
			& $2.17\times10^{-4}$
			& $2.17\times10^{-5}$
			& $2.17\times10^{-6}$
			& $2.17\times10^{-7}$ \\
			\hline
			$\ell=4$
			& $2.85\times10^{-3}$
			& $2.85\times10^{-4}$
			& $2.85\times10^{-5}$
			& $2.85\times10^{-6}$
			& $2.85\times10^{-7}$ \\
			\hline
		\end{tabular}
		\caption{Maximum relative errors between the layered-medium dyadic Green's function and the free-space dyadic Green's function in each layer of the five-layer medium.}
		\label{tab:layered_free_convergence}
	\end{table}

	\subsection{Verification of the interface transmission conditions}
	We next verify the transmission conditions \eqref{eq_interfacecond} satisfied by the elastodynamic dyadic Green's function across each material interface. 
	In order the calculate the traction tensor $\mathbb T=\left[\mathbf T^{\hat{\bs x}},\mathbf T^{\hat{\bs y}},\mathbf T^{\hat{\bs z}}\right]$, we adopt
	using second-order centered difference scheme:
	\begin{equation*}
		\begin{aligned}
			\frac{\partial f}{\partial x}&=\frac{f(x+h,y,z)-f(x-h,y,z)}{2h}+\mathcal O(h^2),\\
			\frac{\partial f}{\partial y}&=\frac{f(x,y+h,z)-f(x,y-h,z)}{2h}+\mathcal O(h^2).
		\end{aligned}
	\end{equation*}
	for the derivatives with respect to the tangential coordinates $x$ and $y$ and the second-order one-sided difference scheme:
	\begin{equation*}
		\begin{aligned}
			\frac{\partial f}{\partial z}&=\frac{-3f(x,y,z)+4f(x,y,z+h)-f(x,y,z+2h)}{2h}+\mathcal O(h^2),\qquad z\geq d_{\ell},\\
			\frac{\partial f}{\partial z}&=\frac{3f(x,y,z)-4f(x,y,z-h)+f(x,y,z-2h)}{2h}+\mathcal O(h^2),\qquad z\leq d_{\ell}.
		\end{aligned}
	\end{equation*}
	for the normal derivative on
	the two sides of the interfaces.
	
	For this test, we return to the settings given at the beginning of section 4. At each interface $z=d_{\ell}$, the Green tensors are evaluated on a uniform $10\times10$ grid over the horizontal domain $(x,y)\in [-3, 3]^2$.
	For the traction matrix, the finite-difference step sizes $h=10^{-3},10^{-4},10^{-5}$ are tested.
	For either $\mathbb K=\mathbb G$ or $\mathbb K=\mathbb T$, we define the relative jump error at an interface point by
	\begin{equation}
		E_{\mathbb K}(\bs r):=\frac{\left\|\mathbb K^{+}(\bs r)-\mathbb K^{-}(\bs r)\right\|_{F}}{\max\left\{\|\mathbb K^{+}(\bs r)\|_{F},\|\mathbb K^{-}(\bs r)\|_{F}\right\}},
		\qquad\mathbb K=\mathbb G,\mathbb T,
	\end{equation}
	where the superscripts $+$ and $-$ denote the one-sided limits from the
	upper and lower layers, respectively. The maximum relative jump errors on
	each interface are reported in Table~\ref{tab:interface_condition_test}.
	\begin{table}[H]
		\centering
		\begin{tabular}{|c|c|c|c|c|}
			\hline
			\multirow{2}{*}{Interface}
			& \multirow{2}{*}{$E_{\mathbb G}$}
			& \multicolumn{3}{c|}{$E_{\mathbb T}$} \\
			\cline{3-5}
			&
			& $h=10^{-3}$
			& $h=10^{-4}$
			& $h=10^{-5}$ \\
			\hline
			$d_0=0\;(\ell=0,1)$
			& $1.97\times10^{-10}$
			& $6.54\times10^{-5}$
			& $6.56\times10^{-7}$
			& $2.03\times10^{-8}$ \\
			\hline
			$d_1=-1.5\;(\ell=1,2)$
			& $6.44\times10^{-11}$
			& $3.78\times10^{-5}$
			& $3.79\times10^{-7}$
			& $2.32\times10^{-8}$ \\
			\hline
			$d_2=-2.5\;(\ell=2,3)$
			& $1.34\times10^{-13}$
			& $3.95\times10^{-5}$
			& $3.96\times10^{-7}$
			& $6.60\times10^{-9}$ \\
			\hline
			$d_3=-4.3\;(\ell=3,4)$
			& $2.06\times10^{-14}$
			& $3.35\times10^{-5}$
			& $4.67\times10^{-7}$
			& $3.32\times10^{-9}$ \\
			\hline
		\end{tabular}
		\caption{Maximum relative jump errors of the displacement
			Green matrix and the traction Green matrix across the four material
			interfaces.}
		\label{tab:interface_condition_test}
	\end{table}
	
	
	As shown in Table~\ref{tab:interface_condition_test}, the displacement Green matrix remains continuous across all four interfaces, with maximum relative jump errors less than $2\times10^{-10}$. For the traction Green matrix, the relative jump errors decrease quadratically with respect to the step size $h$. Although, the convergence rate becomes less uniform at the smaller step size $h=10^{-5}$, possibly due to the numerical stability of the numerical differentiation and round-off errors, the results provide a convincing numerical verification of the transmission conditions in \eqref{eq_interfacecond}.

	\section{Conclusion}\label{sect_conclusion}
	In this paper,  we develop an efficient and stable scheme for computing the dyadic Green’s function for time-harmonic elastic wave in layered media. Using the matrix basis proposed in \cite{bo2022maxwellDGF}, we decouple the SH mode from the coupled P–SV modes and reduce the problem to three independent Helmholtz-like problems involving a total of five unknown functions. For the two mixed-wavenumber Helmholtz problems, we propose a stable recursive algorithm based on the general reflection–transmission theory for layered media. The algorithm is stable in the whole   complex plane  and  scales linearly according   to the number of layers. The dyadic Green's function in the physical domain is then computed via Hankel transforms along appropriate Sommerfeld integration paths. The resulting integral representation directly enables the development of fast multipole methods for boundary integral formulations of elastic wave scattering in layered media.

	\newpage
	\begin{appendix}
		\section{Proof of Theorem \ref{thm_reactfield_matrixbasis}}\label{sect_proof_reactfield_matrixbasis}

		Firstly, we assume that the source point $\bd{r}'$ is not on the interface $z=d_{\ell},\ell=0,1,\cdots,L-1$. The case with a source on the interface should be considered  as  the limit of source approaching the interface.
		Based on the matrix form of the interface conditions \eqref{eq_interfacecont_T1} and the expressions \eqref{Elasticdynamics_layerDGf_matrix_react}, we calculate that
		\begin{equation*}
			\begin{aligned}
				\J{1}\widehat{\mathbb{G}}_{\ell\ell'} =&\left[\J{1}+(k^{s})^{-2}\left(\J{5}+\J{3}\partial_z\right)\right]\hat{{\bs\Phi}}_{\ell\ell'}-(k^{c})^{-2}\left(\J{5}+\J{3}\partial_z\right)\hat{{\bs\Psi}}_{\ell\ell'}\\
				\J{2}\widehat{\mathbb{G}}_{\ell\ell'}
				=&(k^{s})^{-2}\left[\J{2}k_{\rho}^2+\J{4}\partial_z\right]\hat{{\bs\Phi}}_{\ell\ell'}-(k^{c})^{-2}\left(\J{4}\partial_z-\J{2}(k_{z}^{s})^{2}\right)\hat{{\bs\Psi}}_{\ell\ell'},\\
				\mu\left(\J{1}\partial_z+\J{3}\right)\widehat{\mathbb{G}}_{\ell\ell'} 
				=&\mu\left[\J{1}\partial_z+\J{3}+2(k^{s})^{-2}\left(\J{5}\partial_z-(k_{z}^{s})^{2}\J{3}\right)\right]\hat{{\bs\Phi}}_{\ell\ell'}\\
				-&2\mu(k^{c})^{-2}\left[\J{5}\partial_z-(k_{z}^{c})^{2}\J{3}\right]\hat{{\bs\Psi}}_{\ell\ell'}\\
				\left(\gamma\J{2}\partial_z+\lambda\J{4}\right)\widehat{\mathbb{G}}_{\ell\ell'} 
				=&\left[\gamma\J{2}\partial_z+\lambda\J{4}-(k^{s})^{-2}(\gamma (k_{z}^{s})^{2}+\lambda k_{\rho}^2)\left(\J{4}+\J{2}\partial_z\right)\right]\hat{{\bs\Phi}}_{\ell\ell'}\\
				&+(k^{c})^{-2}(\gamma (k_{z}^{c})^{2}+\lambda k_{\rho}^2)\left[\J{4}+\J{2}\partial_z\right]\hat{{\bs\Psi}}_{\ell\ell'}
			\end{aligned}
		\end{equation*}
		Substituting into the expression \eqref{reactionLMGF} and using the result into the interface conditions \eqref{eq_interfacecont_T1} gives
		\begin{equation}\label{interface_forsolid-solidv1}
			\begin{aligned}
				\left\llbracket \left(\J{1}+(k^{s})^{-2}\left[\J{5}+\J{3}\partial_z\right]\right)\hat{\bs\Phi}-(k^{c})^{-2}\left(\J{5}+\J{3}\partial_z\right)\hat{\bs\Psi} \right\rrbracket=0,\\
				\left\llbracket (k^{s})^{-2}\left(\J{2}k_{\rho}^2+\J{4}\partial_z\right)\hat{\bs\Phi}-(k^{c})^{-2}\left(\J{4}\partial_z-\J{2}(k_{z}^{s})^{2}\right)\hat{\bs\Psi} \right\rrbracket=0,\\
				\left\llbracket\mu\left(\J{1}\partial_z+\J{3}+2(k^{s})^{-2}\left[\J{5}\partial_z-(k_{z}^{s})^{2}\J{3}\right]\right)\hat{\bs\Phi}-2\mu(k^{c})^{-2}\left(\J{5}\partial_z-(k_{z}^{c})^{2}\J{3}\right)\hat{\bs\Psi}\right\rrbracket=0,\\
				\left\llbracket \left(\gamma\J{2}\partial_z+\lambda\J{4}-\dfrac{\gamma (k_{z}^{s})^{2}+\lambda k_{\rho}^2}{(k^{s})^{2}}\left[\J{4}+\J{2}\partial_z\right]\right)\hat{\bs\Phi}+\dfrac{\gamma (k_{z}^{c})^{2}+\lambda k_{\rho}^2}{(k^{c})^{2}}\left(\J{4}+\J{2}\partial_z\right)\hat{\bs\Psi} \right\rrbracket=0,
			\end{aligned}
		\end{equation}
		on each interface $z=d_{\ell-1},\ell=1,\cdots,L$. Here, the definitions  \eqref{Xst_freeexpand} and \eqref{layeredDecomp} has been used to rearrange the resulted formulas.
		
		Denoted by 
		\begin{equation}
			\hat{h}_{\ell}(k_{\rho}):=\hat{e}_{\ell}^{\uparrow}(d_{\ell-1})=\hat{e}_{\ell}^{\downarrow}(d_{\ell})=e^{\ri k_{\ell z}^{s}D_{\ell}},\quad\tilde{h}_{\ell}(k_{\rho}):=\tilde{e}_{\ell}^{\uparrow}(d_{\ell-1})=\tilde{e}_{\ell}^{\downarrow}(d_{\ell})=e^{\ri k_{\ell z}^{c}D_{\ell}}
		\end{equation}
		then the system \eqref{interface_forsolid-solidv1} can be rewritten in the following matrix form
		\begin{equation}\label{interface_forsolid-solidv3}
			\begin{split}
				&\mathbb K_{\ell-1}^{(a)}\mathbb X_{\ell-1,\ell'}-\mathbb K_{\ell}^{(b)}\mathbb X_{\ell\ell'}=\mathbb B_{\ell},\quad\ell=1,\cdots,L,
			\end{split}
		\end{equation}
		where the interface coefficient matrices $\mathbb K_{\ell}^{(a)},\mathbb K_{\ell}^{(b)}$, the unknown coefficient matrix $\mathbb X_{\ell\ell'}$ and the free-space contribution $\mathbb B_{\ell}$ from the source layer are defined by 
		\begin{equation*}
			\begin{split}
				\mathbb K_{\ell}^{(a)}:=&\begin{bmatrix}
					\widehat{\mathbb{K}}_{\ell}^{\uparrow} & 
					\widetilde{\mathbb{K}}_{\ell}^{\uparrow} &
					\widehat{\mathbb{K}}_{\ell}^{\downarrow}\hat h_{\ell} & \widetilde{\mathbb{K}}_{\ell}^{\downarrow}\tilde h_{\ell}
				\end{bmatrix},\quad
				\mathbb K_{\ell}^{(b)}:=\begin{bmatrix}
					\widehat{\mathbb{K}}_{\ell}^{\uparrow}\hat h_{\ell} & 
					\widetilde{\mathbb{K}}_{\ell}^{\uparrow}\tilde h_{\ell} &
					\widehat{\mathbb{K}}_{\ell}^{\downarrow} & \widetilde{\mathbb{K}}_{\ell}^{\downarrow}
				\end{bmatrix},\\
				\mathbb X_{\ell\ell'}:=&\begin{bmatrix}
					\hat{{\bs\Phi}}_{\ell\ell'}^{\uparrow} &
					\hat{{\bs\Psi}}_{\ell\ell'}^{\uparrow} &
					\hat{{\bs\Phi}}_{\ell\ell'}^{\downarrow} &
					\hat{{\bs\Psi}}_{\ell\ell'}^{\downarrow}
				\end{bmatrix}^{\rm T},\quad
				\mathbb B_{\ell}=\delta_{\ell\ell'}\mathbb B_{\ell'}^{\downarrow}-\delta_{\ell-1,\ell'}\mathbb B_{\ell'}^{\uparrow},
			\end{split}
		\end{equation*}
		with the four matrices $\widehat{\mathbb{K}}_{\ell}^{\ast},\widetilde{\mathbb{K}}_{\ell}^{\ast},\ast=\uparrow,\downarrow$ explicitly given by
		\begin{equation*}
			\begin{aligned}
				\widehat{\mathbb{K}}_{\ell}^{\uparrow}:=&\begin{bmatrix}
					\ri\gamma_{\ell}k_{\ell z}^{s}\J{2}+\lambda_{\ell}\J{4}-\dfrac{\gamma_{\ell} (k_{\ell z}^{s})^{2}+\lambda_{\ell} k_{\rho}^2}{(k_{\ell}^{s})^{2}}\left(\J{4}+\ri k_{\ell z}^{s}\J{2}\right) \\
					(k_{\ell}^{s})^{-2}(\J{2}k_{\rho}^2+\ri k_{\ell z}^{s}\J{4})\\
					\mu_{\ell}\left(\ri k_{\ell z}^{s}\J{1}+\J{3}+2(k_{\ell}^{s})^{-2}k_{\ell z}^{s}\left[\ri\J{5}-k_{\ell z}^{s}\J{3}\right]\right)\\
					\J{1}+(k_{\ell}^{s})^{-2}\left(\J{5}+\ri k_{\ell z}^{s}\J{3}\right) 
				\end{bmatrix},\\
				\widehat{\mathbb{K}}_{\ell}^{\downarrow}:=&\begin{bmatrix} -\ri\gamma_{\ell}k_{\ell z}^{s}\J{2}+\lambda_{\ell}\J{4}-\dfrac{\gamma_{\ell} (k_{\ell z}^{s})^{2}+\lambda_{\ell} k_{\rho}^2}{(k_{\ell}^{s})^{2}}\left(\J{4}-\ri k_{\ell z}^{s}\J{2}\right)\\ 
					(k_{\ell}^{s})^{-2}(\J{2}k_{\rho}^2-\ri k_{\ell z}^{s}\J{4})\\ 
					\mu_{\ell}\left(-\ri k_{\ell z}^{s}\J{1}+\J{3}-2(k_{\ell}^{s})^{-2}k_{\ell z}^{s}\left[\ri\J{5}+k_{\ell z}^{s}\J{3}\right]\right)\\
					\J{1}+(k_{\ell}^{s})^{-2}\left(\J{5}-\ri k_{\ell z}^{s}\J{3}\right)
				\end{bmatrix},\\
				\widetilde{\mathbb{K}}_{\ell}^{\uparrow}:=&\begin{bmatrix}
					\dfrac{\gamma_{\ell} (k_{\ell z}^{c})^{2}+\lambda_{\ell} k_{\rho}^2}{(k_{\ell}^{c})^{2}}(\J{4}+\ri k_{\ell z}^{c}\J{2}) \\
					-(k_{\ell}^{c})^{-2}k_{\ell z}^{c}\left(\ri\J{4}-k_{\ell z}^{c}\J{2}\right) \\
					-2\mu_{\ell}(k_{\ell}^{c})^{-2}k_{\ell z}^{c}(\ri\J{5}-k_{\ell z}^{c}\J{3}) \\
					-(k_{\ell}^{c})^{-2}\left(\J{5}+\ri k_{\ell z}^{c}\J{3}\right) 
				\end{bmatrix},\quad\widetilde{\mathbb{K}}_{\ell}^{\downarrow}:=\begin{bmatrix} 
					\dfrac{\gamma_{\ell} (k_{\ell z}^{c})^{2}+\lambda_{\ell} k_{\rho}^2}{(k_{\ell}^{c})^{2}}(\J{4}-\ri k_{\ell z}^{c}\J{2})\\ 
					(k_{\ell}^{c})^{-2}k_{\ell z}^{c}\left(\ri\J{4}+k_{\ell z}^{c}\J{2}\right)\\ 
					2\mu_{\ell}(k_{\ell}^{c})^{-2}k_{\ell z}^{c}(\ri\J{5}+k_{\ell z}^{c}\J{3})\\ 
					-(k_{\ell}^{c})^{-2}\left(\J{5}-\ri k_{\ell z}^{c}\J{3}\right)
				\end{bmatrix},
			\end{aligned}
		\end{equation*}
		and the source matrices $\mathbb{B}_{\ell'}^{\ast},\ast=\uparrow,\downarrow$ explicitly given by
		\begin{equation*}\label{righthandmatrix}
			\begin{aligned}
				\mathbb{B}_{\ell'}^{\uparrow}:=&\dfrac{1}{\mu_{\ell'}}\widehat{G}_{k_{\ell'}^{s}}^f(k_{\rho},d_{\ell'-1},z')\widehat{\mathbb{K}}_{\ell'}^{\uparrow}+ \dfrac{1}{\gamma_{\ell'}}\widehat{G}_{k_{\ell'}^{c}}^f(k_{\rho},d_{\ell'-1},z')\widetilde{\mathbb{K}}_{\ell'}^{\uparrow},\\
				\mathbb{B}_{\ell'}^{\downarrow}:=&\dfrac{1}{\mu_{\ell'}}\widehat{G}_{k_{\ell'}^{s}}^f(k_{\rho},d_{\ell'},z')\widehat{\mathbb{K}}_{\ell'}^{\downarrow}+ \dfrac{1}{\gamma_{\ell'}}\widehat{G}_{k_{\ell'}^{c}}^f(k_{\rho},d_{\ell'},z')\widetilde{\mathbb{K}}_{\ell'}^{\downarrow}.
			\end{aligned}
		\end{equation*}

		In order to solve \eqref{interface_forsolid-solidv3}, we shall use the matrix basis introduced in \cite{bo2022maxwellDGF}. In addition to the 5 basis matrices given in \eqref{mat-J}, the other 4 basis matrices are given as follows
		\begin{equation*}\label{mat-J_rem}
			\begin{array}{llll}
				\J{6}  = \begin{bmatrix}
					0 & 0 & 0 \\
					0 & 0 & 0 \\
					-\ri k_y & \ri k_x & 0
				\end{bmatrix}, &
				\J{7}  = \begin{bmatrix}
					0 & 0 & \ri k_y \\
					0 & 0 & -\ri k_x \\
					0 & 0 & 0
				\end{bmatrix}, &
				\J{8}  = \begin{bmatrix}
					k_x k_y & k_y^2 & 0 \\
					-k_x^2 & -k_x k_y & 0 \\
					0 & 0 & 0
				\end{bmatrix}, &
				\J{9}  = \begin{bmatrix}
					0 & 1 & 0 \\
					-1 & 0 & 0 \\
					0 & 0 & 0
				\end{bmatrix}.
			\end{array}
		\end{equation*}
		Due to the completeness of the matrix basis, the solution $\hat{{\bs\Phi}}_{\ell\ell'}^{\ast},\hat{{\bs\Psi}}_{\ell\ell'}^{\ast},\ast=\uparrow,\downarrow$ can be represented by the matrix basis $\J{1},\J{2},\cdots,\J{9}$, i.e.,
		\begin{equation}\label{react_field_matrixbasis_coeffv1}
			\hat{{\bs\Phi}}_{\ell\ell'}^{\ast}=\sum_{j=1}^{9}\alpha_{\ell\ell',j}^{\ast}(k_x,k_y,z')\J{j},\quad \hat{{\bs\Psi}}_{\ell\ell'}^{\ast}=\sum_{j=1}^{9}\beta_{\ell\ell',j}^{\ast}(k_x,k_y,z')\J{j},
		\end{equation}
		where $\{\alpha_{\ell\ell',j}^{\ast}(k_x,k_y,z'), \beta_{\ell\ell',j}^{\ast}(k_x,k_y,z')\}_{j=1}^9$ are unknown coefficients.
		Substituting \eqref{react_field_matrixbasis_coeffv1} into \eqref{interface_forsolid-solidv3} gives
		\begin{equation}\label{interface_forsolid-solidv4}
			\sum_{i=1}^{9}\left[\mathbb K_{\ell-1}^{(a)}\mathbb X_{\ell-1,\ell', i}-\mathbb K_{\ell}^{(b)}\mathbb X_{\ell\ell', i}\right]\J{i}=\mathbb B_{\ell},\quad\ell=1,\cdots,L,
		\end{equation}
		where the unknown expansion coefficient matrix $\mathbb X_{\ell\ell',i}$ is defined by 
		\begin{equation*}
			\begin{split}
				\mathbb X_{\ell\ell', i}:=&\begin{bmatrix}
					\alpha_{\ell\ell',i}^{\uparrow}\bb{I}&
					\beta_{\ell\ell',i}^{\uparrow}\bb{I}&
					\alpha_{\ell\ell',i}^{\downarrow}\bb{I}&
					\beta_{\ell\ell',i}^{\downarrow}\bb{I}
				\end{bmatrix}^{\rm T}.
			\end{split}
		\end{equation*}
		
		Let $\mathbb{F}$ be the field consists of all functions of $k_x$ and $k_y$. 
		The matrix function space $\mathcal L(\mathbb{F}):={\rm span}_{\mathbb{F}}\{\J{1}, \J{2},\cdots, \J{9}\}$ has orthogonal decomposition $\mathcal L=\mathcal R\oplus \mathcal J$ where
		\begin{equation}
			\mathcal{R}(\mathbb{F}):={\rm span}_{\mathbb{F}}\left\{\J{1},\J{2},\J{3},\J{4},\J{5}\right\},\quad 
			\mathcal{J}(\mathbb{F}):={\rm span}_{\mathbb{F}}\left\{\J{6},\J{7},\J{8},\J{9}\right\}.
		\end{equation}
		We define matrix function spaces $[\mathcal{S}(\mathbb{F})]^4, \mathcal{S}=\mathcal{L},\mathcal{R},\mathcal{J}$ whose elements are $12\times 3$ matrix functions such that their four $3\times 3$ square blocks belong to $\mathcal{S}(\mathbb F)$. 
		Obviously, we have 
		\begin{equation}\label{orthogonaldecompose}
			[\mathcal{L}(\mathbb{F})]^4=[\mathcal{R}(\mathbb{F})]^4\oplus[\mathcal{J}(\mathbb{F})]^4.
		\end{equation}

		Note that $\mathbb{B}_{\ell'}^{\uparrow},\mathbb{B}_{\ell'}^{\downarrow}\in[\mathcal{R}(\mathbb{F})]^4$, $\widehat{\mathbb{K}}_{\ell}^{\uparrow},\widetilde{\mathbb{K}}_{\ell}^{\uparrow},\widehat{\mathbb{K}}_{\ell}^{\downarrow},\widetilde{\mathbb{K}}_{\ell}^{\downarrow}\in[\mathcal{R}(\mathbb{F})]^4$. We are able to decompose the summation on the left side of \eqref{interface_forsolid-solidv4} into two parts: the summation from $i=1$ to 5 and from 6 to 9. By the product table (cf. \cite{bo2022maxwellDGF})
		\begin{equation}\label{producttable}
			\begin{array}{c|ccccc|cccc}
				\times &\J{1} & \J{2} & \J{3} & \J{4} & \J{5} & \J{6}& \J{7} & \J{8} & \J{9}\\[7pt]
				\hline
				\J{1} &\J{1} & \bd{0} & \J{3} & \bd{0} & \J{5} & \bd{0} & \J{7} & \J{8} & \J{9} \\[7pt]
				\J{2} &\bd{0} & \J{2} & \bd{0} & \J{4} & \bd{0} & \J{6} & \bd{0} & \bd{0} & \bd{0} \\[7pt]
				\J{3} &\bd{0} & \J{3} & \bd{0} & \J{5} & \bd{0} & \J{8}-\kr^2 \J{9} & \bd{0} & \bd{0} & \bd{0} \\[7pt]
				\J{4} &\J{4} & \bd{0} & -\kr^2\J{2} & \bd{0} & -\kr^2\J{4} & \bd{0} & \bd{0} & \bd{0} & \J{6} \\[7pt]
				\J{5} &\J{5} & \bd{0} & -\kr^2\J{3} & \bd{0} & -\kr^2\J{5} & \bd{0} & \bd{0} & \bd{0} & \J{8}-\kr^2 \J{9} \\[7pt] 
				\hline
				\J{6} &\J{6} & \bd{0} & \bd{0} & \bd{0} & \bd{0} & \bd{0} & \kr^2 \J{2} & -\kr^2 \J{4} & -\J{4} \\[7pt]
				\J{7} &\bd{0} & \J{7} & \bd{0} & -\J{8} & \bd{0} & \kr^2 \J{1}+\J{5} & \bd{0} & \bd{0} & \bd{0} \\[7pt]
				\J{8} &\J{8} & \bd{0} & \kr^2\J{7} & \bd{0} & -\kr^2\J{8} & \bd{0} & \bd{0} & \bd{0} & -\kr^2 \J{1}-\J{5} \\[7pt]
				\J{9} &\J{9} & \bd{0} & \J{7} & \bd{0} & -\J{8} & \bd{0} & -\J{3} & \J{5} & -\J{1}
			\end{array}
		\end{equation}
		we can check that 
		\begin{equation}
			\sum_{i=1}^{5}\mathbb K_{\ell}^{(f)}\mathbb X_{\ell\ell', i}\in[\mathcal{R}(\mathbb{F})]^4,\quad \sum_{i=6}^{9}\mathbb K_{\ell\ell'}^{(f)}\mathbb X_{\ell\ell', i}\in[\mathcal{J}(\mathbb{F})]^4,\quad f=a,b.
		\end{equation}
		Together with the decomposition \eqref{orthogonaldecompose}, \eqref{interface_forsolid-solidv4} can be decoupled into the following two independent systems:
		\begin{equation}\label{interface_forsolid-solidv3_J1to5}
			\sum_{i=1}^{5}\left[\mathbb K_{\ell-1}^{(a)}\mathbb X_{\ell-1,\ell', i}-\mathbb K_{\ell}^{(b)}\mathbb X_{\ell\ell', i}\right]\J{i}=\mathbb B_{\ell},\quad \sum_{i=6}^{9}\left[\mathbb K_{\ell-1}^{(a)}\mathbb X_{\ell-1,\ell', i}-\mathbb K_{\ell}^{(b)}\mathbb X_{\ell\ell', i}\right]\J{i}=\bs 0.
		\end{equation}
		for $\ell=1,\cdots,L$.
		Suppose that we have obtained a solution to the system \eqref{interface_forsolid-solidv4} given by the form \eqref{react_field_matrixbasis_coeffv1}. Let $\alpha_{\ell\ell',i}^{\ast}=\beta_{\ell\ell',i}^{\ast}=0,i=6,7,8,9,\ast=\uparrow,\downarrow,\ell=0,1,\cdots,L$ in \eqref{react_field_matrixbasis_coeffv1}, which obviously satisfies the system of equations \eqref{interface_forsolid-solidv3_J1to5}, that is, it is still the solution to the system of equations \eqref{interface_forsolid-solidv4}.
		Therefore, the unknown coefficient matrices $\hat{{\bs\Phi}}_{\ell\ell'}^{\ast},\hat{{\bs\Psi}}_{\ell\ell'}^{\ast},\ast=\uparrow,\downarrow$ in the reaction field expansion \eqref{react_field_expand_ud} can be represented by the matrix basis \eqref{mat-J}, which completes the proof for Theorem \ref{thm_reactfield_matrixbasis}.
		
	\end{appendix}

	\bibliographystyle{plain}    
	\bibliography{reference/reference}           

\end{document}